\documentclass[3p,10cpt]{elsarticle}

\journal{Journal of Computational Physics}

\usepackage{graphicx}
\usepackage{subcaption}
\usepackage{multirow}
\usepackage{hhline}
\usepackage{soul}
\usepackage{color}
\usepackage{enumerate}
\usepackage{amsmath}
\usepackage{amssymb}
\usepackage[section]{placeins}
\usepackage{textcomp}
\usepackage{fancyhdr}
\usepackage{hyperref}
\hypersetup{
    colorlinks = true
    }

\usepackage{amsthm}
\newtheorem{remark}{Remark}

\hypersetup{pdfauthor={Ashley Scillitoe}}

\graphicspath{{Figs/}}
\usepackage{bm}

  \def\clap#1{\hbox to 0pt{\hss#1\hss}}

\providecommand{\mat}[1]{\bm{#1}}%
\renewcommand{\vec}[1]{\mathbf{#1}}

\providecommand{\mV}{\ensuremath{\mat{V}}}

\providecommand{\mX}{\ensuremath{\mat{X}}}

\providecommand{\vx}{\ensuremath{\vec{x}}}
\providecommand{\vy}{\ensuremath{\vec{y}}}

\newcommand{\av}[1]{\left\langle{#1}\right\rangle}
\newcommand{\diff}[2]{\frac{\partial #1}{\partial #2}}

\newbox\tempbox

\makeatletter
\def\ps@pprintTitle{%
    \def\@oddhead{\hfill \small Copyright \textcopyright\ 2021. This manuscript version is made available under the \href{http://www.creativecommons.org/licenses/by-nc-nd/4.0/}{CC-BY-NC-ND 4.0 license.} \hfill}
    \let\@evenhead\@empty
    \def\@oddfoot{\footnotesize\itshape
         {Manuscript accepted for publication in Journal of Computational Physics, doi: \href{https://doi.org/10.1016/j.jcp.2021.110116}{10.1016/j.jcp.2021.110116}.} \hfill}%
    \let\@evenfoot\@oddfoot
    }
\makeatother

\begin{document}

\begin{frontmatter}

\title{Uncertainty Quantification for Data-driven Turbulence Modelling with Mondrian Forests}

\author[turing]{Ashley Scillitoe\corref{mycorrespondingauthor}}
\ead{ascillitoe@turing.ac.uk}

\author[turing,imperial]{Pranay Seshadri}
\author[turing,cambridge]{Mark Girolami}

\address[turing]{Data-Centric Engineering, The Alan Turing Institute, London, UK}
\address[imperial]{Department of Mathematics, Imperial College London, London, UK}
\address[cambridge]{Department of Engineering, University of Cambridge, Cambridge, UK}

\cortext[mycorrespondingauthor]{Corresponding author}

\begin{abstract}
Data-driven turbulence modelling approaches are gaining increasing interest from the CFD community. Such approaches generally aim to improve the modelled Reynolds stresses by leveraging data from high fidelity turbulence resolving simulations. However, the introduction of a machine learning (ML) model introduces a new source of uncertainty, the ML model itself. Quantification of this uncertainty is essential since the predictive capability of a data-driven model diminishes when predicting physics not seen during training. In this work, we explore the suitability of Mondrian forests (MF's) for data-driven turbulence modelling. MF's are claimed to possess many of the advantages of the commonly used random forest (RF) machine learning algorithm, whilst offering principled uncertainty estimates. An example test case is constructed, with a turbulence anisotropy constant derived from  high fidelity turbulence resolving simulations. A number of flows at several Reynolds numbers are used for training and testing. MF predictions are found to be superior to those obtained from a linear and non-linear eddy viscosity model. Shapley values, borrowed from game theory, are used to interpret the MF predictions. Predictive uncertainty is found to be large in regions where the training data is not representative. Additionally, the MF predictive uncertainty is found to exhibit stronger correlation with predictive errors compared to an a priori statistical distance measure, which indicates it is a better measure of prediction confidence. The MF predictive uncertainty is also found to be better calibrated and less computationally costly than the uncertainty estimated from applying jackknifing to random forest predictions. Finally, Mondrian forests are used to predict the Reynolds discrepancies in a convergent-divergent channel, which are subsequently propagated through a modified CFD solver. The resulting flowfield predictions are in close agreement with the high fidelity data. A procedure for sampling the Mondrian forests' uncertainties is introduced. Propagating these samples enables quantification of the uncertainty in quantities of interest such as velocity or a drag coefficient, due to the uncertainty in the Mondrian forests' predictions. This work suggests that uncertainty quantification can be incorporated into existing data-driven turbulence modelling frameworks by replacing random forests with Mondrian forests. This would also open up the possibility of online learning, whereby new training data could be added without having to retrain the Mondrian forests. 
\end{abstract}


\end{frontmatter}

\pagestyle{fancy}
\chead{\small Copyright \textcopyright\ 2021. This manuscript version is made available under the \href{http://www.creativecommons.org/licenses/by-nc-nd/4.0/}{CC-BY-NC-ND 4.0 license.}}
\lhead{}
\rhead{}
\cfoot{\thepage} 
\renewcommand{\headrulewidth}{0pt}
\renewcommand{\footrulewidth}{0pt}


\section{Introduction}
\label{sec:intro}

Turbulence is a key characteristic of fluid flows across many different industries. For aircraft, delaying the transition to turbulence over the wing surfaces can reduce drag \cite{Wild2015}, while large turbulent structures can increase the effectiveness of cooling systems in aircraft engines \cite{Tucker2014a}, both of which lead to reduced fuel consumption. For wind turbines, turbulence can increase the power output, at the expense of increased fatigue loading \cite{Brand2011}. In oil refineries, turbulence plays a key role in critical processes such as fluid catalytic cracking \cite{Raynal2016}. These examples, as well as many others, indicate the importance of being able to accurately predict the effects of turbulence for a range of industrial flows.

The characterisation of turbulence is challenging due to its chaotic nature, and the broad range of spatio-temporal scales involved. The continued growth of computing power has enabled the unsteady Navier-Stokes equations to be computed directly, so that all the scales of turbulence are simulated. Unfortunately, the extreme computational cost of these direct numerical simulations (DNS) limits their application to relatively simple flows. To mitigate the high computational cost, large eddy simulation (LES) techniques only simulate the larger energy containing turbulent eddies, whilst the smaller unresolved turbulent scales are modelled. LES type approaches are gaining popularity in many industrial applications, for example for turbo-machinery \cite{Scillitoe2019}, wind turbines \cite{Mehta2014}, and urban flows \cite{Blocken2018}. However, due to their high cost, such simulations are often simplified representations of real industrial flows. As discussed by \citet{Tucker2014a}, even with the ever-increasing computing power available, LES methods are unlikely to be routinely used in the design process of complex engineering systems over the next decade and beyond.

Instead, computations based on Reynolds-averaged Navier–Stokes (RANS) models are still the workhorse for predicting turbulent flows in many industries. RANS computations are orders of magnitude cheaper than LES techniques. However, RANS models are known to perform poorly in many flows of engineering relevance, including those with swirl, pressure gradients, and streamline curvature \cite{Hunt2005,Tucker2013}. Many researchers have attempted to use LES or DNS to better understand the physics of turbulence in various flows, with the aim of developing better turbulence models. As summarised in the recent review paper by \citet{Duraisamy2019}, data-driven turbulence modelling is emerging as a promising way to inform turbulence models with data in a more systematic way.

Many of the data-driven turbulence modelling strategies fall under what is known as supervised machine learning (ML). This involves learning a function based on \emph{training data} consisting of a set of input-output pairs (see Sec.~\ref{sec:supervised_ml}). The learned function can then be used to make predictions on the \emph{test data}. \citet{Ling2015} were one of the first to apply ML to turbulence modelling, using a random forest classifier to predict when RANS modelling assumptions would fail. \citet{Ling2016} further used neural networks to predict Reynolds stress anisotropy. \citet{Singh2017} used field inversion to determine functional discrepancies in existing RANS models, which are then reconstructed as functions of local flow features using the Adaboost ML algorithm. \citet{Wu2018} investigated the use of random forests to predict the discrepancies of RANS modelled Reynolds stresses in separated flows. \citet{Kaandorp2020} pursue similar lines, but they modify the random forest algorithm to accept a tensor basis in order to guarantee Galilean invariance. 

These studies demonstrate the growing interest in applying ML techniques to turbulence modelling. However, for such methods to be employed in industry more trust is needed in the ML predictions. All ML algorithms have an inductive bias, which is the assumptions the learner uses to predict outputs given inputs that it has not encountered. The \emph{no free lunch} theorem \cite{Wolpert1996} states that if a bias is correct on some cases, it must be incorrect on equally many cases, thus in the context of data-driven turbulence modelling it is unlikely for one learner to be able to generalise to all unseen flows. This is evidenced in \cite{Ling2015} and \cite{Wu2018}, where the authors report that their data-driven closures performed poorly on flows that were significantly different from the ones on which they were trained. It is desirable for the learning algorithm to be able to generalise well in situations with a significant \emph{dataset shift}, where the test and training data have significantly different distributions. Yet, a more realistic and equally important requirement is that the algorithm can at least quantify the uncertainty in its own predictions. Then the user can at least know when the training data is suitable, and whether the ML predictions can be trusted. This is a pertinent issue for the field of data-driven turbulence modelling, since many of the available LES and DNS datasets are on simplified geometries at low Reynolds numbers\footnote{As noted in \cite{Tucker2013}, the grid resolution required for DNS and LES increases dramatically with Reynolds number.}. But, we wish to know whether we trust ML models trained on these flows to make predictions on the more complex higher Reynolds number flows seen in industry. 
 
The random forest algorithm is commonly used for turbulence modelling applications \cite{Ling2015,Wu2018,Kaandorp2020} for its robustness and high predictive accuracy. Jackknife re-sampling can be used to provide uncertainty estimates for random forest predictions \cite{Wager2014}. However, these estimates do not correctly account for the uncertainty arising from differences between the training and test data. \citet{Ling2015} and later \citet{Wu2017} proposed the use of statistical distance metrics to measure co-variate shift between the test and training datasets. Although informative, these metrics don't provide uncertainty estimates for the ML predictions. A number of authors have explored the use of ML techniques with inbuilt uncertainty estimates. \citet{Parish2016} used Gaussian processes (GPs) to predict the turbulence intermittency and correction terms for the turbulence transport equations. To predict the turbulence anisotropy tensor \citet{Geneva2019} used Bayesian neural networks (BNN), while \citet{blauw2019} used Bayesian additive regression trees (BART). All three approaches show promise, but also have disadvantages. GPs are known to deliver high quality uncertainty estimates, however they can be challenging to scale to large datasets since their cost scales cubically with dataset size \cite{Hensman2013}. Meanwhile, neural networks can be challenging and costly to train, and Bayesian neural networks are particularly computationally expensive \cite{Lakshminarayanan2017}. For BART, Metropolis-Hastings type samplers commonly used to perform inference can struggle with large high dimensional datasets \cite{Lakshminarayanan2015}.

An alternative supervised ML technique is the Mondrian forest (MF), first conceived for classification \cite{Lakshminarayanan2016a}, and extended to regression by \citet{Lakshminarayanan2016b}. This relatively new technique was demonstrated to be competitive in computational cost and accuracy to popular tree ensemble approaches such as random forests. Unlike random forests, the trees in a Mondrian forest are probabilistic; they use a hierarchical Gaussian prior, and the posterior parameters are efficiently computed using Gaussian belief propagation. Lakshminarayanan et al.~claim that this endows MF's with principled uncertainty estimates, more akin to those of a GP, with predictions smoothly returning to the prior and exhibiting higher uncertainty far from training data. To the authors' knowledge the use of Mondrian forests for data-driven turbulence modelling is yet to be explored. The present paper aims to explore this, to understand whether Mondrian forests offer useful uncertainty estimates in addition to accurate predictions in data-driven turbulence frameworks. The paper is structured as follows: In Section~\ref{sec:supervised_ml}, commonly used frameworks of supervised ML for turbulence modelling are introduced; the Mondrian forest algorithm is then introduced in Section~\ref{sec:methods}, along with the procedures for generating training data and for propagating ML predictions through a CFD solver. In Section~\ref{sec:MF_vs_RF} random and Mondrian forests are used to predict a turbulence field variable, in order to understand their relative performances in more detail. Finally, in Section~\ref{sec:DDRANS} an example of the use of Mondrian forests for a data-driven turbulence modelling task is explored.

\section{Supervised machine learning for turbulence modelling}
\label{sec:supervised_ml}

\subsection{RANS modelling uncertainties}

The difficulty of RANS modelling arises from the fundamental closure problem that is introduced when the Navier–Stokes equations are averaged with respect to time. The incompressible RANS momentum equation in the $i^{th}$ direction is

\begin{equation} \label{eqn:NS}
\av{u_j}\diff{\av{u_i}}{z_j}=\diff{}{z_j} \left[-\frac{\av{p}}{\rho}\delta_{ij} + \nu \left(\diff{\av{u_i}}{z_j} + \diff{\av{u_j}}{z_i} \right) + \frac{\tau_{ij}}{\rho} \right] + \av{g_i},
\end{equation}

where $\av{\cdot}$ indicates a time-averaged quantity and $\boldsymbol{z}=(z_1,z_2,z_3)$ is the spatial coordinate vector. The RANS equations are unclosed, with the turbulent (or Reynolds) stress term, $\tau_{ij}$, unknown. It is common for the Boussinesq linear eddy-viscosity hypothesis to be used

\begin{equation} \label{eqn:evm}
\tau_{ij} = - \rho \av{u'_iu'_j} =  \nu_t \left( \diff{\av{u_i}}{z_j} + \diff{\av{u_j}}{z_i} \right) - \frac{2}{3}k\delta_{ij},
\end{equation}

where $k = \av{u'_iu'_i}/2$ is the turbulent kinetic energy. However, the turbulent eddy viscosity $\nu_t$ is still unknown, and RANS models must be used to estimate this term. \citet{Xiao2019} divide the uncertainties arising from RANS modelling into two categories: (i) structural (or model-form) uncertainties, which arise from the structure of the RANS equations solved to provide $\nu_t$, or from the eddy-viscosity assumption in \eqref{eqn:evm}, and (ii) parametric uncertainties, which are caused by a lack of generalisation of closure coefficients in the RANS model. As discussed in the following section, machine learning can be used to inform both categories of uncertainties.

\subsection{Machine learning to inform RANS}
\label{sub:ML_RANS_review}

This paper deals only with the general framework of supervised machine learning. In this context, we define $\vx \in \mathbb{R}^{d}$ to be a vector of $d$ non-dimensional flow-features obtained at a given location in a flow-field. These flow features, described further in Section~\ref{sub:preproc}, are intended to uniquely describe the RANS flow-field. Let $y \in \mathbb{R}$ be a scalar-valued output quantity of interest also obtained at each location within the flow-field. Assuming a flow-field is specified by $N$ spatial locations, the following matrix-vector notation is adopted to characterise the data
\begin{equation}
\mX_{N}=\left[\begin{array}{c}
\vx_{1}^{T}\\
\vdots\\
\vx_{N}^{T}
\end{array}\right], \; \; \; \; \vy_{N}=\left[\begin{array}{c}
y_{1}\\
\vdots\\
y_{N}
\end{array}\right],
\end{equation}
where the subscript denotes the $N$ \emph{training samples}. Analogously, predictions of the spatially-varying scalar quantity of interest for $M$ different feature vectors are given by
\begin{equation}
\tilde{\mX}_{M}=\left[\begin{array}{c}
\tilde{\vx}^{T}_{1}\\
\vdots\\
\tilde{\vx}_{M}^{T}
\end{array}\right], \; \; \; \;  \tilde{\vy}_{M}=\left[\begin{array}{c}
\tilde{y}_{1}\\
\vdots\\
\tilde{y}_{M}
\end{array}\right],
\end{equation}
where the superscript $\tilde{\cdot}$ indicates \emph{testing samples}. These definitions will be important for the supervised ML exposition.

\begin{figure}[ht]
  \centering
  \includegraphics[width=0.85\textwidth]{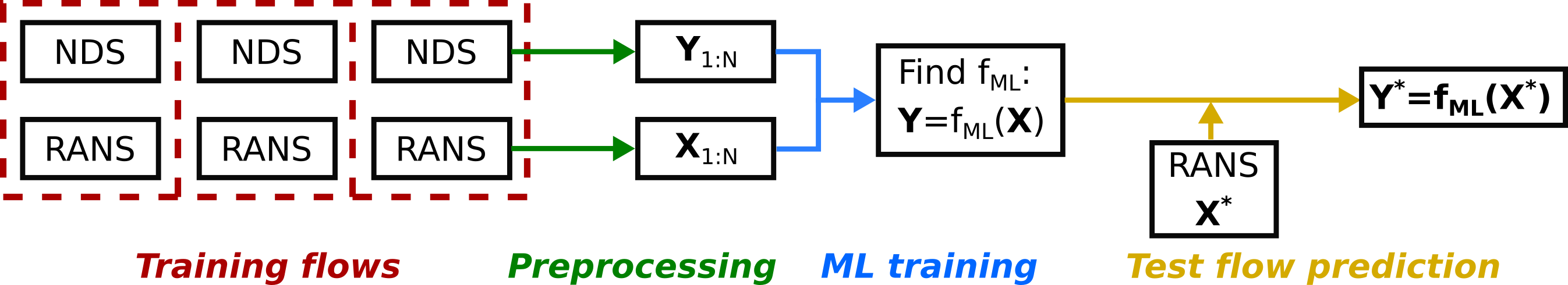}
  \caption{An example framework using supervised machine learning data-driven turbulence modelling.}
  \label{fig:supervised_ML}
\end{figure}

An example of supervised ML in the context of turbulence modelling is shown in Figure~\ref{fig:supervised_ML}. A number of DNS and LES datasets are obtained, and since only well-validated LES data is used it is assumed to be well resolved. For the remainder of this paper the LES and DNS will both be referred to as near direct simulations (NDS). For each NDS dataset, a companion RANS solution is obtained with the same boundary conditions and geometry. Each NDS and RANS data pair are then pre-processed to form the training data set $\mX_{N}, \vy_{N}$, where $N$ samples are taken at matching locations between the RANS and NDS flows. It is the task of the ML algorithm to learn a functional mapping $f_{ML}$ between the training inputs $\mX_{N}$ and training outputs $\vy_{N}$, so that $\tilde{\vy}_{M}$ can be predicted for a new \emph{test} flow where only the RANS data inputs $\tilde{\mX}_{M}$ is available. An important point to note here is that the NDS data is taken as the \emph{truth} data, hence errors in this data are not accounted for in the uncertainty estimates discussed later on. Such errors are assumed to be small in the present work, since only well-validated NDS datasets are selected. However, numerous errors do exist in LES and DNS simulations \cite{Tucker2014}, and accounting for these in data-driven modelling frameworks may be an important area of future research. 

Various authors have used the aforementioned framework to predict different turbulent quantities.  \citet{Ling2015} used a machine learning classifier, where the response $\vy$ is a binary label which indicates whether the RANS model can be trusted in different flow regions. For example, since the Boussinesq linear eddy-viscosity assumption (Eq.~\ref{eqn:evm}) presumes a positive eddy-viscosity $\nu_t$, the response
\begin{equation} \label{eqn:trust_region}
y =
\begin{cases}
	0, & \text{if}\ \nu_{t} < 0 \\
	1, & \text{otherwise}
\end{cases}
\end{equation}
can be used to suggest regions where \eqref{eqn:evm} can not be trusted. More recently, many authors \cite{Singh2017,Parish2016,Ling2016,Kaandorp2020,Geneva2019,Wu2018} have used machine learning regressors to predict the Reynolds stress discrepancy ${\Delta\boldsymbol\tau} = \boldsymbol{\tau}^{RANS} - \boldsymbol{\tau}^{NDS}$, which can be used to correct $\boldsymbol{\tau}$ before it is injected back into a CFD solver. The raw discrepancy ${\Delta\boldsymbol\tau}$ is not Galilean invariant, so it is common to instead perform the eigenspace decomposition
 \begin{equation}  \label{eqn:stress_decompose}
\boldsymbol{\tau} = 2k\left(\frac{1}{3}\mathbf{I} + \mathbf{B}\right) = 2k \left(\frac{1}{3}\mathbf{I} + \mathbf{V}\boldsymbol{\Lambda}\mathbf{V}^T\right),
 \end{equation}
where $\mathbf{I}$ is the second order identity tensor, $\mathbf{B}$ is the normalised turbulent anisotropy tensor, and 
\begin{equation}
\mV=\left[\begin{array}{ccc}
| & | & |\\
v_{1} & v_{2} & v_{3}\\
| & | & |
\end{array}\right] \; \; \;\text{and} \; \; \; \boldsymbol{\Lambda}=\left[\begin{array}{ccc}
\lambda_{1}\\
 & \lambda_{2}\\
 &  & \lambda_{3}
\end{array}\right],
\end{equation}
are the eigenvectors and eigenvalues of $\mathbf{B}$, respectively, indicating its shape and orientation. The eigenvalues are sorted such that $\lambda_1 \geq \lambda_2 \geq \lambda_3$. These eigenvalues can be transformed to the Barycentric coordinates $C_1$, $C_2$, and $C_3$, and then to Cartesian coordinates $\xi$ and $\eta$. \citet{Wu2018} use machine learning to predict the discrepancy $\Delta \boldsymbol{\tau} = (\Delta \log k, \Delta \xi, \Delta \eta, h_1, h_2, h_3)$, where $h_1$, $h_2$ and $h_3$ are the first three unit quaternions used to represent the transformation from the RANS eigenvectors $\mathbf{V}^{RANS}$ to the target eigenvectors $\mathbf{V}^{NDS}$.
 
The prediction of $\Delta \boldsymbol{\tau}$ accounts for the overall uncertainty due to the RANS model. However, \citet{Wu2019} show that the injection of Reynolds stresses directly into a CFD solver can sometimes lead to the RANS equations becoming ill-conditioned. Instead, field inversion can be used \cite{Parish2016,Singh2017} to find a field parameter $\beta(\boldsymbol{z})$ which improves the RANS model's predictions. For example, \citet{Singh2017} use $\beta(\boldsymbol{z})$ to scale the production term in the $\omega$ equation of the $k\text{-}\omega$ RANS model:

\begin{equation} \label{eqn:omega_w_beta}
\frac{D\omega}{Dt}=\beta( \boldsymbol{z} )P\left(k,\omega,\av{\mathbf{u}}\right) - D\left(k,\omega,\av{\mathbf{u}}\right) + T\left(k,\omega,\av{\mathbf{u}}\right)
\end{equation}
This field parameter accounts only for the parametric uncertainties in the RANS model. The parameter is problem-specific since it is a function of the problem's coordinates $\boldsymbol{z}$, but machine learning can then be used to infer the relationship between $\beta\left(\boldsymbol{z} \right)$ and the flow features $\mX$.

\section{Data-driven framework}
\label{sec:methods}

In this section, the data-driven framework we explore in this paper is introduced. A python package implementing the framework is available from \url{github.com/ascillitoe/mondrian_turbulence}.

\subsection{Mondrian forests}
\label{sub:methods_mondrian}

\begin{figure}[ht]
	\centering
	\begin{subfigure}[b]{0.4498\linewidth}
       	\includegraphics[width=\linewidth]{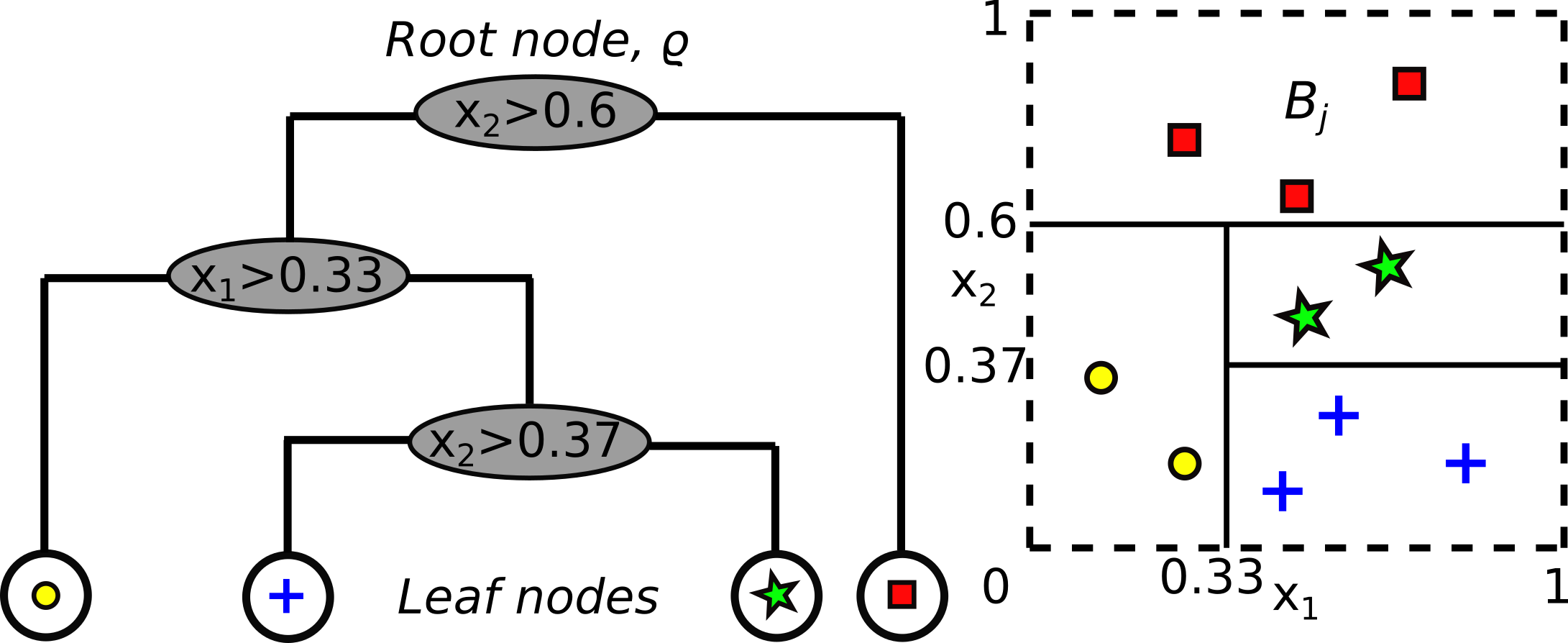}
        \caption{Decision tree}
        \label{subfig:tree_decision}
    \end{subfigure} \hspace{10pt}
    \begin{subfigure}[b]{0.5002\linewidth}
        \includegraphics[width=\linewidth]{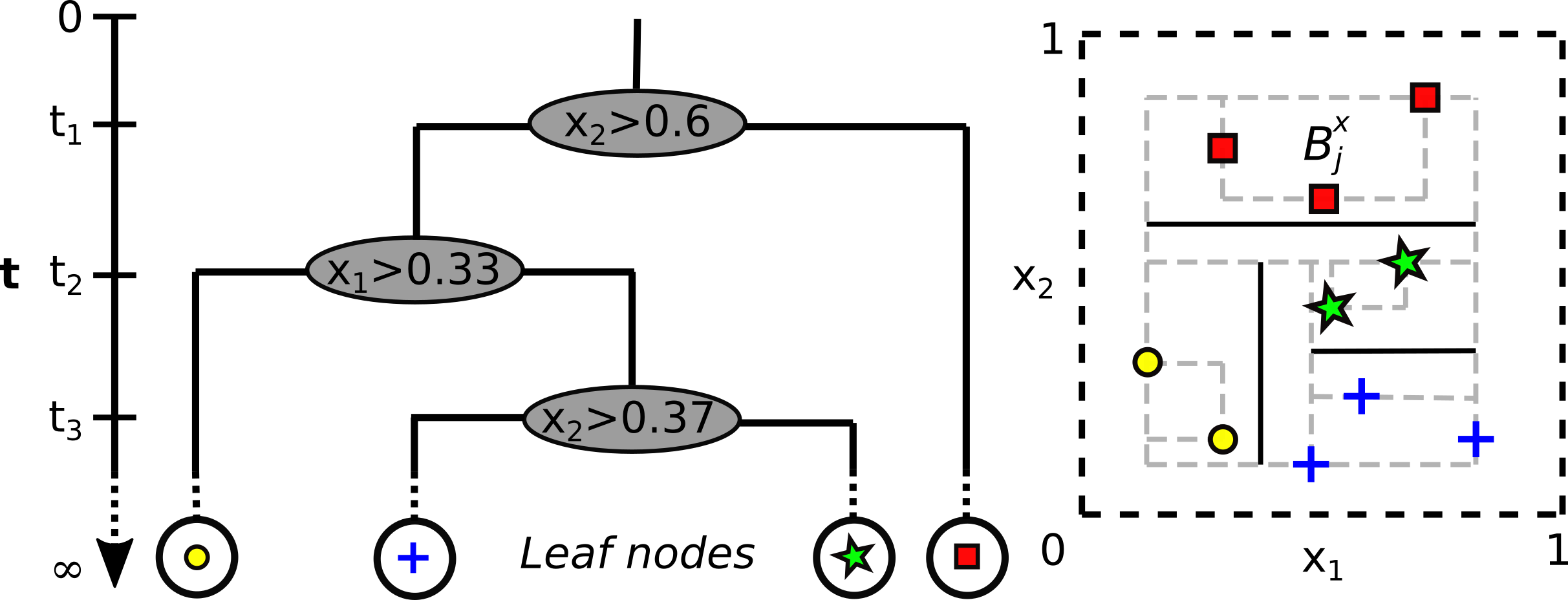}
        \caption{Mondrian tree}
        \label{subfig:tree_mondrian}
    \end{subfigure}
	\caption{Example of classification trees trained on data with four classes over eight data points in $[0, 1]^2$. Trees are shown on the left, and the partitions in feature space are shown on the right. Regression trees are similar, except the leaf nodes contain continuous values instead of classes.}
	\label{fig:tree}
\end{figure}

This paper explores the possibility of using Mondrian forests to replace the random forests used as the functional mapping $f_{ML}$ in many data-driven turbulence modelling frameworks, such as that proposed by \citet{Wu2018}. Random forests \citep{Breiman2001} are a popular supervised machine learning algorithm, which can often achieve good accuracy even with minimal tuning of their hyper-parameters. They consist of an ensemble of decision trees, akin to the one shown in Figure~\ref{subfig:tree_decision}. At each node in the tree, the data is split in a binary fashion, with the split locations chosen to optimise an appropriate criterion, such as mean squared error (MSE). Decision trees have low bias, but high variance (they over-fit to training data). By taking an average of each tree's prediction, random forests are able to reduce this variance.

Like random forests, a Mondrian forest \cite{Lakshminarayanan2016a,Lakshminarayanan2016b} is also an ensemble of trees. However, in this case, the trees are Mondrian trees. Figure~\ref{subfig:tree_mondrian} visualises a Mondrian tree, which is a restriction of a Mondrian process to a finite set of points.   A Mondrian tree can be thought of as a decision tree, but with a number of important differences:

\begin{enumerate}[(i)]
\item Splits are committed only within the range of training data, hence the split represented by an internal node $j$ holds only within $B^x_j$ and not the whole of $B_j$. The data block $B^x_j$, shown in Figure~\ref{subfig:tree_mondrian}, represents the smallest rectangle enclosing the input data $\mX$ at node $j$:
\begin{equation}
B^x_j = \left(l^x_{j1},u^x_{j1}\right] \times \dots \times \left(l^x_{jd},u^x_{jd}\right],
\end{equation}
where $l^x_{jd}$ and $u^x_{jd}$ denote the lower and upper bounds of the training data points in node $j$ along dimension $d$. 

\item The CART algorithm \cite{Breiman1984}, often used to build decision trees, samples splits based on the MSE of the predicted response data $\vy$. Whereas Mondrian tree splits are sampled independently of $\vy$. 
\end{enumerate}

Unlike standard decision trees, Mondrian trees are probabilistic models, which determine $p_{\mathcal{T}}(\tilde{y} | \tilde{\vx} , \left(\mX, \vy \right) )$. This is a predictive distribution for the response $\tilde{y}$, tree $\mathcal{T}$, training dataset $ \left(\mX, \vy \right)$ and the test point $\tilde{\vx}$. As with other Bayesian approaches, a \emph{prior} must be chosen, and the \emph{posterior} is obtained via marginalisation. The prediction of a Mondrian forest is then the average prediction from the $L$ number of Mondrian trees
\begin{equation}
p(\tilde{y} |  \tilde{\vx} , \left(\mX, \vy \right) ) = \frac{1}{L}\sum_j^{L} p_{j}( \tilde{y} |  \tilde{\vx} , \left(\mX, \vy \right) ).
\end{equation}
This predictive posterior over $\tilde{y}$ is a mixture of Gaussians, and it is straightforward to calculate the predictive mean and variance from this. The predictive variance is said to provide principled uncertainty estimates. The bounded nature of the splits in Mondrian trees, noted in (i) above, means that Mondrian trees  (and forests) exhibit higher uncertainty as $\tilde{\vx}$ moves further away from the training data $\left( \mX, \vy \right)$. 

To obtain uncertainty estimates for random forests, \citet{Wager2014} propose the use of jackknife re-sampling to provide confidence intervals for random forest predictions (see \ref{sub:jackknife} for more details). However, preliminary investigations (\ref{sub:hyperparams}) into using re-sampling for turbulence field predictions suggested the computational cost is prohibitively high, hence they are not explored in detail in this paper. For a more detailed exposition of Mondrian forests and random forests, see \ref{appendixA}.

\subsection{Database of turbulent flows}
\label{sub:database}

\begin{figure}[ht]
	\centering
	\includegraphics[width=0.35\linewidth]{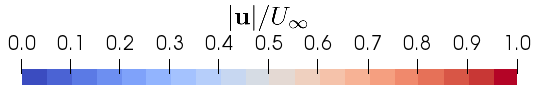} \\
	\begin{subfigure}[b]{0.69\linewidth}
       	\includegraphics[width=\linewidth]{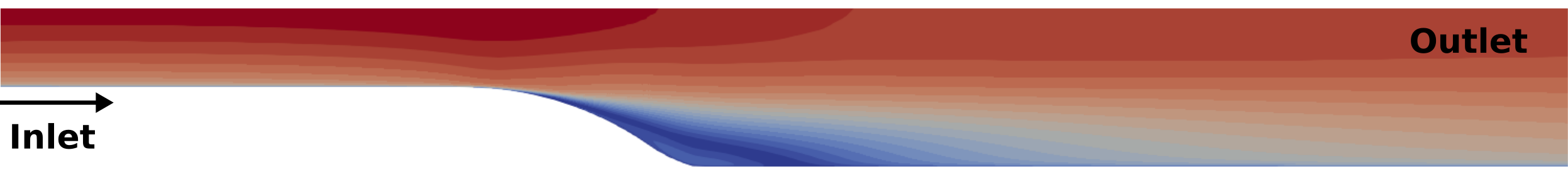}
        \caption{Case 1}
        \label{subfig:flow_1}
    \end{subfigure} \hfil
    \begin{subfigure}[b]{0.29\linewidth}
        \includegraphics[width=\linewidth]{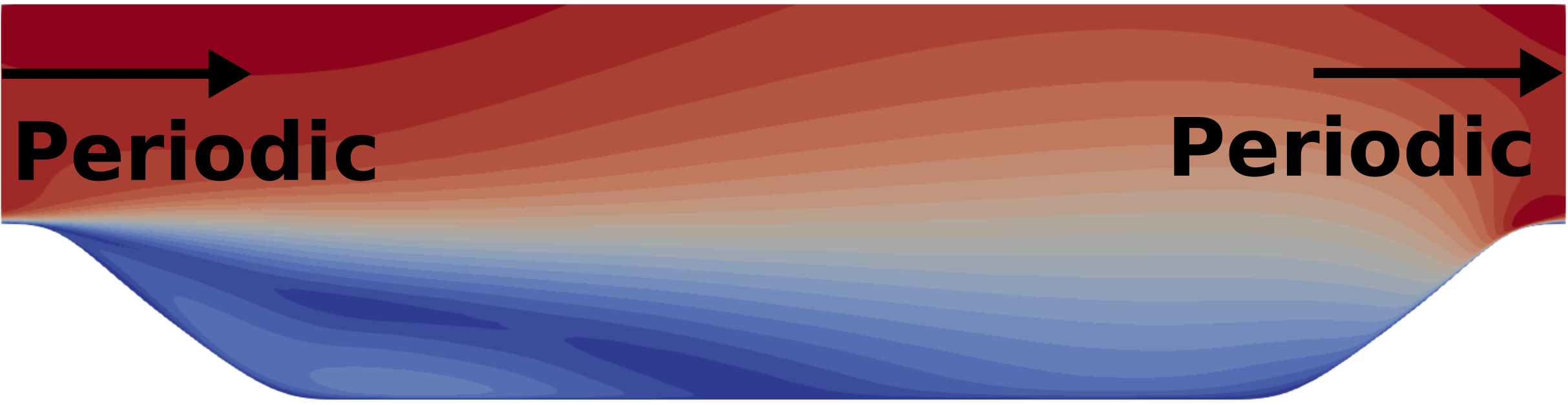}
        \caption{Case 2} 
        \label{subfig:flow_3}
    \end{subfigure} \vspace{5pt} \\
    \begin{subfigure}[b]{0.49\linewidth}
        \includegraphics[width=\linewidth]{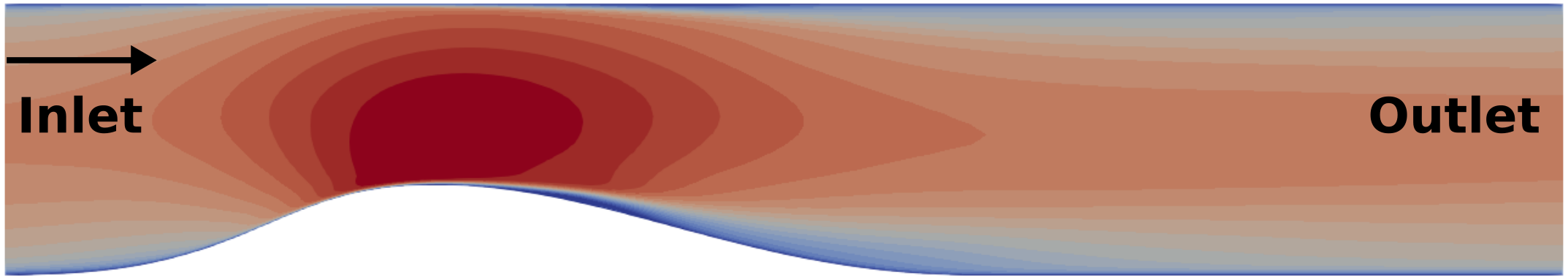}
        \caption{Case 3a}
        \label{subfig:flow_2a}
    \end{subfigure} \hfil 
     \begin{subfigure}[b]{0.49\linewidth}
=        \includegraphics[width=\linewidth]{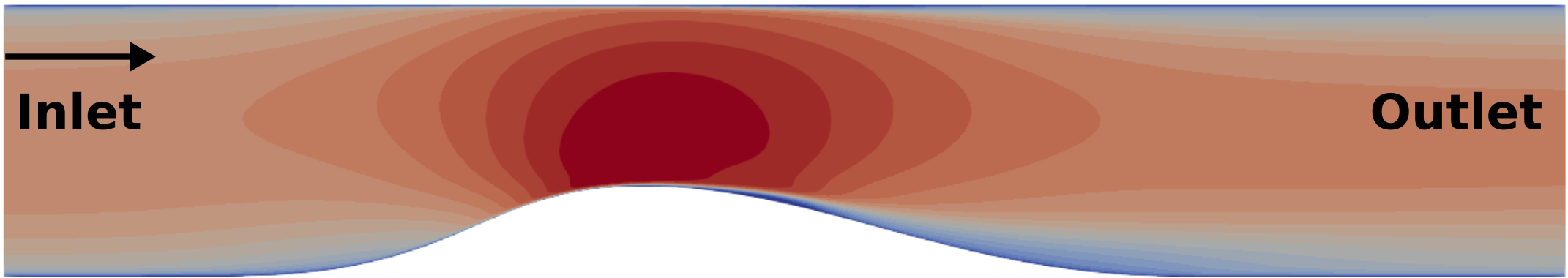}
        \caption{Cases 3b/c}
        \label{subfig:flow_2b}
    \end{subfigure} \vspace{5pt} \\
   	\begin{subfigure}[b]{0.1533\linewidth}
        \includegraphics[width=\linewidth]{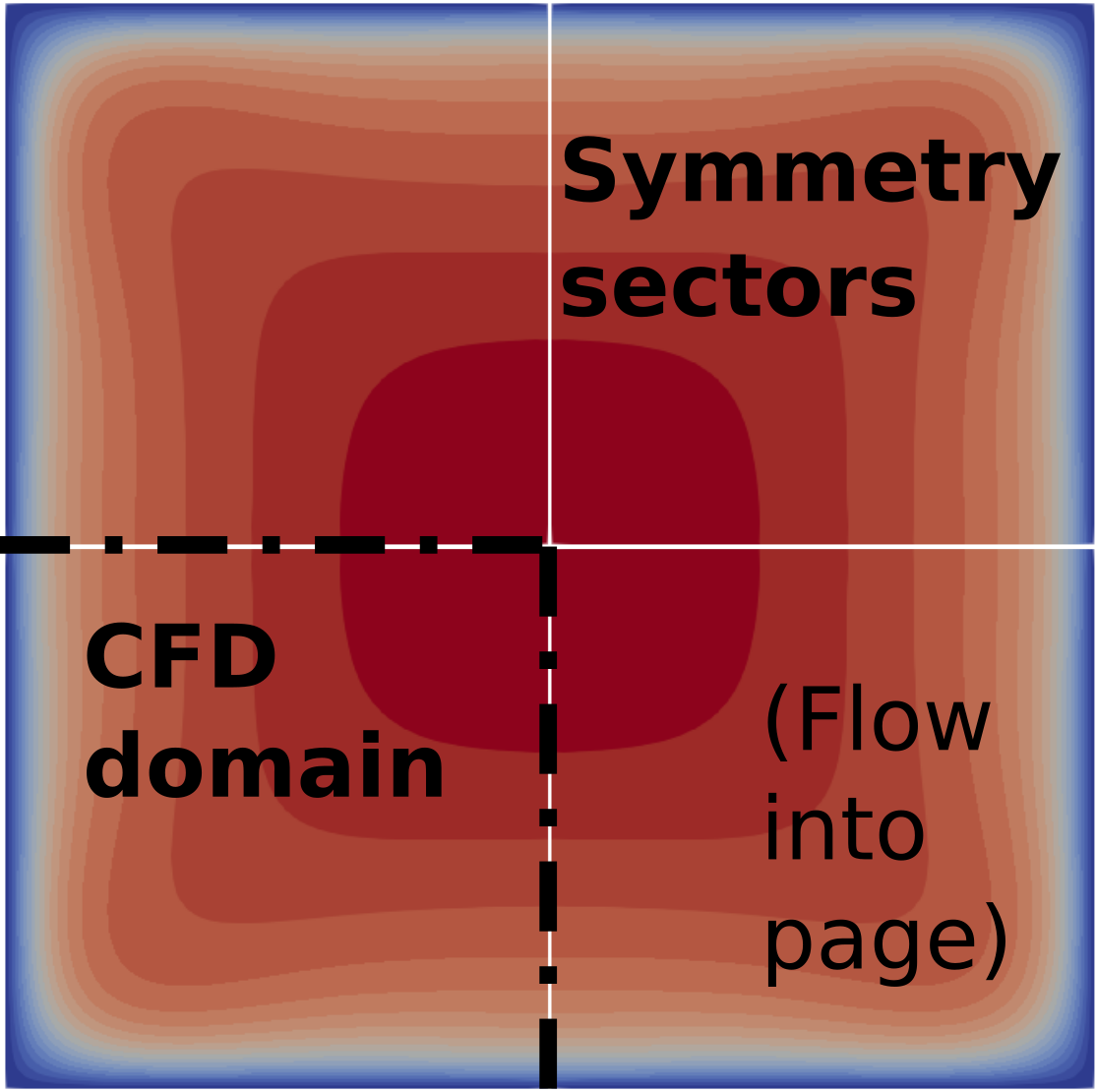}
        \caption{Cases 4a/b}
        \label{subfig:flow_5a}
    \end{subfigure} \hfil
   	\begin{subfigure}[b]{0.46\linewidth}
        \includegraphics[width=\linewidth]{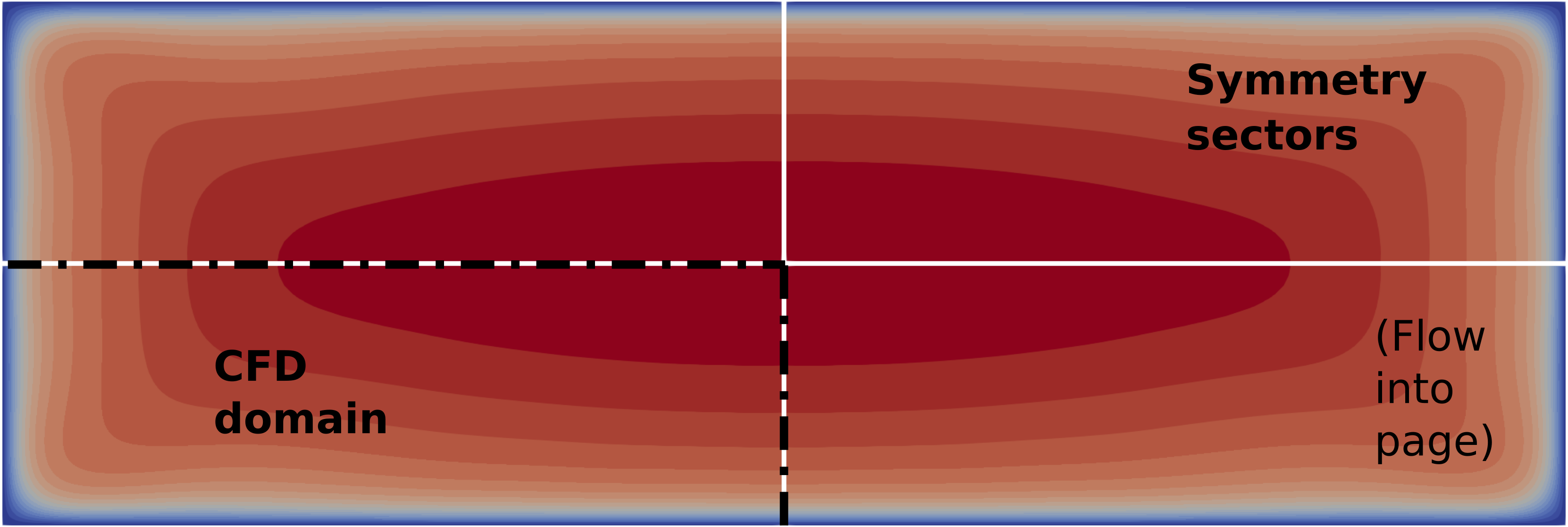}
        \caption{Cases 4c/d}
        \label{subfig:flow_5c}
    \end{subfigure} \hfil 
  	\begin{subfigure}[b]{0.32\linewidth}
        \includegraphics[width=\linewidth]{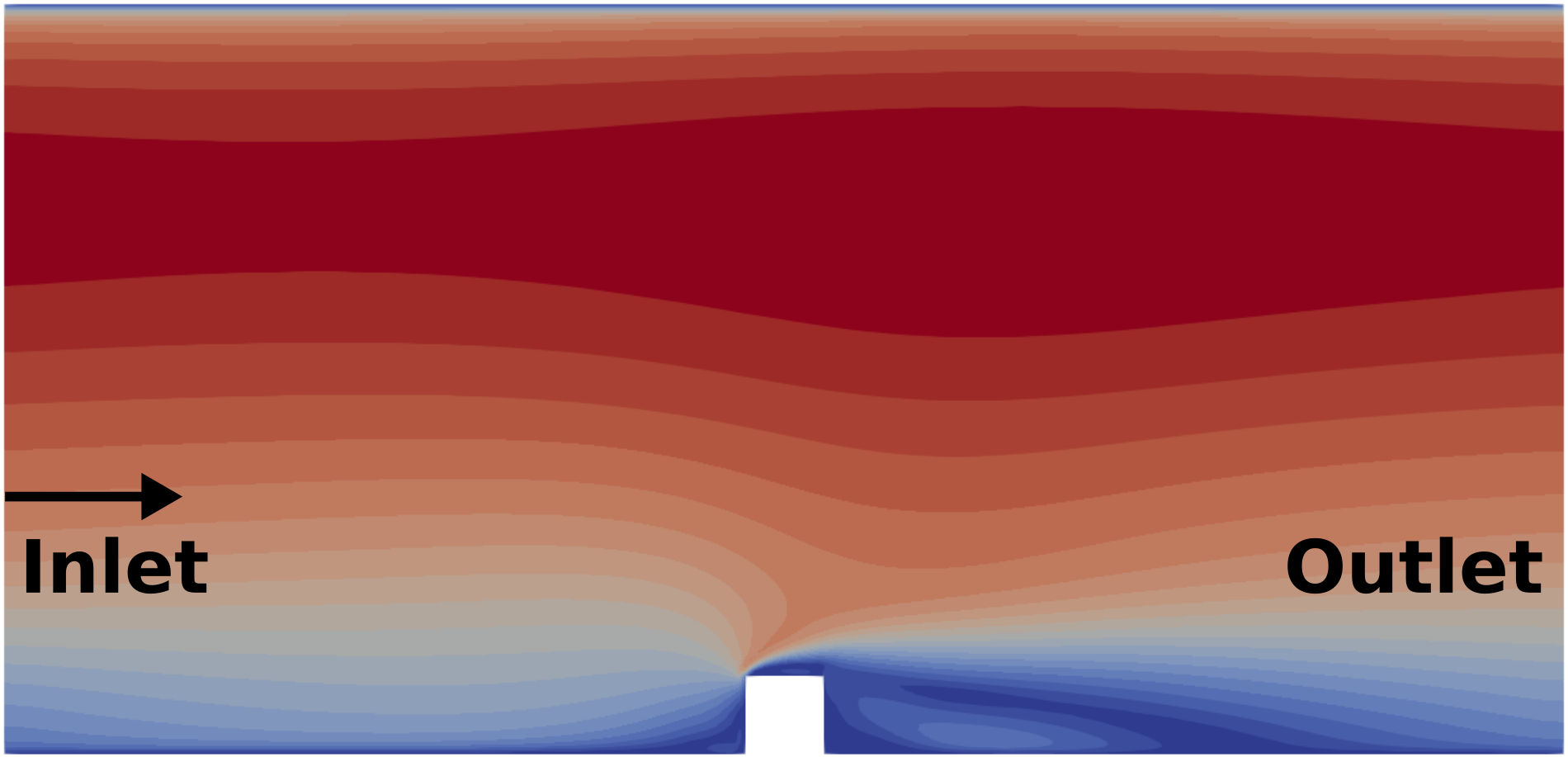}
        \caption{Case 5}
        \label{subfig:flow_8}
    \end{subfigure} \\
	\caption{Contours of normalised velocity magnitude for the LES/DNS datasets listed in Table~\ref{tab:cases}.}
	\label{fig:flow}
\end{figure}

Training and testing data are generated from the ten turbulent flow cases outlined in Table~\ref{tab:cases} and visualised in Figure~\ref{fig:flow}. Each case is a well-validated near direct simulation (NDS), with the majority obtained from publicly accessible databases\footnote{\url{turbmodels.larc.nasa.gov}, \url{turbase.cineca.it} and \url{flow.kth.se/flow-database}.}. For each NDS case a companion RANS dataset is generated using the SU2 open source CFD solver (version 7.0.6) available from \url{github.com/su2code/SU2}. For all cases the incompressible solver is used, which spatially discretises the incompressible Navier-Stokes equations using a finite volume method on unstructured grids, and solves them in a coupled manner with a custom preconditioning approach \cite{Economon2018}. For all cases, grid refinement is carried out to ensure mesh independence, and implicit time-stepping is used with an FGMRES linear solver. For turbulence modelling, the SST-2003 model \cite{Menter2003a} is used, which solves two equations for the turbulent kinetic energy $k$ and specific dissipation rate $\omega$

\begin{equation} \label{eqn:SST_k}
\diff{k}{t} + \diff{\av{u_i}k}{z_i} = \frac{1}{\rho}\tilde{P}_k - \beta^* k \omega + \diff{}{z_i} \left[ \left( \nu + \sigma_k \nu_t \right) \diff{k}{z_i} \right]
\end{equation}
\begin{equation} \label{eqn:SST_w}
\diff{\omega}{t} + \diff{\av{u_i}\omega}{z_i} = \frac{1}{\rho}P_{\omega} - \beta \omega^2 + \diff{}{z_i} \left[ \left(\nu + \sigma_{\omega} \nu_t \right) \diff{\omega}{z_i} \right] + 2\left(1-F_1\right)\sigma_{\omega 2} \frac{1}{\omega} \diff{k}{z_i} \diff{\omega}{z_i}
\end{equation}

The production of $k$ term in \eqref{eqn:SST_k} is limited by $\tilde{P}_k = \min \left( P_k,20\beta^*\rho k \omega \right)$, where the raw production term is obtained from the Boussinesq hypothesis
\begin{equation}  \label{eqn:SST_Pk}
P_k = \mu_t \diff{\av{u_i}}{z_j} \left( \diff{\av{u_i}}{z_j} + \diff{\av{u_j}}{z_i} \right).
\end{equation}
The production term in the $\omega$ equation is then given by $P_{\omega}=\rho \alpha \tilde{P}_k/\mu_t$, where $\alpha$ is a term used to blend between the $k-\epsilon$ and $k-\omega$ models in the SST model (see \cite{Menter2003a} for more details).

For cases 1 and 3 a 1D inlet velocity profile is set at the inlet to match the NDS data, and a uniform static pressure outlet boundary condition is enforced to match the area-averaged static pressure from the NDS at the same location. For Cases 2, 4 and 5 streamwise periodicity is enforced, with the streamwise forcing calibrated to match the NDS mass flow rate. All cases are assumed to be fully turbulent, with no transition model used. All RANS cases are 2D, and the NDS cases are span-wise averaged to match the RANS cases. A locally adaptive CFL number is used, with the minimum set at 2 and the maximum at 20.

\begin{table}[ht]
  \centering
  \caption{Summary of flow cases used for training and testing.}
\begin{tabular}{lllll} \hline
Case & 	Description & Re & LES/DNS & Ref. \\
\hline 
1 	&	Curved backwards step 	& $Re_{\tau}=618$ & LES &  \cite{Bentaleb2012} \\
2& Periodic hills								& $Re_{\tau}=160$ & LES & \cite{Frohlich2005} \\
3a/b/c&	Convergent-divergent channel 	& $Re_{\tau}=395/617/950$ & DNS &  \cite{Laval2011}, \cite{Schiavo2015}  \\
4a/b& Duct with aspect ratio $AR=1$	& $Re_{\tau}=180, 360$ & DNS & \cite{Vinuesa2018} \\
4c/d& Duct with aspect ratio $AR=3$	& $Re_{\tau}=180,360$ & DNS & \cite{Vinuesa2018} \\
5 & Ribbed channel flow & $Re_{\tau} = 360$ & LES & \citep{Tyacke2012}
\end{tabular}
\label{tab:cases}
\end{table}

\subsection{Pre-processing of CFD data}
\label{sub:preproc}

The RANS and NDS data is pre-processed according to the framework in Figure~\ref{fig:supervised_ML}. To form the training dataset $\mX_{N}, \vy_{N}$, the mean flow data at each mesh point in the RANS cases is converted to a feature vector $\vx \in \mathbb{R}^d$. To reduce the amount of data, the input data is filtered to remove non-unique data points (to the $4^{th}$ significant figure), and 20\% of the points are randomly sub-sampled. After filtering and sub-sampling all ten cases together yield 25000 \emph{observations}. At the same spatial locations the responses $\vy$ are then calculated from the NDS data, with the quantity stored in $\vy$ depending on what we wish to predict with the framework later on.

The $d$ flow features utilised in this paper are the same as in \cite{Wu2018}. Following \cite{Ling2015}, the raw features $\alpha$ are non-dimensionalised by a local normalisation factor $\beta$
\begin{equation}
\hat{\alpha}=\frac{\alpha}{|\alpha| + |\beta|},
\end{equation}
where $\alpha$ and $\beta$ are given in Table~\ref{tab:feature_nondim}. The strain-rate tensor $\boldsymbol{S}$ and rotation-rate tensor $\boldsymbol{\Omega}$ are included as it has been recognised that all algebraic Reynolds stress and eddy viscosity models can be written in the general form $\boldsymbol{\tau}=f(\boldsymbol{S},\boldsymbol{\Omega})$ \cite{Pope2000}. The pressure gradient is included since it is known that turbulence is also influenced by this, while the turbulent kinetic energy (TKE)  gradient is added in an attempt to account for non-local turbulent transport effects. Finally, near wall viscous effects have an important influence on turbulence so a non-dimensional wall distance feature is included, along with features to describe the length and time-scales of turbulence. 

\begin{table}[ht]
 \caption{Non-dimensional features used to represent the flow fields. Notations are as follows: $\mathbf{u}$ is the mean velocity vector, $k = \overline{u_iu_i}/2$ is the turbulence kinetic energy (TKE), $\omega$ is the specific turbulent dissipation rate, $|\cdot|$ and $||\cdot||$ indicate vector and matrix norm.}
  \centering
  \begin{tabular}{ p{6cm} p{2.5cm} p{2.5cm} }
  \hline
  	Description & Raw input, $\alpha$ & Normalisation, $\beta$ \\
  	\hline
  	Strain-rate tensor 		& $\boldsymbol{S} $ 			& $\omega$ \\
  	Rotation-rate tensor 	& $\boldsymbol{\Omega}$ 	& $\boldsymbol{||\Omega}||$ \\
	Pressure gradient 		& $\nabla p$							& $\rho |D\mathbf{u}/Dt|$ \\
	TKE gradient 			& $\nabla k$							& $\omega \sqrt{k}$ \\
	Wall-distance based Reynolds number, $Re_d$ & $\min \left( \frac{\sqrt{k}d}{50\nu},2 \right)$ & - \\
	Turbulence intensity, $Ti$ & $k$									& $\frac{3}{2}|\mathbf{u}|^2$ \\
	Ratio of turbulent to mean strain time-scales, $t_t/t_s$ & $\omega$ & $\frac{1}{||\boldsymbol{S}||}$
  \end{tabular}
  \label{tab:feature_nondim}
\end{table}

To achieve Galilean and rotational invariance of the machine learning function, the approach of \cite{Wu2018} is adopted. The normalised gradients $\widehat{\nabla p}$ and $\widehat{\nabla k}$ are transformed into the anti-symmetric tensors $\widehat{\mathbf{A}}_p$ and $\widehat{\mathbf{A}}_k$, and the minimal integrity bases for the tensorial set $\{ \widehat{\boldsymbol{S}},\widehat{\boldsymbol{\Omega}},\widehat{\mathbf{A}}_p,\widehat{\mathbf{A}}_k\}$ are calculated. Each of the 47 invariant bases is the trace of a tensor such as $\widehat{\boldsymbol{S}}^2$ and $\widehat{\boldsymbol{\Omega}}\widehat{\mathbf{A}}_p \widehat{\mathbf{A}}_k \widehat{\boldsymbol{S}}$. Combining the remaining 3 scalar features in Table~\ref{tab:feature_nondim} with the invariant bases leads to a 50 dimensional feature space $\vx \in \mathbb{R}^{50}$. For further details see Appendix B in Ref. \cite{Wu2018}. 

\begin{remark}
By learning a mapping between $\vx$ and $y$, we have assumed the Reynolds stress $\boldsymbol{\tau}$ can be described by purely local (point-wise) quantities $(k,\omega,\nu)$, and the local gradients $(\nabla \boldsymbol{u}, \nabla k, \nabla p)$ at a given point. This assumption is likely to be questionable in regions where the turbulence scales are large or transport of turbulence is strong. Additionally, since $y$ is derived from NDS data and $\vx$ from RANS data, we have implicitly assumed the RANS and NDS mean velocity and pressure fields match. This assumption is likely to be invalid if more complex flow cases were used for training.
\end{remark}

As noted in point (ii) in \ref{sub:methods_mondrian}, Mondrian forests choose splits independent of the response data $\vy$. This means a Mondrian forest's predicitve accuracy can be degraded if features are present in $\mX$ which are irrelevant to the response variable $\vy$. To check for the effect of irrelevant features, backward elimination was performed, with prediction accuracy monitored as the least important features were iteratively removed. This study was performed on Mondrian forest's trained to predict the $C_{aniso}$ constant, with further details given in \ref{sub:feature_select}. Surprisingly, irrelevant features were not found to be an issue in this case, perhaps because all non-zero features are important and many of the 47 invariant terms are zero for two-dimensional flows. Nevertheless, a 12 dimensional subset of the original 50 features is used in the remainder of this paper:
\begin{equation} \label{eqn:feature_set}
\boldsymbol{x} = \left\lbrace \widehat{\boldsymbol{S}}^2,\widehat{\boldsymbol{\Omega}}^2,\widehat{\mathbf{A}}_p^2, \widehat{\mathbf{A}}_k^2, \widehat{\boldsymbol{\Omega}}\widehat{\mathbf{A}}_k, \widehat{\boldsymbol{\Omega}}\widehat{\mathbf{A}}_k \widehat{\boldsymbol{S}}^2, \widehat{\boldsymbol{\Omega}}^2\widehat{\mathbf{A}}_k \widehat{\boldsymbol{S}}, \widehat{\mathbf{A}}_k^2\widehat{\mathbf{A}}_p\widehat{\boldsymbol{S}}, \widehat{\mathbf{A}}_k^2\widehat{\boldsymbol{S}}\widehat{\mathbf{A}}_p\widehat{\boldsymbol{S}}^2,Re_d, \hat{k},\hat{\omega} \right\rbrace^T.
\end{equation}
This reduced feature space allows for reduced training/prediction times and data storage requirements, with no appreciable loss in predictive accuracy. This pruning down of features via backward elimination constitutes a departure from the methodology pursued in \cite{Wu2018}.

\subsection{Propagation of machine learning predictions}
\label{sub:propagating_mean}

If one uses the ML framework to predict the discrepancy vector $\Delta \boldsymbol{\tau}$, it is desirable to propagate this through the RANS solver so that revised flow field predictions can be obtained. A number of authors propose strategies to achieve this, for example, see \cite{Ling2016,Wu2018,Kaandorp2020,Geneva2019}. We choose a variation of that proposed by \citet{Kaandorp2020}. Here, to achieve favourable numerical convergence, the ML derived anisotropy tensor $\mathbf{B}_{ML}$ is under-relaxed against the original Boussinesq tensor $\mathbf{B}_{BL} := \nu_t \mathbf{S}$. We replace the Reynolds stress term in the open source SU2 CFD code with
\begin{equation} \label{eqn:DDRANS}
\boldsymbol{\tau} = \frac{2}{3}k\mathbf{I} + 2k\left[\left(1-\gamma_n\right)\mathbf{B}_{BL} + \gamma_n \mathbf{B}_{ML} \right],
\end{equation}
where $\mathbf{B}_{ML}$ is reconstructed from the ML predicted eigenvalues and eigenvectors (recall Eqn.~\ref{eqn:stress_decompose}), with $ \xi_{ML} = \xi_{RANS} + \Delta \xi$. The blending parameter $\gamma_n$ is defined with a linear ramp
\begin{equation} \label{eqn:ramp}
\gamma_n = \gamma_{max} \min\left(1, \frac{n}{n_{max}}  \right),
\end{equation}
which iteratively converges to $n=n_{max}$, where $\gamma_n = \gamma_{max}$. After this a sufficient number of iterations are performed to achieve full convergence, i.e., $n > n_{max}$. As will be described later in Section~\ref{sec:DDRANS}, $n_{max}$ here is the user-defined number of iterations used for ramping up the blending of $\mathbf{B}_{ML}$. Further, \citet{Kaandorp2020} replace the limited production term $\tilde{P}_k$ in \eqref{eqn:SST_Pk} by the exact production term
\begin{equation} \label{eqn:SST_Pk_mod}
P_k = -\tau_{jj} \diff{\av{u_i}}{z_j},
\end{equation}
and the $\omega$ production term in \eqref{eqn:SST_w} is then given by $P_{\omega}=\rho \alpha P_k/\mu_t$. Boundary conditions are easily satisfied in the above implementation since the anisotropy term in \eqref{eqn:DDRANS} is scaled by $k$, which equals zero at the wall for the low Reynolds number RANS model used here. If wall functions were to be used, a blending such as that proposed by \citet{Geneva2019} would be necessary. Since $k$ is allowed to vary here, we no longer require its discrepancy $\Delta \log k$, and so instead of requiring an ML prediction for the Reynolds stress discrepancy term $\Delta \boldsymbol{\tau} = (\Delta \log k, \Delta \xi, \Delta \eta, h_1, h_2, h_3)$, we only require a prediction for $\Delta \mathbf{B} = (\Delta \xi, \Delta \eta, h_1, h_2, h_3)$. The modified SU2 code described here is available at \url{github.com/ascillitoe/SU2/tree/feature_DDRANS}. 

\subsection{Propagation of Mondrian forest's predictive uncertainty}
\label{sub:propagating_uq}

As mentioned in Section~\ref{sub:methods_mondrian}, the Mondrian forest framework outputs the mean and variance for the five quantities in $\mathbf{B}$. To propagate the variance estimates through the RANS solver, we need a way of generating spatially consistent samples that satisfy the mean and variances for these five quantities. To reiterate, these mean and variance values are spatially varying, and thus care must be taken to ensure that random samples adhering to these statistical quantities preserve the underpinning spatial correlations. The challenge, however, is that we do not have a parametric definition for this correlation.

One approach is to assume that a single discrepancy quantity, say, $\Delta k$ can be modelled as a spatial Gaussian random field, defined in terms of its spatial coordinates $\boldsymbol{z}$. Then, we use standard Gaussian process machinery to generate random realisations that adhere to the aforementioned mean $\boldsymbol{\mu}_{\Delta k} \left( \boldsymbol{z} \right) $ and variance $\boldsymbol{\sigma}_{\Delta k}^2 \left( \boldsymbol{z} \right)$.

There are three key challenges that arise when doing so. First, one has to assume a certain covariance kernel. Here the squared exponential (radial basis function) is the default option, where the covariance function is set by the distance between two coordinates and not their spatial location within the domain. Second, within the Gaussian process framework, we need to invert (via Cholesky) a square matrix of dimension set by the number of mesh nodes, which can be prohibitive for large meshes. However, this can be abated by coarsening the flow-field. Third, and perhaps of greatest importance to highlight, is that many of our discrepancy fields have large variance values. Thus, when optimising using either maximum likelihood or Markov chain Monte Carlo, the Gaussian process model sought to explain the spatial distribution with a very flat mean and large variance posterior spatial model---violating the Mondrian forest yielded statistical output. 

This motivated our heuristic, where we construct a custom covariance function of the form
\begin{equation}
\boldsymbol{K}_{\Delta k}  \left(\boldsymbol{z}_i, \boldsymbol{z}_j \right) = \phi \sigma_{\boldsymbol{z}_i} \sigma_{\boldsymbol{z}_j} , \; \;  \text{with} \; \; \phi  = \frac{1}{\omega_1  + \omega_2 \text{exp} \left( \left\Vert \boldsymbol{z}_{i}- \boldsymbol{z}_{j}\right\Vert _{2}^{2} - \omega_3 \right) }.
\label{equ:covariance_function}
\end{equation}
Assuming the values of constants $\omega_1, \omega_2$ and $\omega_3$ are selected such that 
\begin{equation}
0 \leq \phi \leq 1
\label{equ:correlation_constraint}
\end{equation}
holds, \eqref{equ:covariance_function} explicitly sets $\phi$ to be the correlation between any $\boldsymbol{z}_{i}$ and $\boldsymbol{z}_{j}$, whilst preserving the underlying variances. Although the subscript notation in \eqref{equ:covariance_function} is shown for $\Delta k$'s covariance function, we use the same expression for all the remaining five spatial discrepancy fields. Random spatial samples for say $\Delta k$ can then be generated by sampling from the multivariate normal distribution identified by 
\begin{equation}
\mathcal{N} \left( \boldsymbol{\mu}_{\Delta k}  \left( \boldsymbol{z} \right), \boldsymbol{K}_{\Delta k}  \left( \boldsymbol{z}, \boldsymbol{z}' \right) \right).
\end{equation}
We iterated over different values of the constants ensuring that \eqref{equ:correlation_constraint} holds and identified $\omega_1=0.58$, $\omega_2 = 1.4$ and $\omega_3 = 1.2$ to offer sufficient spatial smoothing. It should be noted that these values yield a covariance matrix that is positive definite, which is another constraint governing these constants. 

\section{Predicting turbulence variables with Mondrian forests}
\label{sec:MF_vs_RF}

Prior to examining the use of Mondrian forests for a data-driven turbulence model application, we compare them to random forests on a simple problem. The training response $\vy$ is set to be the turbulence anisotropy constant proposed by \citet{Banerjee2007},
\begin{equation} \label{eqn:Caniso}
C_{aniso} = -3\lambda_3,
\end{equation}
where $\lambda_3$ is calculated from the NDS Reynolds stress fields. The constant $C_{aniso}$ describes the turbulent anisotropy, with $C_{aniso}=0$ indicating isotropic turbulence and $C_{aniso}=1$ indicating fully one component turbulence.

\subsection{Convergent-divergent channel}
\label{sub:codi_channel}

\begin{figure}[ht]
  \centering
  \includegraphics[width=0.9\linewidth]{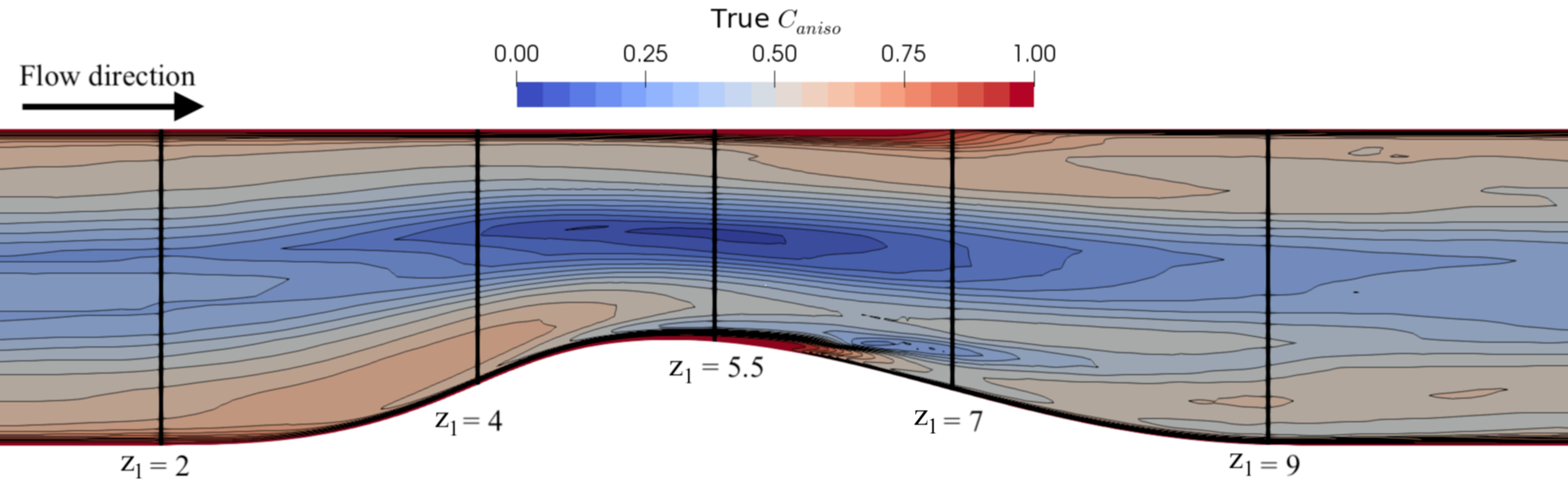}
  \caption{Contours of the \emph{true} turbulent anisotropy constant $C_{aniso}$ for the convergent-divergent channel at $Re_{\tau}=617$ (case 3b). The inflow is located at $z_1=0$, and $\boldsymbol{z}$ is non-dimensionlised by the channel half-height.}
  \label{fig:2b_Caniso_contours}
\end{figure}

The turbulence anisotropy constant $C_{aniso}$ obtained from the DNS data \cite{Laval2011} for the convergent-divergent channel at $Re_{\tau}=617$ is shown in Figure~\ref{fig:2b_Caniso_contours}. As expected, the turbulence anisotropy is generally high near the upper and lower viscous walls. Near the lower wall, downstream of the bump ($5.5<z_1<7$), a strong adverse pressure gradient leads to flow separation and a small recirculation region. The production of $\av{u'_1u'_1}$ is an order of magnitude larger than $\av{u'_2u'_2}$ in this region \cite{Laval2011}, leading to highly anisotropic turbulence. Near the upper surface at the same horizontal location, a weaker adverse pressure gradient is observed, implying that the flow is on the verge of separation, and turbulence anisotropy is also high. The turbulence returns towards isotropy in the center of the channel around $z_1=5.5$ due to the favourable pressure gradient (leading to flow acceleration) in this region.

\subsubsection{Initial Mondrian forest predictions}

\begin{figure}[ht]
	\centering
	\includegraphics[width=0.8\linewidth]{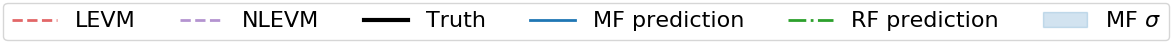}
	\begin{subfigure}[b]{0.9\linewidth}
       	\includegraphics[width=\linewidth]{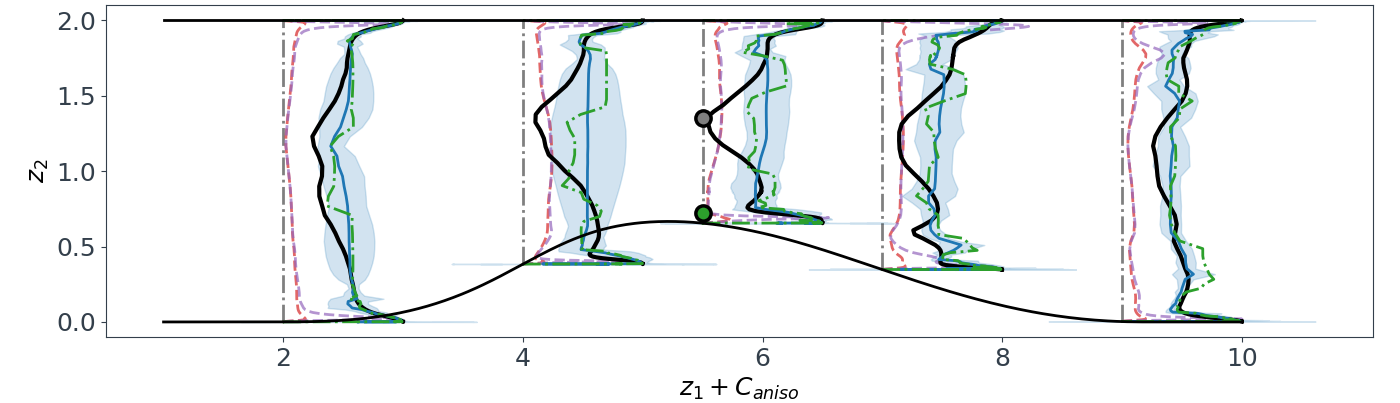}
        \caption{Only curved backward step and periodic hill cases (1 and 2) used for training}
        \label{subfig:2b_Caniso_slices_c12}
    \end{subfigure} \vspace{8pt} \\
    \begin{subfigure}[b]{0.9\linewidth}
        \includegraphics[width=\linewidth]{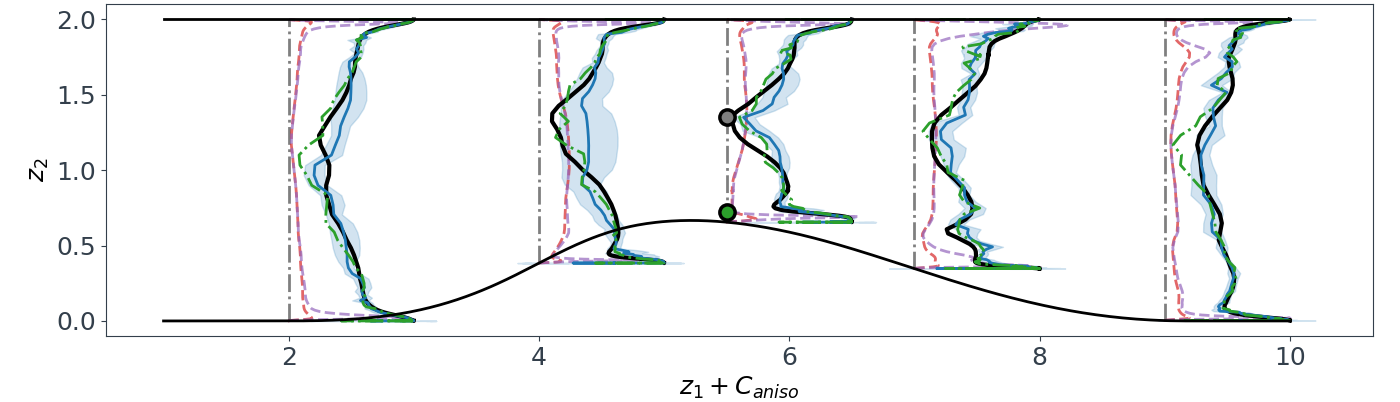}
        \caption{All cases (except for 3b) used for training}
        \label{subfig:2b_Caniso_slices_call}
    \end{subfigure}
	\caption{Mondrian forest and random forest predicted turbulent anisotropy constant $C_{aniso}$ for the convergent-divergent channel at $Re_{\tau}=617$ (case 3b), at the slices marked in Figure~\ref{fig:2b_Caniso_contours}.}
	\label{fig:2b_Caniso_slices}
\end{figure}

In Figure~\ref{fig:2b_Caniso_slices} the Mondrian forest (MF) predictions for $C_{aniso}$ and the \emph{truth} (DNS) are plotted across the $z_1$-planes shown in Figure~\ref{fig:2b_Caniso_contours}. When judging the Mondrian forest predictions here it is important to note that the Mondrian forest has not seen this flow case during training, so it is making predictions based on the physics it has learnt from the other flow cases. In Figure~\ref{subfig:2b_Caniso_slices_c12}, the Mondrian forest is trained only on the curved backward step (case 1, $Re_{\tau}=618$) and periodic hills (case 2, $Re_{\tau}=160$). Hence, a degree of extrapolation in feature space is likely here. Generally, the MF is predicting the physics seen in the DNS well, with high anisotropy predicted in the near-wall regions, and a further increase in anisotropy in the adverse pressure gradient, downstream of the bump (slices 3 and 4). Where predictions are less accurate, for example the return to isotropy region in the centre of the channel at slices 2-4, the Mondrian forest returns large predictive uncertainty values. Comparing the Random forest (RF) and MF predictions, they are generally about equal when measured against the Truth data. The RF is more accurate at the $z_1=4$ plane, but the MF predictions are superior at the $z_1=9$ plane. Overall, the MF appear to be slightly less noisy than the RF's predictions. 

The linear eddy viscosity model (LEVM) in \eqref{eqn:evm}, and the quadratic terms which extend the LEVM to form a non-linear (NL) EVM, can be generalised as
\begin{equation} \label{eqn:NLEVM}
\begin{split}
B_{ij} = \overbrace{-\frac{\nu_t}{k}S_{ij}}^{\text{LEVM}} & + c_1 \left(S_{ik}S_{jk} - \frac{1}{3}S_{mk}S_{mk}\delta_{ij}\right) \\
&+ \underbrace{c_2\left( \Omega_{ik}S_{kj} + \Omega_{jk}S_{ki} \right) + c_3\left(\Omega_{ik}\Omega_{jk} - \frac{1}{3} \Omega_{lk}\Omega_{lk}\delta_{ij}\right)}_{\text{Quadratic NLEVM}}.
\end{split}
\end{equation}
The values of $C_{aniso}$ obtained when the LEVM is used are shown in Figure~\ref{fig:2b_Caniso_slices}, along with the values obtained when the quadratic NLEVM terms are included\footnote{The constants $c_1,C_2,c_3$ are chosen to give Shih's quadratic non-linear eddy viscosity model \cite{Shih1995}.}. In most regions, the LEVM is seen to predict an insufficient level of turbulent anisotropy. Near the viscous walls, the NLEVM predicts high levels of turbulent anisotropy in closer agreement with the NDS, however it doesn't significantly improve predictions towards the centre of the channel. The clear superiority of the MF predictions here suggest that MFs offer the potential to augment LEVM and NLEVM based RANS models for such flows. As discussed by \citet{Hellsten2009}, many more elaborate RANS approaches exist, such as explicit algebraic Reynolds stress models (EARSM's) and Reynold stress transport models (RSTM's). However, \citet{Wu2018} find a random forest based framework is able to offer substantial improvements over a RSTM model for duct and periodic hill flows, suggesting MF's could also offer improvements for more advanced RANS models.

\begin{remark}
The uncertainty bounds given by $\mu \pm \sigma$ do not entirely encompass the truth data here. However, this is expected, since we are plotting $\pm 1 \sigma$, so would only expect $\approx68\%$ of truth values to fall within these uncertainty bounds. Additionally, \citet{Lakshminarayanan2016b} examine the \emph{probability calibration measures} for MF vs RF, and although MFs perform signficantly better than RFs with jackknifing, they still tend to be slightly over-confident with their uncertainty estimates---an artifact that could perhaps be fixed with calibration of their uncertainty estimates. That said, it should be noted these uncertainty estimates do not account for uncertainty due to the assumptions discussed in Remark 1.
\end{remark}

\subsubsection{More representative training data}
\label{sub:2b_moretrain}

In Figure~\ref{subfig:2b_Caniso_slices_call} all cases except the present test case (case 3b) are used for training. The effect of this additional training data is to improve the accuracy of the MF's mean predictions, whilst reducing its predictive uncertainty. This is especially noticeable in the return to isotropy region near the channel centre at slices 2-4. To understand why, the MF's predictions are interpreted using \textbf{SH}apley \textbf{A}dditive ex\textbf{P}lanation (SHAP) values, recently proposed by \citet{Lundberg2017}. This is an additive feature attribution method based on Shapley values from coalitional game theory. They are defined as Shapley values of a conditional expectation function of the original model \cite{Lundberg2017}; effectively the $j$-th SHAP value $\phi_j$ represents the contribution to the prediction $\tilde{y}$, compared to the average prediction for the dataset, when conditioning on the $j$-th feature. Thus, the sum of the SHAP values satisfies
\begin{equation}
\sum_{j=1}^D \phi_j (\tilde{y} ) = \tilde{y} - \mathbb{E}( \vy).
\end{equation}

\begin{figure}[ht]
	\centering
	\begin{subfigure}[b]{0.47\linewidth}
       	\includegraphics[width=\linewidth]{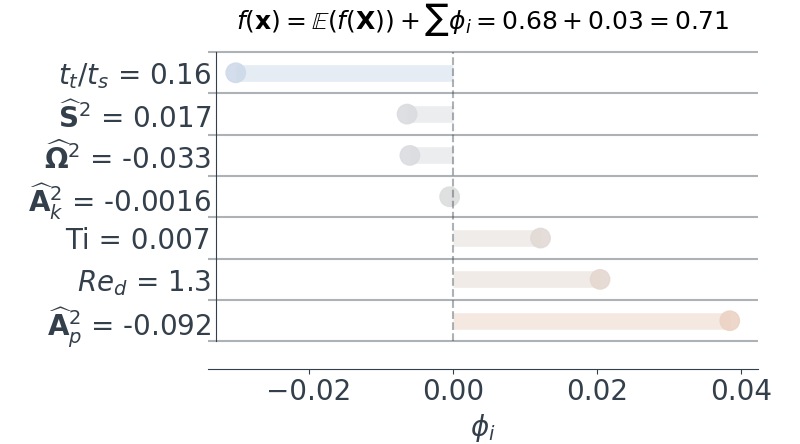}
        \caption{Near lower wall, $z_2=0.72$}
        \label{subfig:SHAP_low}
    \end{subfigure} \hfil
    \begin{subfigure}[b]{0.47\linewidth}
        \includegraphics[width=\linewidth]{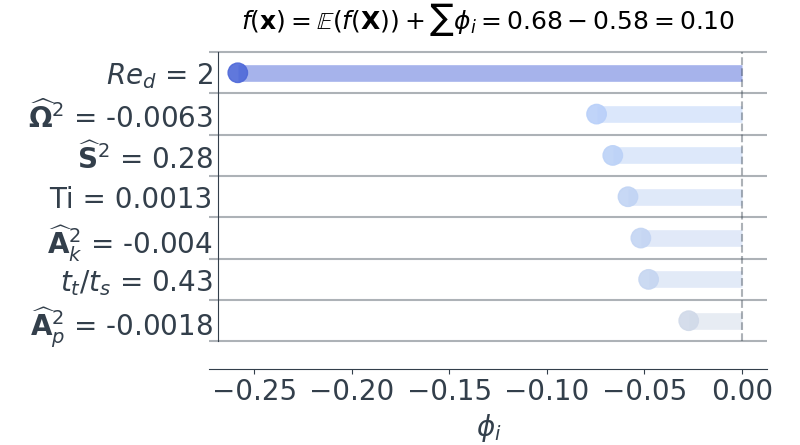}
        \caption{Mid-channel, $z_2=1.35$}
        \label{subfig:SHAP_mid}
    \end{subfigure}
	\caption{The seven largest SHAP values at two different $z_2$ locations across $z_1=5.5$, for the convergent-divergent channel at $Re_{\tau}=617$ (case 3b). Computed using the Mondrian forest trained on all cases (except for case 3b).}
	\label{fig:SHAP}
\end{figure}

To efficiently compute exact SHAP values for tree-based machine learning models, \citet{Lundberg2020} propose TreeSHAP, available from \url{github.com/slundberg/shap}\footnote{The code is slightly modified to be compatible with the Mondrian forests, and is made available at \url{github.com/ascillitoe/shap}.}. This is used to compute SHAP values for predictions at the two locations marked by the circles in Figure~\ref{fig:2b_Caniso_slices}. The seven SHAP values with greatest magnitude are plotted in Figure~\ref{fig:SHAP}. Near the wall (Fig.~\ref{subfig:SHAP_low}), the relatively small turbulent-to-strain timescale ratio ($t_t/t_s=0.16$) decreases the predicted $C_{aniso}$ value compared to the average, whereas the axi-symmetric pressure gradient tensor value, $\widehat{\mathbf{A}}_p^2=-0.092$, increases the predicted turbulent anisotropy. However, the sum of the SHAP values is relatively small here ($\sum \phi_i=0.03$), meaning the predicted anisotropy is close to the average value ($\tilde{y}=0.71$). In the center of the channel (Fig.~\ref{subfig:SHAP_mid}), the predicted anisotropy is considerably lower than the average mainly due to the high $Re_d$, but the other six features shown also contribute to the $C_{aniso}$ constant tending towards isotropy here. 

\begin{figure}[ht]
	\centering
	\includegraphics[width=0.4\linewidth]{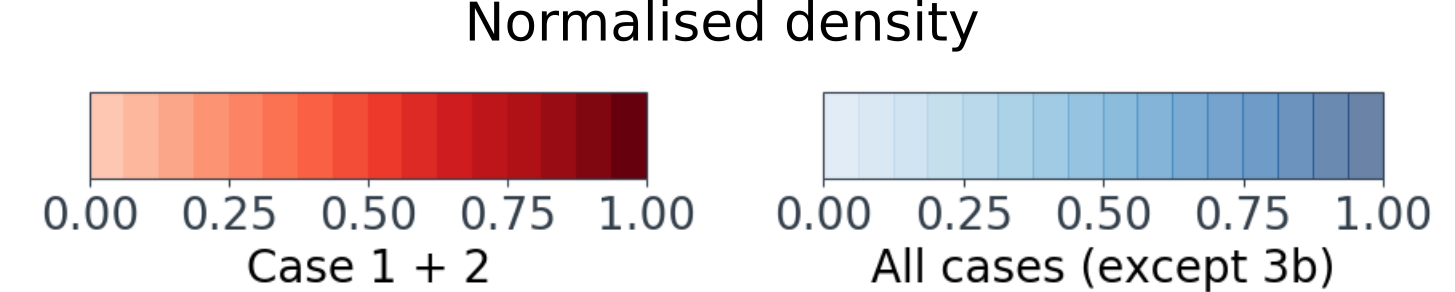} \\ \vspace{2pt}
	\begin{subfigure}[b]{0.31\linewidth}
       	\includegraphics[width=\linewidth]{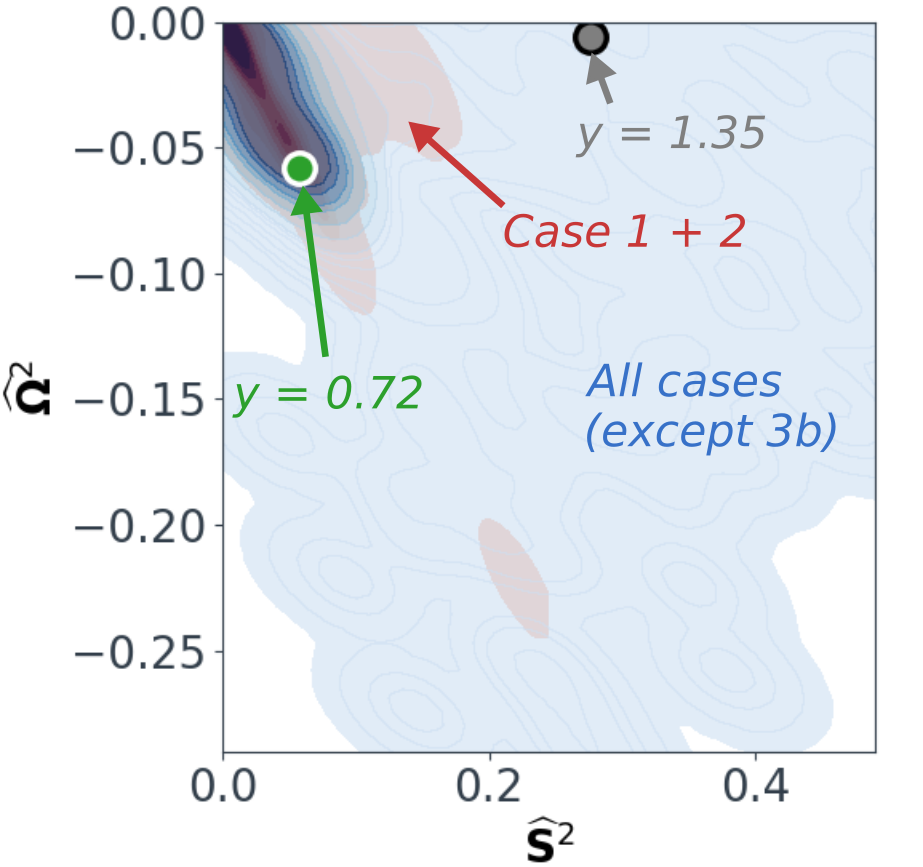}
        \caption{}
        \label{subfig:training_kde_1}
    \end{subfigure} \hfil
    \begin{subfigure}[b]{0.31\linewidth}
        \includegraphics[width=\linewidth]{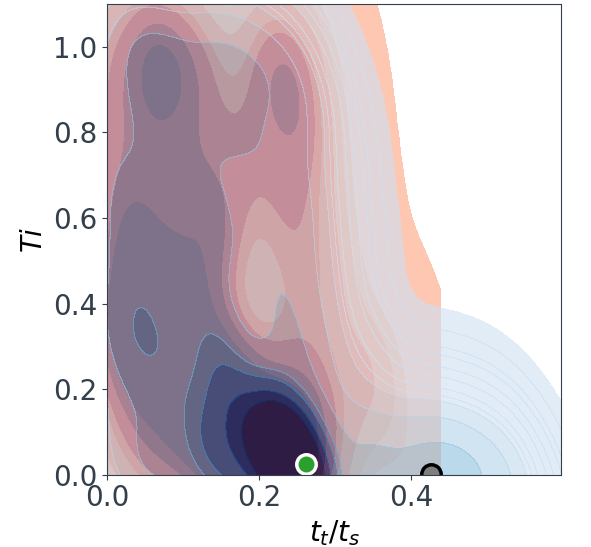}
        \caption{}
        \label{subfig:training_kde_2}
    \end{subfigure} \hfil
     \begin{subfigure}[b]{0.305\linewidth}
        \includegraphics[width=\linewidth]{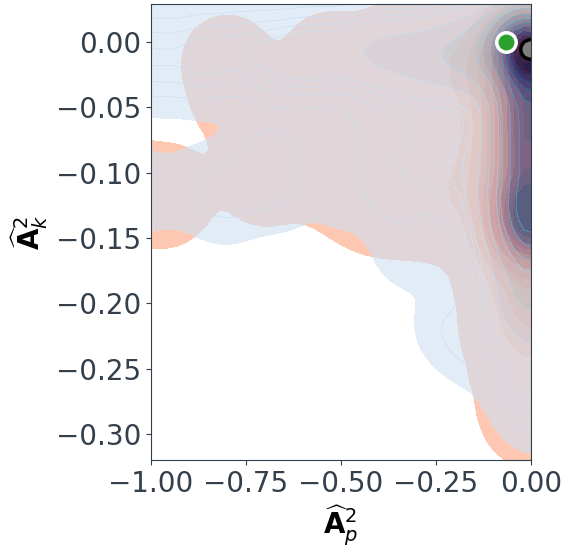}
        \caption{}
        \label{subfig:training_kde_3}
    \end{subfigure}
	\caption{Bivariate kernel density estimation plots showing distributions of the two training data sets used for the convergent-divergent channel test case at $Re_{\tau}=617$ (case 3b). The six most important features (after $Re_d$) are shown. The circular markers indicate the feature values at the two points plotted in Figure~\ref{fig:SHAP}. Density is normalised by the maximum density.}
	\label{fig:training_kde}
\end{figure}

In Figure~\ref{fig:training_kde} kernel density estimates (KDE) show the distribution of training data in feature space, for the two training data sets used for the predictions in Figure~\ref{fig:2b_Caniso_slices}. The six features identified as important for the two locations in question are shown, with $Re_d$ ignored since $Re_d=2$ for all mid-channel locations. The near-wall test point ($z_2=0.72$, green circle) lies within both of the training data distributions. However, the mid-channel test point ($z_2=1.35$, grey circle) lies outside the first training data distribution (red contours), indicating that this training data isn't representative of the test point. The second training data set includes a training case (case 3c) with the same geometry as the test case, and the blue contours show that this training data better represents the mid-channel test point. The reduced extrapolation in feature space (for the six locally important features) explains why the predictive error and uncertainty are reduced in the return to isotropy region (fig.~\ref{subfig:2b_Caniso_slices_call}), when more training data is added. 

\subsection{Periodic hill}
\label{sub:periodic_hill}

\begin{figure}[ht]
  \centering
  \includegraphics[width=0.9\linewidth]{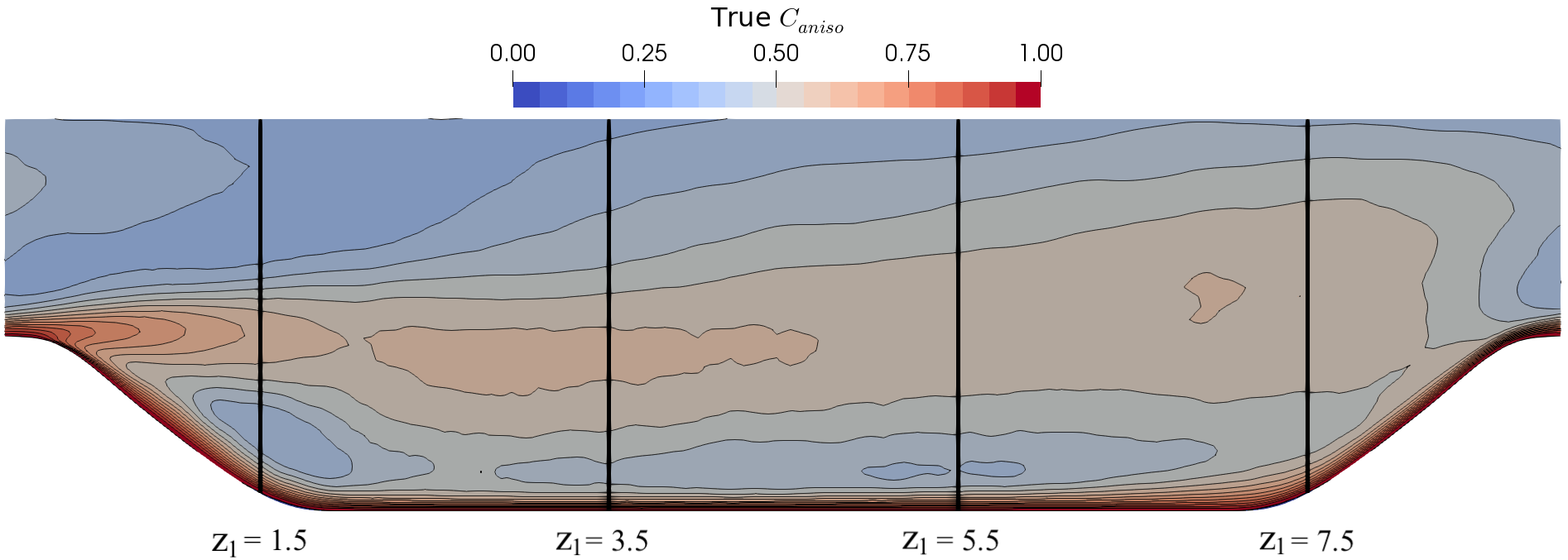}
  \caption{Contours of the \emph{true} turbulent anisotropy constant $C_{aniso}$ for the periodic hill (case 2). The stream-wise periodic boundaries are located at $z_1=0$ and $z_1=9$, with $\boldsymbol{z}$ non-dimensionlised by the hill height.}
  \label{fig:3_Caniso_contours}
\end{figure}

The second case considered is the periodic hill (case 2), for which contours of $C_{aniso}$ from the DNS \cite{Frohlich2005} are shown in Figure~\ref{fig:3_Caniso_contours}. The flow physics here is similar to the previous case, with high turbulence anisotropy observed in the shear layer where the flow separates at the hill crest ($z_1<1.5$). After the flow reattaches to the lower surface around $z_1=2$ the turbulence anisotropy decreases near the wall. When the windward side of the next hill is reached around $z_1=7.5$, the spanwise turbulent stress $\av{u'_3u'_3}$ increases significantly compared to the $\av{u'_1u'_1}$ and $\av{u'_2u'_2}$ stresses near the wall, leading to high $C_{aniso}$ values here. This effect is caused by pressure–strain interactions in this region \cite{Frohlich2005}, and is interesting as many pressure-strain models assume that pressure–strain interactions isotropise the normal Reynolds stresses.

\begin{figure}[ht]
	\centering
	\includegraphics[width=0.8\linewidth]{slices_legend.png}
	\begin{subfigure}[b]{0.9\linewidth}
       	\includegraphics[width=\linewidth]{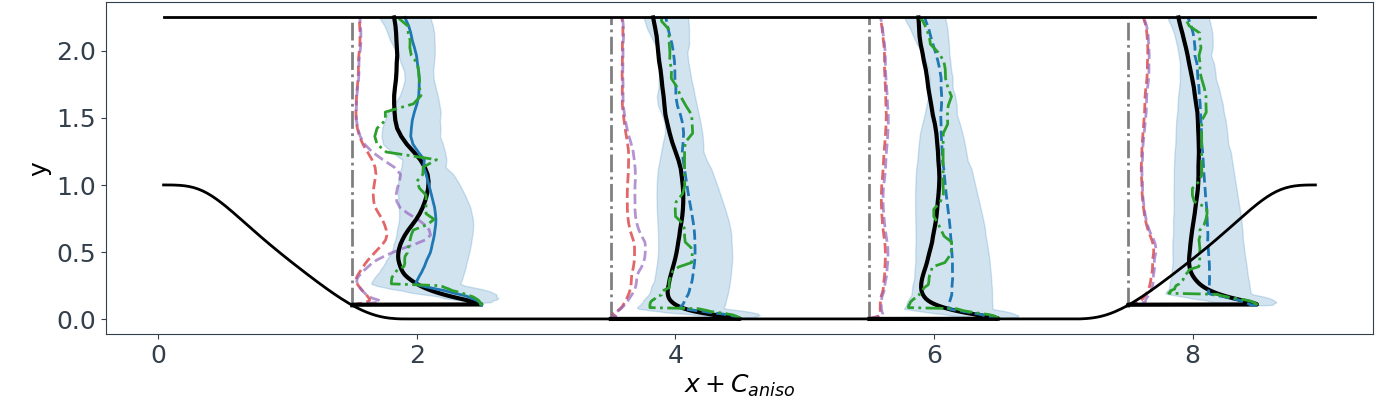}
        \caption{Only duct cases (4a/b/c/d) used for training}
        \label{subfig:3_Caniso_slices_c4}
    \end{subfigure} \vspace{8pt} \\
    \begin{subfigure}[b]{0.9\linewidth}
        \includegraphics[width=\linewidth]{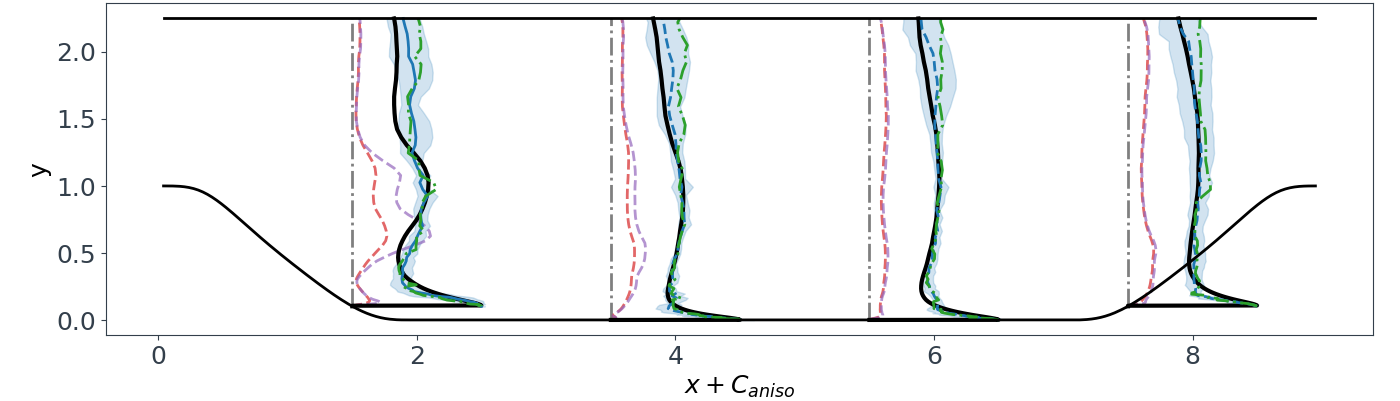}
        \caption{All cases (except for case 2) used for training}
        \label{subfig:3_Caniso_slices_call}
    \end{subfigure}
	\caption{Mondrian forest and random forest predicted turbulent anisotropy constant $C_{aniso}$ for the periodic hill (case 2), at the slices marked in Figure~\ref{fig:3_Caniso_contours}.}
	\label{fig:3_Caniso_slices}
\end{figure}

In Figure~\ref{fig:3_Caniso_slices} the Mondrian forest predictions for $C_{aniso}$ and the \emph{truth} are plotted across the $z_1$-planes shown in Figure~\ref{fig:3_Caniso_contours}. In Figure~\ref{subfig:3_Caniso_slices_c4}, the Mondrian forest is trained only on the four duct flow cases (cases 4a/b/c/d). No pressure induced separation is present in the duct flows, hence these predictions represent a considerable extrapolation. The anisotropy close to the lower wall is well captured, with small uncertainty bounds. However, away from the wall predictive errors are larger and the MF is less confident in its predictions. When the other flow cases (except for case 2) are added to the training data (Fig.~\ref{subfig:3_Caniso_slices_call}) the predictions are significantly more accurate, and the uncertainty bounds are much tighter. Again, the MF achieves significantly better accuracy compared to the LEVM and NLEVM predictions. Additionally, the MF and RF predictions are generally close here, with the MF predictions appearing less noisy.

\subsection{Calibration of predictive uncertainty}
\label{sub:calibration}

\begin{figure}[ht]
	\centering
	\begin{subfigure}[b]{0.365\linewidth}
       	\includegraphics[width=\linewidth]{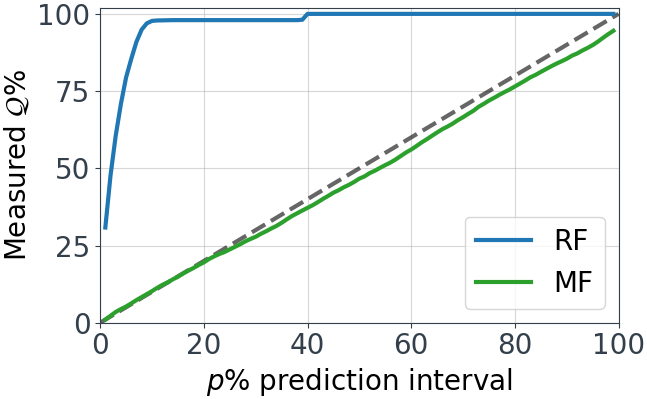}
        \caption{Case 2: Periodic hill}
        \label{subfig:reliability_3}
    \end{subfigure} \hspace{5pt}
    \begin{subfigure}[b]{0.365\linewidth}
        \includegraphics[width=\linewidth]{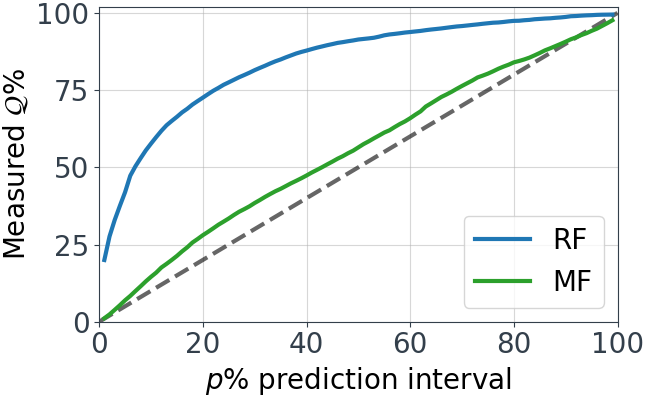}
        \caption{Case 3b: Convergent-divergent channel}
        \label{subfig:reliability_2b}
    \end{subfigure}
	\caption{Probability calibration curves for random forest and Mondrian forest predictions.} 
	\label{fig:reliability}
\end{figure}

In the preceding sections, the MF predictive mean $\mu$ and variance $\sigma^2$ are used to produce prediction intervals where, for example, the $1\sigma$ interval says we expect 68.27\% of the true test values to lie within the interval $\mu \pm 1.0 \sigma$. For these prediction intervals to be trusted, it is important to determine whether they are well calibrated. Following Lakshminarayanan et al.~\cite{Lakshminarayanan2016b}, \textit{probability calibration curves} are plotted in Figure~\ref{fig:reliability}. For each $p\%$ (e.g. $90\%$), the prediction interval $\mu \pm q_p \sigma$ is calculated assuming Gaussian quantiles
\begin{equation}
q_p = \sqrt{2}\text{ erf}^{-1}(p).
\end{equation}
The percentage $\mathcal{Q}\%$ of true test points $\tilde{y}$ that lie within each prediction interval $\mu \pm q_p \sigma$ is then measured. If the model is perfectly calibrated, $\mathcal{Q}\%$ would equal $p\%$ for all $p$. The MF calibration curves for cases 2 and 3b are plotted in Figure~\ref{fig:reliability}, and RF jackknife uncertainty estimates are included for comparison. For both cases, the MF uncertainty estimates are well calibrated, with the MF calibration curves lying close to the ideal dashed line. This means that for a given $p\%$ interval, approximately $p\%$ of the test predictions are expected to lie within the interval. On the other hand, the RF jackknife uncertainty estimates are poorly calibrated, with the confidence intervals\footnote{The jackknife estimate returns confidence intervals not prediction intervals for the random forest (see Sec.~\ref{sub:jackknife}).} displaying significant under-confidence. This suggests the random forest confidence intervals are not suitable to be used as prediction intervals. Additionally, as mentioned previously, preliminary (studies(\ref{sub:hyperparams}) indicated that jackknife re-sampling is prohibitively expensive for use in the framework explored here.

\subsection{Comparison to distance based prediction confidence}
\label{sub:distance_comparision}

A number of other approaches have been suggested to supplement data-driven turbulence modelling frameworks with measures of prediction confidence. \citet{Ling2015} suggest using the Mahalanobis distance, while \citet{Wu2017} suggest a measure based on KDE. Both are statistical \emph{distance measures}, which are used to measure the distance between a test point and a distribution of training data. The Mahalanobis distance is defined as the distance between a point $\boldsymbol{x}$ in feature space and the mean of the training points $\boldsymbol{\mu}$,
\begin{equation} \label{eqn:Dm}
D_m = \sqrt{(\boldsymbol{x}-\boldsymbol{\mu})^T \Sigma^-1 (\boldsymbol{x}-\boldsymbol{\mu})},
\end{equation}
where $\Sigma$ is the covariance matrix of the training points. The normalised Mahalanobis distance for the point $\boldsymbol{x}$ is then defined as $\hat{D}_m=1-\psi_{D_m}$, where $\psi_{D_m}$ is the fraction of training points with a larger raw Mahalanobis distance than the point $\boldsymbol{x}$. It follows that a prediction at a point where $\hat{D}_m=0$ involves no extrapolation, whereas $\hat{D}_m=1$ involves very high extrapolation.

\begin{figure}[ht]
	\centering
	\begin{subfigure}[b]{0.24\linewidth}
       	\includegraphics[width=\linewidth]{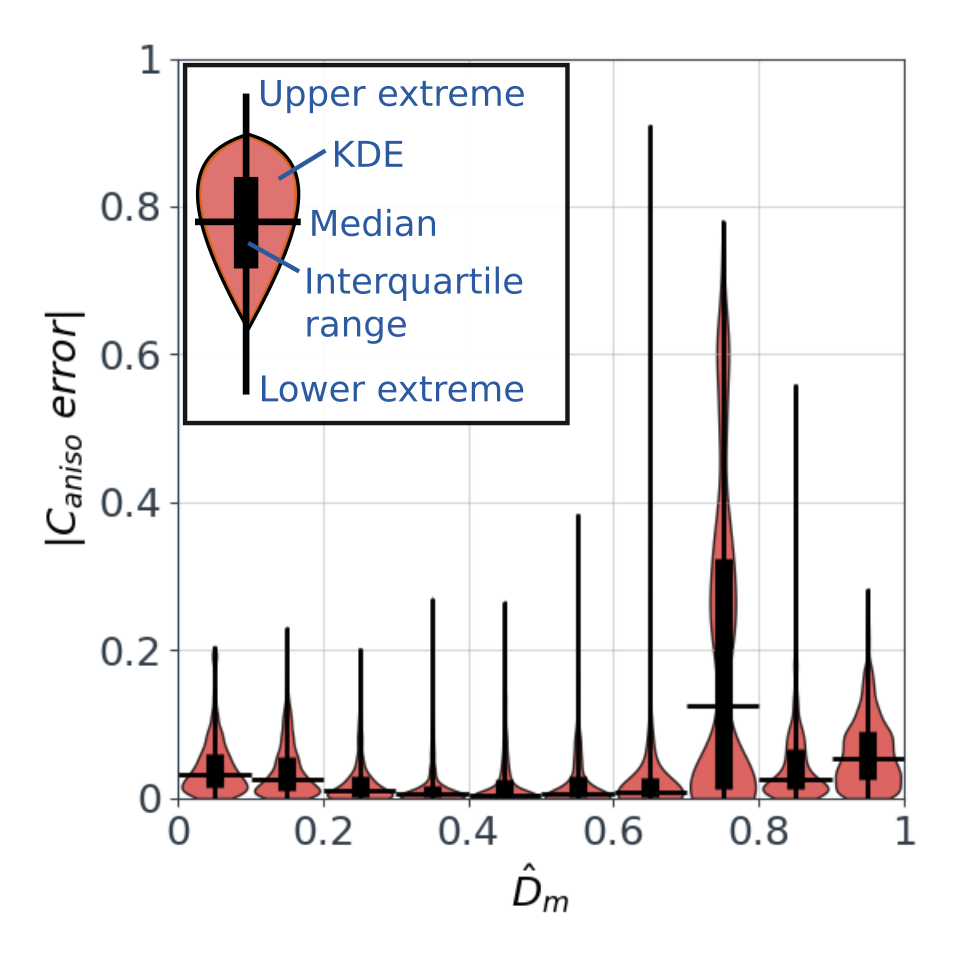}
        \caption{Less extrapolation}
        \label{subfig:2b_error_Dm_violin_alltrain}
            \end{subfigure} \hspace{5pt}
    	\begin{subfigure}[b]{0.24\linewidth}
       	\includegraphics[width=\linewidth]{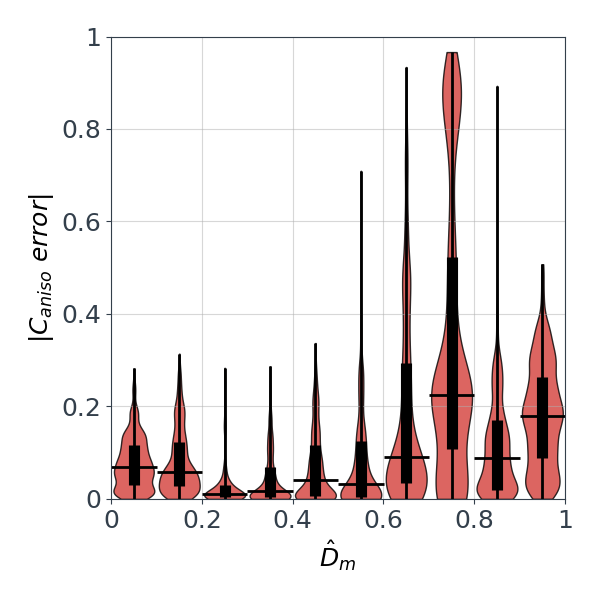}
        \caption{More extrapolation}
        \label{subfig:2b_error_Dm_violin_lowtrain}
    \end{subfigure} \hfil
    	\begin{subfigure}[b]{0.24\linewidth}
       	\includegraphics[width=\linewidth]{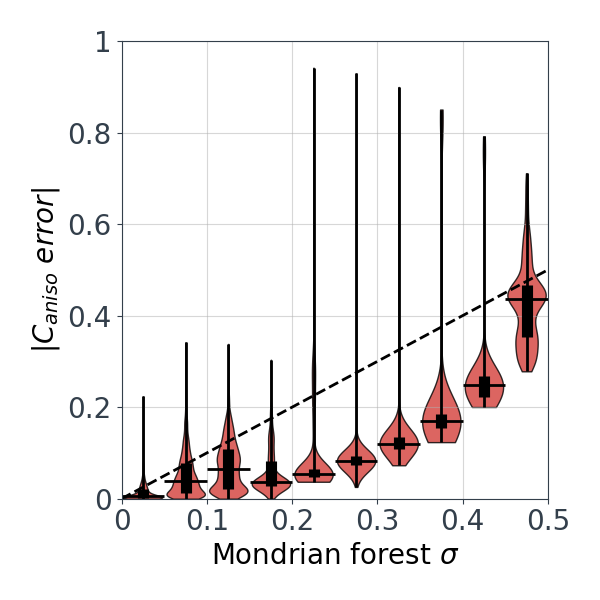}
        \caption{MF, less extrapolation}
        \label{subfig:2b_error_std_violin_MF_alltrain}
    \end{subfigure} \hfil
    \begin{subfigure}[b]{0.24\linewidth}
        \includegraphics[width=\linewidth]{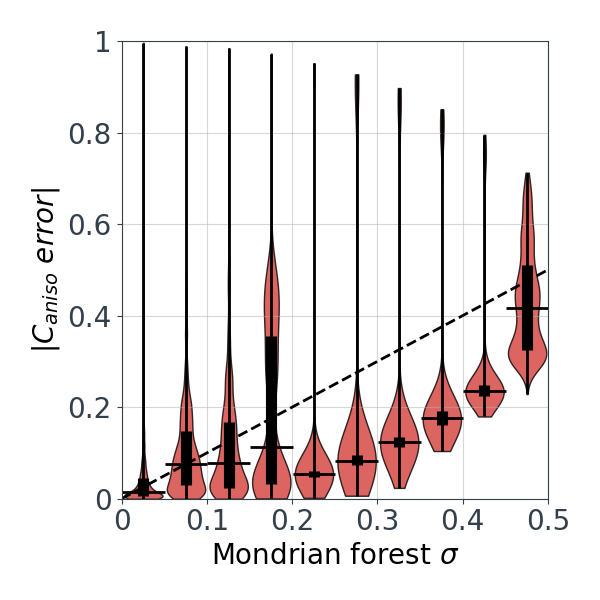}
        \caption{MF, more extrapolation}
        \label{subfig:2b_error_std_violin_MF_lowtrain}
    \end{subfigure}
	\caption{Violin plot of error in random forest predicted turbulent anisotropy constant $C_{aniso}$ versus normalised Mahalanobis distance $\hat{D}_m$ ((a) and (b)), and Mondrian forest error versus predictive uncertainty ((c) and (d)), for the convergent-divergent channel at $Re_{\tau}=617$ (case 3b). For (a) and (b), all the cases except for 3b are used as training data. For (c) and (d), the convergent-divergent channel cases at $Re_{\tau}=395$ and $950$ (3a and 3c) are also removed from the training data, implying a greater extent of extrapolation from the training data.}
	\label{fig:2b_error_std_violin}
\end{figure}

To assess how well $\hat{D}_m$ performs as a measure of prediction confidence, \emph{violin plots} are used to inspect the correlation between $C_{aniso}$ prediction error and $\hat{D}_m$. The normalised Mahalanobis distance $\hat{D}_m$ is discretised into ten bins, and a kernel density plot\footnote{Scott's rule is used to estimate kernel bandwidth.} is generated for each bin. Figures~\ref{subfig:2b_error_Dm_violin_alltrain} and~\ref{subfig:2b_error_Dm_violin_lowtrain} consist of violin plots\footnote{For the RF in this section, $L=1280$ is chosen with no maximum depth to offer slightly more converged $\sigma$ estimates.} of a random forest's predictions for the convergent-divergent channel at $Re_{\tau}=617$ (case 3b). Comparing Figures~\ref{subfig:2b_error_Dm_violin_alltrain} and~\ref{subfig:2b_error_Dm_violin_lowtrain} it is apparent that when there is more extrapolation (Fig.~\ref{subfig:2b_error_Dm_violin_lowtrain}) a greater proportion of the test points have high predictive errors and large $\hat{D}_m$ values. Within the test data there is some correlation between error and $\hat{D}_m$. However, there are many data points where $\hat{D}_m$ is large while the error is small. 

In Figures.~\ref{subfig:2b_error_std_violin_MF_alltrain} and~\ref{subfig:2b_error_std_violin_MF_lowtrain}, the same comparison is shown for the Mondrian forest's predictive uncertainty. Correlation between the MF's predictive error and its predictive uncertainty $\sigma$ is generally much better here, compared to the correlation between error and $\hat{D}_m$. This implies the Mondrian forest's $\sigma$ is a more reliable measure of uncertainty compared to $\hat{D}_m$. However the MF is not infallible, and there are a small number of outliers where the MF is overconfident, with small $\sigma$ yet high error. The addition of an outlier filter like that proposed by \citet{Kaandorp2020} might help remove the small number of outliers seen here.

\section{Data-driven turbulence modelling}
\label{sec:DDRANS}

The study of a Mondrian forest's predictions for the anisotropy constant suggests that Mondrian forests are capable of predicting a turbulent field variable with comparable predictive accuracy to random forests. The predictive uncertainty was also shown to be valuable in informing us whether we can trust the predictions. However, in practice, the end user is usually interested in other flow quantities such as velocity, or lift and drag coefficients. Therefore, a more useful application of the machine learning predictions is to augment an existing RANS solver. It may also be desirable to propagate the Mondrian forests' uncertainties through the RANS solver, to obtain uncertainties in the flow quantities of interest.

To explore the above, we train five Mondrian forests on the five variables in the anisotropy tensor discrepancy vector $\Delta \mathbf{B}= (\Delta \xi, \Delta \eta, h_1, h_2, h_3)$. The test case selected is the convergent-divergent channel at $Re_{\tau}=395$ (case 3a), and training data for $\Delta \mathbf{B}$ is obtained by taking the differences between the RANS and NDS fields from the curved backward step (case 1) and periodic hills (case 2).

\begin{figure}[ht]
	\centering
	\includegraphics[width=0.5\linewidth]{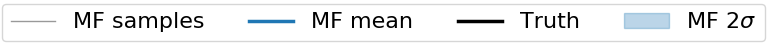}
	\begin{subfigure}[b]{0.9\linewidth}
       	\includegraphics[width=\linewidth]{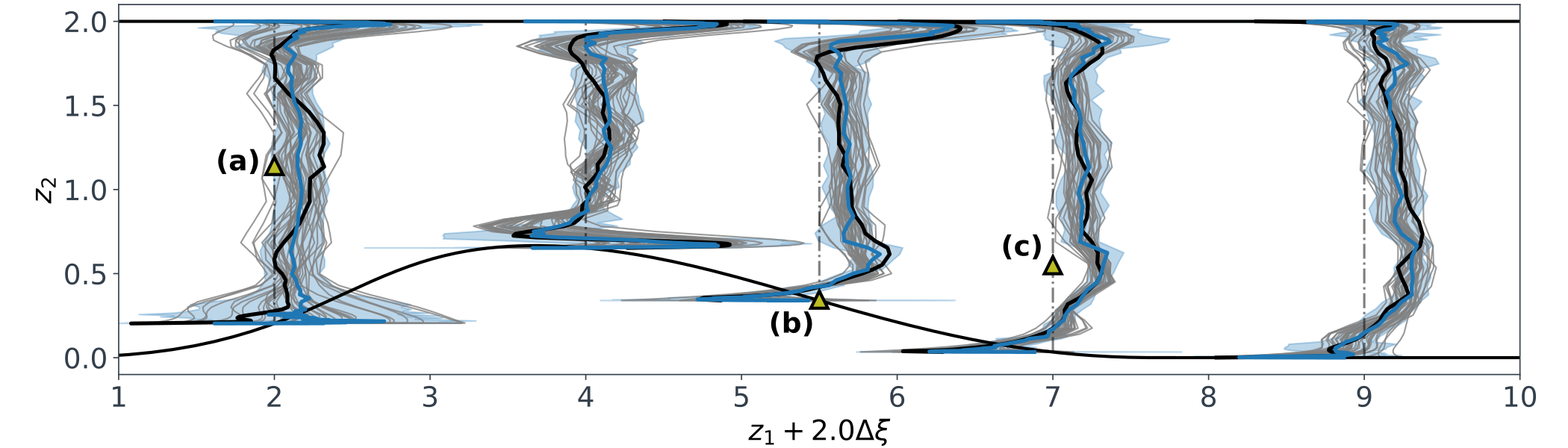}
        \caption{$\xi$ coordinate}
        \label{subfig:2a_samples_in_x}
    \end{subfigure} \vspace{8pt} \\
    \begin{subfigure}[b]{0.9\linewidth}
        \includegraphics[width=\linewidth]{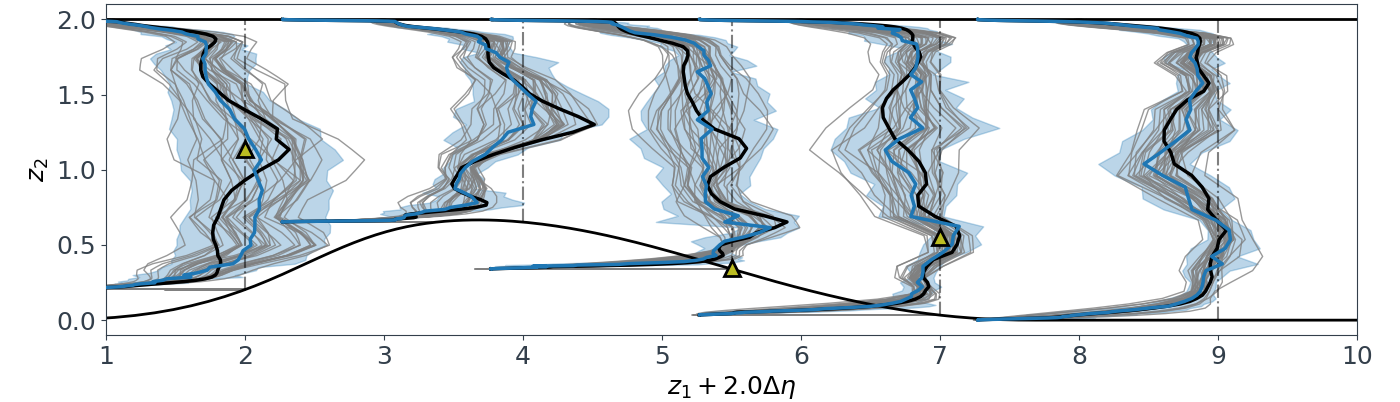}
        \caption{$\eta$ coordinate}
        \label{subfig:2a_samples_in_y}
    \end{subfigure}
	\caption{Mondrian forest predictions of the eigenvalue discrepancies $\Delta \xi$ and $\Delta \eta$, for the convergent-divergent channel at $Re_{\tau}=395$ (case 3a). Also shown are $N_s=50$ samples generated from MF uncertainty using the procedure described in Section~\ref{sub:propagating_uq}.}
	\label{fig:2a_samples_in}
\end{figure}

The resulting Mondrian forest predictions for the eigenvalue discrepancies $\Delta \xi$ and $\Delta \eta$ are visualised in Figure~\ref{fig:2a_samples_in}. Generally, the MF predictions are in good agreement with the true discrepancies derived from comparing the NDS and RANS. However, in some regions the predictive uncertainty is high, especially for $\Delta \eta$. This could be improved by increasing the amount of representative training data, but our focus here is instead on propagating this uncertainty through the CFD solver. This is achieved by generating $N_{s}=200$ sample fields for $\Delta \xi$ and $\Delta \eta$, using the  procedure described in Section~\ref{sub:propagating_uq}. The resulting samples are shown in Figure~\ref{fig:2a_samples_in}, where they are seen to be relatively smooth, whilst correlating closely with the MF uncertainty bounds also shown here.

\begin{figure}[ht!]
	\centering
	\includegraphics[width=0.45\linewidth]{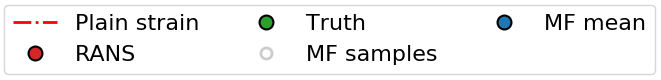} \\ \vspace{2pt}
	\begin{subfigure}[b]{0.28\linewidth}
       	\includegraphics[width=\linewidth]{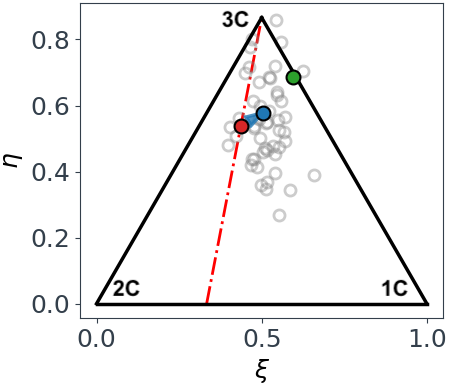}
        \caption{}
        \label{subfig:bary_samples_a}
    \end{subfigure} \hfil
    \begin{subfigure}[b]{0.28\linewidth}
        \includegraphics[width=\linewidth]{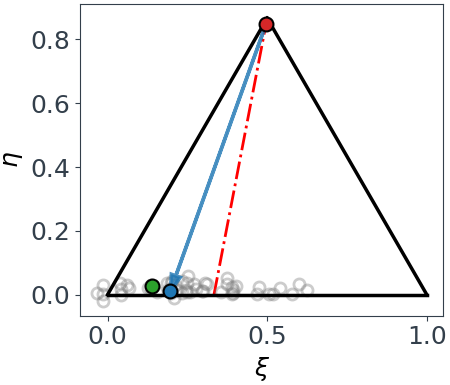}
        \caption{}
        \label{subfig:bary_samples_b}
    \end{subfigure} \hfil
     \begin{subfigure}[b]{0.28\linewidth}
        \includegraphics[width=\linewidth]{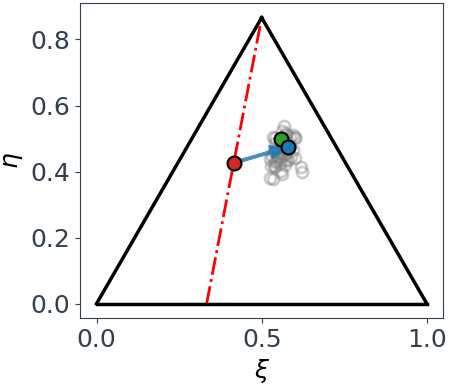}
        \caption{}
        \label{subfig:bary_samples_c}
    \end{subfigure}
	\caption{Turbulent anisotropy visualised on the Barycentric triangle \cite{Banerjee2007} for the points labelled (a), (b) and (c) in Figure~\ref{fig:2a_samples_in}. The true anisotropy from NDS, the MF mean and LEVM predictions, and 50 MF samples are shown.}
	\label{fig:bary_samples}
\end{figure}

In Figure~\ref{fig:bary_samples} fifty of the samples are visualised on the Barycentric triangle (see \cite{Banerjee2007}), for the points labelled (a), (b) and (c) in Figure~\ref{fig:2a_samples_in}. At point (c), the uncertainty in $\Delta \xi$ and $\Delta \eta$ is small and so the samples are clustered closely around the MF mean prediction. Whereas at (a) an (b) there is a high spread in the samples due to the high uncertainty here. When the uncertainty is high, there is a danger of samples being outside of the Barycentric triangle, leading to unrealisable Reynolds stresses. To prevent this, the modified SU2 solver (see Sec.~\ref{sub:propagating_mean}) constrains the Cartesian coordinates $(\xi,\eta)$ to lie within the triangle:
\begin{equation}
\begin{split}
& \eta^* = \max \left[0, \min \left(\frac{\sqrt{3}}{2}, \eta \right) \right] \\
& \xi^* = \max \left[\frac{1}{3}\eta, \min \left( 1-\frac{1}{3}\eta,\xi \right) \right]
\end{split}
\end{equation}

\begin{figure}[ht!]
	\centering
	\includegraphics[width=0.27\linewidth]{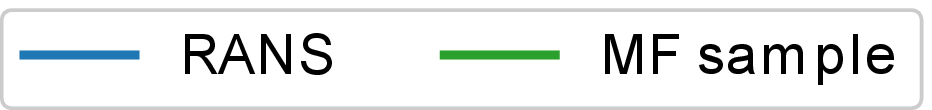} \\ \vspace{2pt}
	\begin{subfigure}[b]{0.45\linewidth}
       	\includegraphics[width=\linewidth]{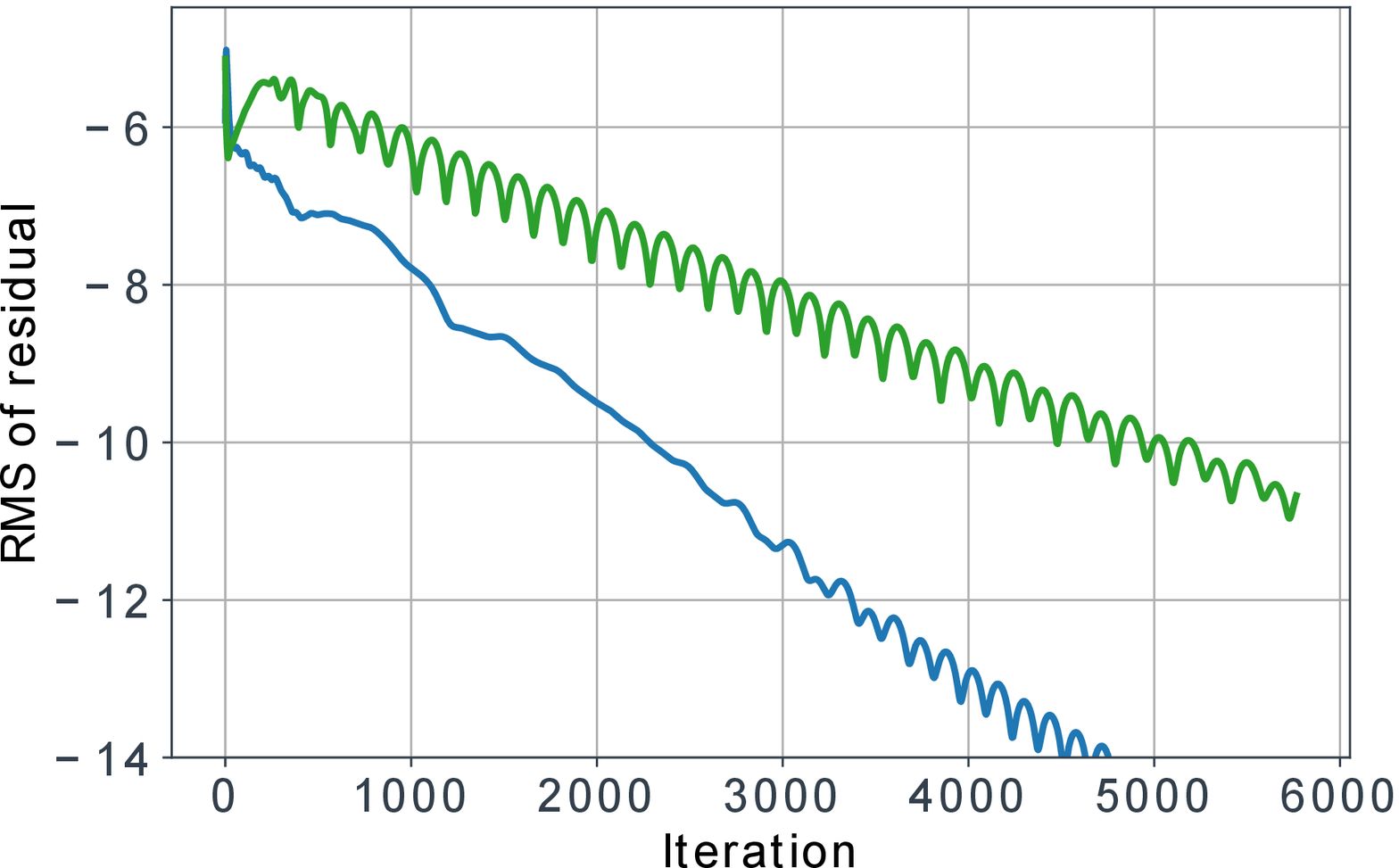}
        \caption{Pressure equation}
        \label{subfig:convergence_p}
    \end{subfigure} \hfil
    \begin{subfigure}[b]{0.45\linewidth}
        \includegraphics[width=\linewidth]{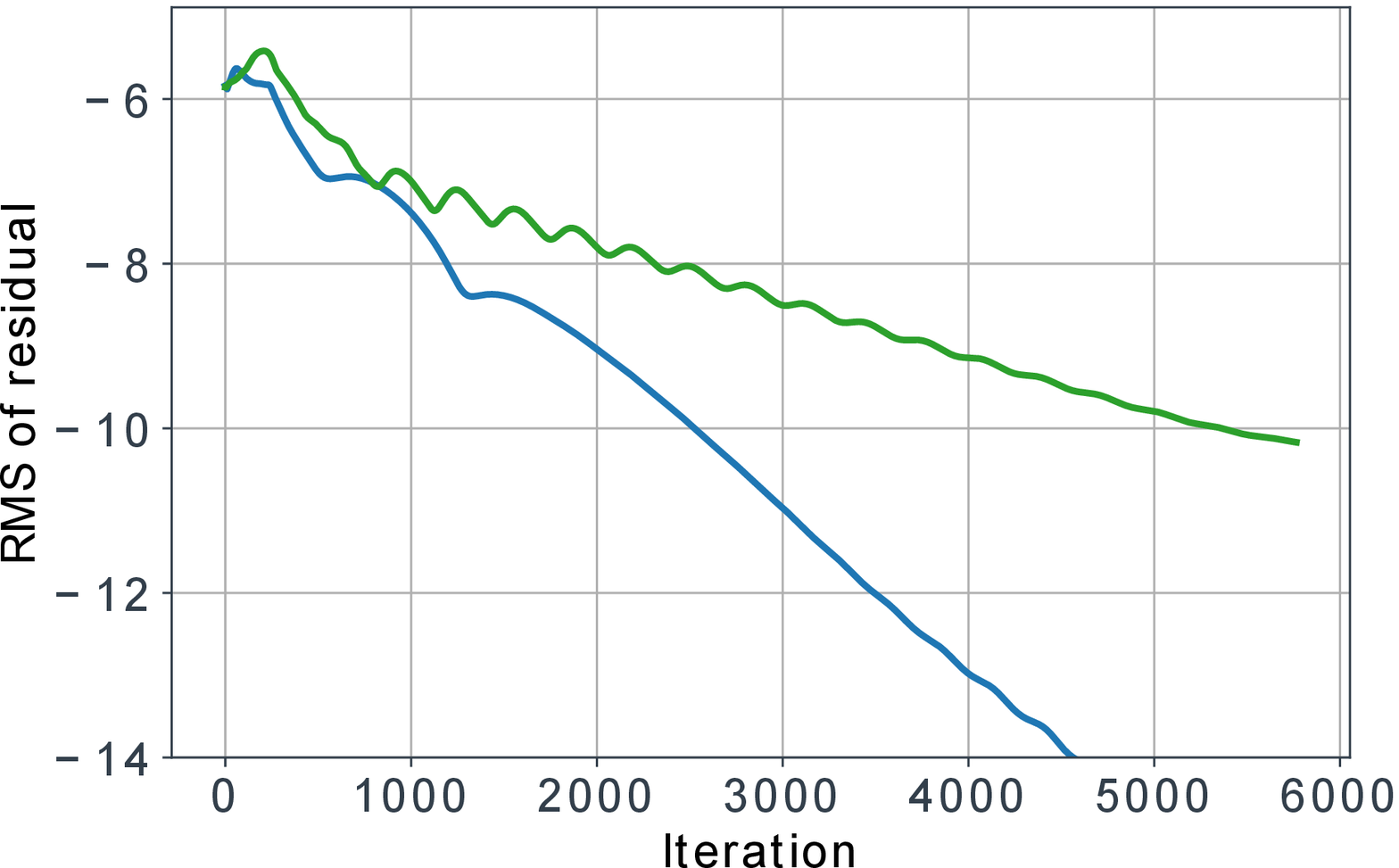}
        \caption{Turbulent kinetic energy equation}
        \label{subfig:convergence_k}
    \end{subfigure} \hfil
	\caption{Convergence histories for case 3a. Shown are the convergence history for the original RANS solution, and the convergence history when a randomly selected MF sample is propagated through the modified solver.}
	\label{fig:convergence}
\end{figure}

The samples can now be propagated through the modified SU2 solver, to obtain flowfield predictions for each sample. Samples are generated for $\Delta \xi$ and $\Delta \eta$ only, and for the eigenvector discrepancies $h_1, h_2, h_3$ we propagate the Mondrian forests' mean predictions. The modified solver is run for $N_s=200$ samples, with the converged baseline RANS solution for case 2a used as the initial condition. The CFL number is reduced to 20, and the ramp parameters set to $\gamma_{max}=0.9$ and $n_{max} = 100$. As seen in Figure~\ref{fig:convergence}, convergence is slower than for the original RANS simulation, but it is still acceptable. 

\begin{figure}[ht]
  \centering
  \includegraphics[width=0.98\linewidth]{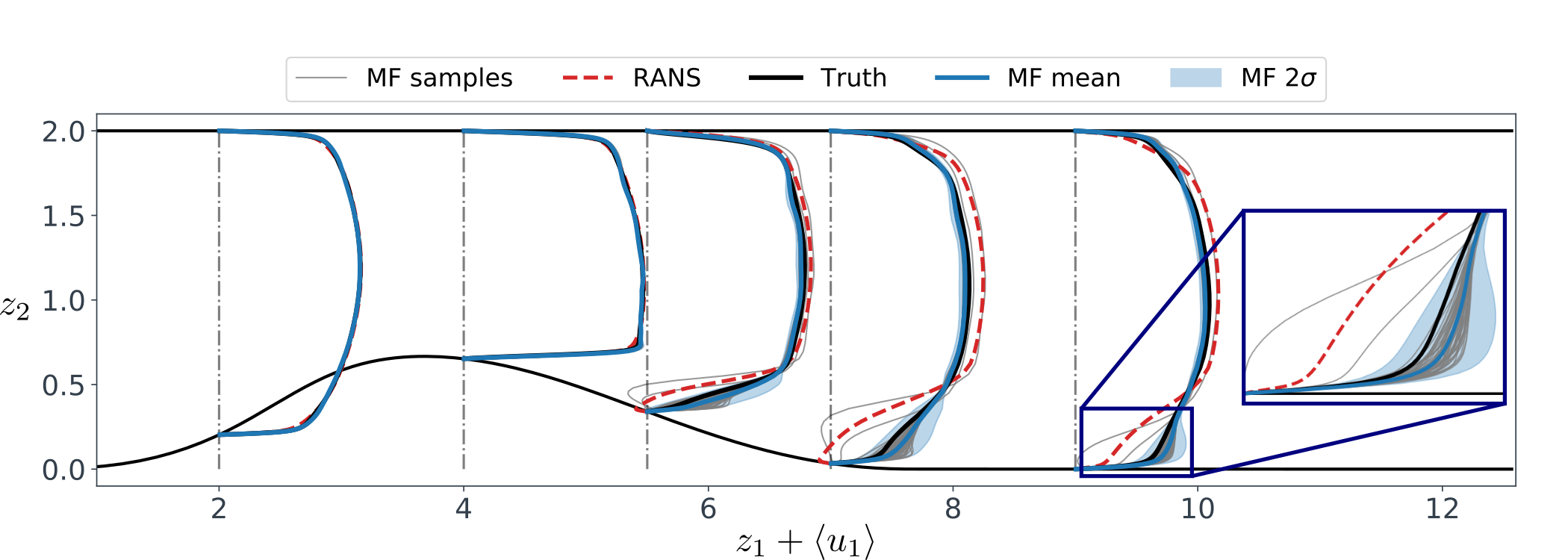}
  \caption{Profiles of axial velocity for the convergent-divergent channel at $Re_{\tau}=395$ (case 3a), when the Mondrian forest correction for $\Delta \mathbf{B}$ is propagated through the CFD solver. The original RANS prediction, and $N_s=50$ samples are also shown. The MF $2\sigma$ uncertainty bounds are obtained by assuming the $N_s=200$ samples are normally distributed about the MF mean.}
  \label{fig:2a_samples_out_U}
\end{figure}

The resulting profiles of axial velocity are shown in Figure~\ref{fig:2a_samples_out_U}. There are slight differences between the MF mean and the \emph{truth} from the NDS solution, which might be due to the small amount of the original RANS anisotropy tensor which was blended in for stability (recall $\gamma_{max}=0.9$). This parameter could potentially be increased, at the expense of less favourable convergence behaviour. However, the predictions are generally in good agreement with the NDS, and are a significant improvement over the original RANS. Also shown are a selection of the samples obtained from the aforementioned sampling procedure. Despite the relatively high uncertainties in $\Delta \xi$ and $\Delta \eta$ observed in Figure~\ref{fig:2a_samples_in}, the uncertainty bounds obtained for the velocity are relatively small overall. This is likely to be because the large uncertainties in $\Delta \xi$ and $\Delta \eta$ are located away from the wall, where the Reynold stresses are relatively small. The uncertainty in velocity is larger in the separated flow region downstream of the divergent section. This suggests that the onset of flow separation, and the flow reattachment, are sensitive to the uncertainty in the ML predicted eigenvalues. This is in agreement with the studies of \citet{Gorle2019}, who quantify the epistemic uncertainty of RANS models by perturbing the eigenvalues and eigenvectors of the anisotropy tensor.

\begin{figure}[ht]
  \centering
  \includegraphics[width=0.5\linewidth]{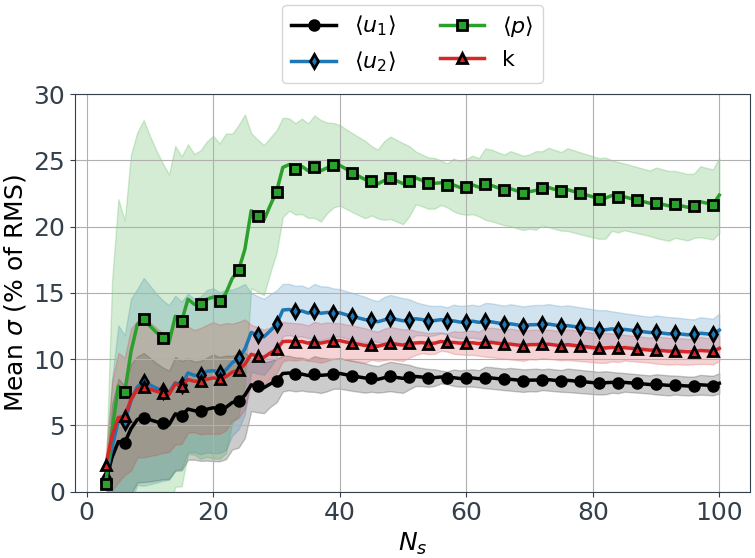}
\caption{Spatial average of standard deviation in flow quantities across different numbers of samples. For each number of samples, ten different random selections of samples are taken from the full $N_s=200$ set. The solid lines depict the mean of the spatial averages across the ten trials, whilst the shaded areas depict the standard deviation of the spatial averages.}
  \label{subfig:sigma_vs_Nsamp}
\end{figure}

A final important aspect of the proposed data-driven framework is its computational cost. Training and prediction times for the Mondrian forests are of the order of seconds for the datasets used here (see Fig.~\ref{subfig:feature_selection_time}), hence the cost of propagating the MF predictions through the CFD solver is the primary concern. With the propagation of each sample taking approximately as long to converge as the original RANS solution, there remains the question of how many samples must be propagated in order to achieve well converged uncertainty estimates for the output quantities of interest. To explore this, we monitor the spatially averaged standard deviation of $\av{u_1}$, $\av{u_2}$, $\av{p}$ and $k$, whilst randomly selecting different numbers of samples from the 200 converged CFD solutions. As seen in Figure~\ref{subfig:sigma_vs_Nsamp}, satisfactory convergence is generally achieved with $N_s \ge 30$, although the uncertainty bounds for the pressure field are slightly more sensitive to the exact samples taken. If the uncertainties for the eigenvector rotations were also propagated, it is likely that more samples would be required due to the increased number of degrees of freedom. 

To provide some points of comparison, in order to quantify the epistemic uncertainty of a RANS model, \citet{Gorle2019} and \citet{Xiao2017} use $N_s=5$ and $N_s=100$ CFD evaluations respectively. \citet{Gorle2019} are able to only run $N_s=5$ evaluations, because they consider the corners of the Barycentric triangle to be the limiting states of $(\eta,\xi)$, and therefore only perturb $(\eta,\xi)$ towards these three states. A similar approach cannot be taken in our case, since we are quantifying the uncertainty arising from augmenting the RANS model with ML predictions, and not the uncertainty of the RANS model itself. Figures~\ref{subfig:bary_samples_a} and~\ref{subfig:bary_samples_b} show that the MF's uncertainty can be strongly anisotropic, which implies uniformly perturbing towards the corners isn't appropriate. On a similar note, Figure~\ref{fig:bary_samples} shows that the \emph{true} turbulence anisotropy state is often not located at one of the corners. It follows that simply perturbing to a single corner state is unlikely to be a reliable way to improve RANS predictions. This provides further motivation for data-driven approaches such as the one proposed in this paper.

\section{Conclusions}

For the emerging field of data-driven turbulence modelling to enter mainstream use, methods to quantify the uncertainties arising from such approaches will be necessary. To this end, the present work explores the use of a recently proposed machine learning algorithm, Mondrian regression forests, for data-driven turbulence modelling. 

In Section~\ref{sec:MF_vs_RF} Mondrian forests were found to offer comparable accuracy to random forests when predicting the turbulence anisotropy constant derived from near direct simulations (NDS). Interestingly, feature selection indicated that irrelevant features did not degrade the Mondrian forests' accuracy in this case. However, this may be specific to the flow cases and feature set used in this paper, and it would be prudent to perform a similar feature selection procedure on any new training data. The Mondrian forests' uncertainty estimates appeared to provide a good measure of prediction confidence, with high uncertainty in regions of the flow where predictions were far from the truth data. Adding more representative training data generally reduced both predictive errors and uncertainty. Comparing to an a priori distance measure suggested by \citet{Ling2015}, the Mondrian forest uncertainty estimates showed better correlation with predictive errors with fewer regions of significant under or over-confidence. Additionally, probability calibration curves suggest the Mondrian forest uncertainty estimates are well calibrated, which is important in order for the prediction intervals to be trusted. Furthermore, hyperparameter tuning demonstrated that the Mondrian forests achieve converged uncertainty estimates with a considerably computational cost compared to random forests with jackknife estimates.

To explore the suitability of Mondrian forests in a data-driven turbulence modelling framework, in Section~\ref{sec:DDRANS} they were trained to predict the turbulent anisotropy tensor discrepancies. Using a modified RANS solver similar to that described by \citet{Kaandorp2020}, the Mondrian forest predicted mean discrepancy vector was successfully propagated forward to obtain converged flowfields. Depending on the use case propagating only the mean prediction might be sufficient, and the Mondran forests' uncertainties can still be used to decide whether it is worth proceeding with this, or whether more suitable training data is required. However, in other cases it might be desirable to propagate the uncertainties themselves. As demonstrated in Section~\ref{sec:DDRANS}, this allows for the quantification of uncertainty in quantities of interest such as velocity and pressure due to the uncertainty in the Mondrian forest's predictions. 

The above quantification of uncertainties is an important step if such approaches are to enter mainstream use, but future work must also address other sources of uncertainty in the ML predictions. For example, the point-wise locality assumption made in the present framework, and any differences between the RANS and NDS mean velocity and pressure fields. The first error source could be mitigated by introducing additional flow features to account for non-locality. The second error source could perhaps be reduced by modifying the present framework so that input-output training data is derived purely from NDS data. Improved accuracy could also be achieved by improving the Mondrian forest implementation itself. 

Nevertheless, based on the aforementioned findings, for data-driven turbulence modelling Mondrian forests appear to offer a promising alternative to random forests, and other probabilistic methods such as BART trees or Bayesian neural networks. A final potential area of future work is online learning. Mondrian forests and trees can be updated in an online fashion as new data becomes available, and the resulting models should be identical to their batch trained counterparts. This capability can save time, as the model must not be completely retrained when new data is obtained. Additionally, the Mondrian trees are only altered in the region of feature space populated by the new online data. As an example, this would allow for new higher Reynolds number training data to be added, without affecting the model's predictions at a lower Reynolds number. Such capability might be important in an industrial setting where repeatability is crucial.

\section*{Acknowledgement}
This work was supported by Wave 1 of The UKRI Strategic Priorities Fund under the EPSRC Grant EP/T001569/1, particularly the \emph{Digital twins for complex systems engineering} theme within that grant and The Alan Turing Institute, and by the Lloyd’s Register Foundation-Alan Turing Institute programme on Data-Centric Engineering under the LRF grant G0095. The authors thank the anonymous reviewers for their detailed and insightful comments, which helped improve the quality and clarity of the manuscript.

\appendix
\section{Mondrian forests and random forests}
\label{appendixA}

\subsection{Decision trees}
\label{sub:decision_tree}

Following the notation of \cite{Lakshminarayanan2016b}, we describe a decision tree by the tuple $(\mathsf{T},\boldsymbol{\delta},\boldsymbol{\xi})$, where $\mathsf{T}$ is the tree, $\boldsymbol{\delta}_j\in \left\lbrace 1,\dots,d\right\rbrace$ is the \emph{split dimension} and $\boldsymbol{\xi}_j\in\mathbb{R}$ is the \emph{split location} for the $j^{th}$ node in the tree. A decision tree trained on the training data $\left( \mX_{N}, \vy_{N} \right)$ is a hierarchical partitioning of the input data. At each node in the tree, the data is split in a binary fashion
\begin{equation}
\begin{split}
B_{left(j)}   &:=\left\lbrace \mathbf{x} \in B_j:x_{\delta_j} \le \zeta_j \right\rbrace \\ 
B_{right(j)} &:=\left\lbrace \mathbf{x} \in B_j:x_{\delta_j} > \zeta_j \right\rbrace
\end{split},
\end{equation}
where the $j^{th}$ block of data is
\begin{equation}
B_j = \left(l_{j1},u_{j1}\right] \times \dots \times \left(l_{jd},u_{jd}\right],
\end{equation}
with $l_{jd}$ and $u_{jd}$ the lower and upper bounds of the rectangular block $B_j$ along dimension $d$. As an example, for the decision tree fitted to the input data $\boldsymbol{x} \in [0,1]^2$ in Figure~\ref{subfig:tree_decision}, the root node $\varrho$ splits with $\delta_1 = 2$ and $\xi_1 = 0.6$, leading to its right child node $j=right(\varrho)$ having the data block $B_j = \left(0,1 \right] \times  \left(0.6,1 \right]$.

There are many induction algorithms available to learn a decision tree structure from training data, such as the popular CART algorithm \citep{Breiman1984} for classification and regression. Generally, these algorithms learn the tree structure $\mathsf{T}$ and leaf node parameters $\boldsymbol{\theta}$ by greedily optimising an appropriate criterion, such as mean squared error (MSE) for regression. The parameter $\boldsymbol{\theta}_j$ parametrises the conditional distribution $p(y| \vx \in B_j)$, and for regression is simply the mean of the $K_j$ number of responses $y_K$ residing in the leaf node's block $B_j$:
\begin{equation}
\boldsymbol{\theta}_j = \frac{1}{|K_j|}\sum_{q\in N_j} y_q
\end{equation}
For a given test data point $\tilde{\vx}$, a prediction is made by walking through the decision tree to identify its corresponding leaf node $j$ and then returning the parameter $\boldsymbol{\theta}_{j}$. 

\subsection{Random forests}
\label{sub:forests}

Let $\boldsymbol{\varphi}$ be a random forest consisting of $L$ decision trees $ \mathsf{T}_1,\dots, \mathsf{T}_L$. For the data point $\tilde{\vx}$, the random forest's prediction is an average of each tree's prediction
\begin{equation}
\boldsymbol{\varphi}\left( \tilde{\vx} \right) = \frac{1}{L}\sum_{i=1}^L w_i \mathsf{T}_i \left(  \tilde{\vx} \right)
\end{equation}
where $w_i$ is the $i$-th weighting term\footnote{All $w_i$'s are usually set to unity for random forests.}. The individual trees in a random forest are randomised with bootstrap aggregation (\emph{bagging}), where each tree is trained on a slightly different subset of the training data, and additionally the set of candidate splits within each node are randomly sub-sampled. This randomisation may slightly increase bias, but it significantly decreases over-fitting (prediction variance). The random forest regressor algorithm from \url{github.com/scikit-learn/scikit-learn} is used in this paper. Unless otherwise stated, all hyper-parameters are kept at their defaults (as of version 0.22.1).

\subsubsection{Jackknife variance estimates}
\label{sub:jackknife}

\citet{Wager2014} propose the use of jackknife re-sampling to provide confidence intervals for random forest predictions. The \emph{infinitesimal jackknife}, provides a variance measure, i.e.,  $V_{IJ}= \text{var}\left(\boldsymbol{\varphi}( \tilde{\vx} )\right)$ for a random forest's predicted response to the input $\tilde{\vx}$. For a random forest $\boldsymbol{\varphi}$ trained on the data $\left( \mX_{N}, \vy_{N} \right)$, the infinitesimal jackknife variance is given by
\begin{equation}
V_{IJ} = \sum_{i=1}^N \left[ \frac{1}{L} \sum_{j=1}^L \left( | \kappa_i |_j -1 \right) \left( \mathsf{T}_j( \tilde{\vx} ) - \overline{\mathsf{T}}( \tilde{\vx} ) \right)\right]
\label{equ:infinite_jack}
\end{equation}
where $| \kappa_i |_j$ denotes the number of times $\kappa_i$ appears in the $j^{th}$ bootstrap sample, $\mathsf{T}_j( \tilde{\vx})$ is the predicted response of the $j^{th}$ tree, and $ \overline{\mathsf{T}}( \tilde{\vx})$ is the mean of $\mathsf{T}_j( \tilde{\vx})$ over $j=1, \ldots, L$. The variance in \eqref{equ:infinite_jack} can be biased upwards when $L$ is small, therefore a bias corrected version is suggested
\begin{equation}
V_{IJ-U} = V_{IJ} - \frac{N}{L^2} \sum_{j=1}^L \left( \mathsf{T}_j( \tilde{\vx} ) - \overline{\mathsf{T}}( \tilde{\vx} )  \right)^2.
\end{equation}
A \emph{Jackknife-after-bootstrap} variance estimate is also defined, and \citet{Wager2014} suggest that the arithmetic mean of this and $V_{IJ-U}$ provide more unbiased variance estimates. However, since the infinitesimal jackknife $V_{IJ-U}$ tends to overestimate the variance, this is used alone here as a conservative estimate. The standard deviation $\sigma= \sqrt{V_{IJ-U}}$ provides \emph{confidence intervals}. They measure how far the random forest prediction $\boldsymbol{\varphi}( \tilde{\vx} )$ -- obtained by building $M$ trees on a sample -- is from its \emph{expected value}, which is the average random forest prediction obtained by building $M$ trees across different samples. They do not provide \emph{prediction intervals}, meaning they do not tell us how far the predicted value $\boldsymbol{\varphi}( \tilde{\vx} )$ is from the true value $\tilde{y}$. 

\subsection{Mondrian forests}
\subsubsection{Mondrian trees}
\label{sub:Mondrian_trees}

Mondrian trees are restrictions of Mondrian processes to a finite set of points. Mondrian processes are families ${\mathcal{M}_t : t \in [0,\infty)}$ of random hierarchical binary partitions of $\mathbb{R}^d$, where $\mathcal{M}_t$ is a refinement of $\mathcal{M}_s$ wherever $t > s$.\footnote{$t$ is referred to as a \emph{time}, but this should not be confused with a physical time related to the data, or to discrete time in an online learning setting.} A Mondrian tree $\mathcal{T}$ is a tuple $(\mathsf{T},\boldsymbol{\delta},\boldsymbol{\xi},\boldsymbol{\zeta})$, where $(\mathsf{T},\boldsymbol{\delta},\boldsymbol{\xi})$ is a decision tree. The additional parameter $\boldsymbol{\zeta}=\{\zeta\}_{j\in \mathsf{T}}$ specifies the split time $\zeta$ for each node $j$. Split times increase with depth of the tree, and are involved in the setting of a hierarchical prior (amongst other things).

\subsubsection{Hierarchical prior and predictive posterior}

Mondrian trees are probabilistic models, which determine $p_{\mathcal{T}}(\tilde{y} | \tilde{\vx} , \left(\mX, \vy \right) )$. \citet{Lakshminarayanan2016b} assume the responses in each leaf node are Gaussian distributed, so that every node $j \in \mathsf{T}$ has a mean parameter $\mu_j$. A hierarchical Gaussian prior is then used for $\boldsymbol{\mu} = \left\lbrace \mu_j:j \in \mathsf{T} \right\rbrace$, such that
%
\begin{equation}
\mu_{\varrho} | \mu_H \sim \mathcal{N} \left(\mu_H, \phi_{\varrho}\right)\text{,}\;\;\; \mu_j | \mu_{ \text{parent}(j)} \sim \mathcal{N} \left( \mu_{\text{parent} (j)},\phi_j \right),
\end{equation}
where $\varrho$ denotes the root node, and $\phi_j = \gamma_1 \sigma (\gamma_2 \zeta) - \gamma_1 \sigma (\gamma_2 \zeta{parent(j)})$. The sigmoid function $\sigma(t) = \left( 1 + e^{-t}\right)^{-1}$ encodes the prior assumption that children are expected to be more similar to their parent nodes as tree depth increases. The hyperparameters $\mu_H, \gamma_1, \gamma_2$ are set according to Appendix B in Ref.~\cite{Lakshminarayanan2016b}.

As discussed in \ref{sub:decision_tree}, for a typical decision tree the predicted response $\tilde{y}$ for a test point $\tilde{\vx}$ is simply the average of the responses in $B^x_{\text{leaf}( \tilde{\vx})}$. With Mondrian trees, the test point can \emph{branch off} the existing tree at any point along the path from the root node to $\text{leaf}(\tilde{\vx})$. Hence, the predictive posterior over $y$ is a weighted mixture of Gaussians along the path from the root node to $\text{leaf}( \tilde{\vx} )$
\begin{equation}
p_{\mathcal{T}} \left( \tilde{y} |  \tilde{\vx} , \left(\mX, \vy \right) \right) = \sum_{j \in \text{path}(\text{leaf}( \tilde{\vx} ))} w_j \mathcal{N} \left( m_j, v_j \right),
\end{equation}
where the weight $w_j$ describes the probability of branching off just before reaching the $j^{th}$ node, and $m_j$ and $v_j$ are the predictive mean and variance at the $j^{th}$ node. As the test point $\tilde{\vx}$ moves further away from the training data at a given node $j$, i.e. as $\tilde{\vx}$ moves away from the block $B_j^x$ in Figure~\ref{subfig:tree_mondrian}, the probability $w_j$ increases. This causes Mondrian forests to exhibit higher uncertainty as $\tilde{\vx}$ moves further away from the training data $\left( \mX, \vy \right)$. Additionally, the Mondrian forest's predicted response $\tilde{y}$ approaches the prior as $\tilde{\vx}$ moves further away from the training data. To obtain the predictive mean and variance at each node, we require $p_{\mathcal{T}} \left( \boldsymbol{\mu} | \left(\mX, \vy \right) \right)$, the posterior over $\boldsymbol{\mu}$. This is computed using Gaussian belief propagation. \citet{Lakshminarayanan2016b} note that, since a hierarchical tree structure is used, posterior inference can be performed with a computational cost of $O(N)$ for $N$ training samples. This is compared to a Gaussian process whose computational cost is typically $O(N^3)$. 

\subsubsection{Forests of Mondrian trees}
\label{sub:mondrian_forests}

\citet{Lakshminarayanan2016a} propose combining Mondrian trees to form a Mondrian forest as a way to reduce over-fitting behaviour. The prediction of a Mondrian forest is then the average prediction from the $L$ number of Mondrian trees
\begin{equation}
p(\tilde{y} |  \tilde{\vx} , \left(\mX, \vy \right) ) = \frac{1}{L}\sum_j^{L} p_{j}( \tilde{y} |  \tilde{\vx} , \left(\mX, \vy \right) ).
\end{equation}
Again, this predictive posterior over $\tilde{y}$ is a mixture of Gaussians, and it is straightforward to calculate the predictive mean and variance from this. For further details see Section 3 in \cite{Lakshminarayanan2016b}.

Following the approach of other practitioners, such as \citet{Mourtada2019}, bootstrap aggregation is not used when training Mondrian forests in this paper. Due to the randomisation involved in the construction of Mondrian trees, each tree is still likely to be different. The Mondrian forest regressor algorithm from \url{github.com/scikit-garden/scikit-garden} is used in this paper. \citet{Lakshminarayanan2016b} control the depth of their Mondrian trees by stopping the splitting of nodes which have less than a given number of data points. To provide a fairer comparison with the random forest algorithm, in this paper we set this hyperparameter to two and additionally limit the depth of Mondrian trees with the $\mathcal{D}_{max}$ hyperparameter.

\section{Feature selection and hyperparameter tuning}
\label{appendixB}

This section documents preliminary studies done to select training features and suitable hyperparameters. In the studies here, random forests and Mondrian forests are trained on all of the flow cases listed in Table~\ref{tab:cases}, with the turbulent anisotropy constant $C_{aniso}$ defined in \eqref{eqn:Caniso} the response variable to be predicted.  

\subsection{Measuring predictive accuracy}
\label{sub:accuracy}

To quantify predictive accuracy two metrics are used. The coefficient of determination, or $R^2$ score, given by
\begin{equation} \label{eqn:R2}
R^2 = 1 - \frac{\sum_{n=1}^N (y_n-\varphi(\boldsymbol{x}_n))^2}{\sum_{n=1}^N (y_n-\bar{y})^2}
\end{equation}
measures how well the true data $y$ is replicated by the model's predictions, $\varphi(\boldsymbol{x}_n)$. The $R^2$ score spuriously increases with the dimension $d$ (recall $\boldsymbol{x} \in \mathbb{R}^D$), therefore score is adjusted to correct for this:
\begin{equation}
\bar{R}^2 = 1 - (1-R^2)\frac{N-1}{N-D-1}.
\end{equation}
The mean absolute error (MAE) provides a measure of error in the model's predictions

\begin{equation} \label{eqn:MAE}
MAE = \frac{1}{N} \sum_{n=1}^N |y_n - \varphi(\boldsymbol{x}_n)|,
\end{equation}
and is given as a percentage of the true mean response $\bar{y}$  in this paper. MAE is chosen over root-mean-square error (RMSE) due to RMSE's sensitivity to sample size and exaggeration of small numbers of large errors. 

When MAE is calculated based on the training response $\vy$ and training data predictions $\varphi(\boldsymbol{x})$ it is referred to as the \emph{training error}, meanwhile when it is based on the test data $y^*$ and $\varphi(\boldsymbol{x}^*)$ it is referred to as the \emph{test error}. The out-of-sample error, or test error, is important as it assesses how well the machine learning model generalises to data it has not seen during training. As is standard practice amongst machine learning practitioners, cross validation is used to provide fair error measures. The commonly used $k$-fold cross validation involves randomly partitioning $\boldsymbol{D}_{1:N}$ into $k$ folds. Training is then performed on $k-1$ folds, with the left-out fold used as test data. This is performed $k$ times so that all folds are used as test data, and the averaged errors over the $k$ folds are taken. \citet{Roberts2017} notes that performing such a strategy on spatial data results in serious underestimation of predictive errors due to the dependence between neighbouring observations. To avoid this problem leave-one-group-out (LOGO) cross validation is used in this paper, where each flow case in Table~\ref{tab:cases} is used as a fold. 

\subsection{Feature selection}
\label{sub:feature_select}

Generally, increasing the dimension $d$ of the input feature space increases the computational cost of training and predictions, and as noted by \citet{Lakshminarayanan2016b}, irrelevant features can be detrimental to Mondrian forest predictions. Therefore, backward elimination is used to remove less important features. At each iteration the least important feature $x_d$ in $\boldsymbol{x} = (x_{1},\dots,x_{D})^T$ is removed, and the forest is retrained on the reduced data. Feature importance is measured using permutation importance, proposed by \citet{Breiman2001}. The permutation importance $PI(x_d)$ quantifies the change in predictive error due to permuting the feature $x_d$ in the input data. 
 
\begin{figure}[ht]
	\centering
	\begin{subfigure}[b]{0.4\linewidth}
       	\includegraphics[width=\linewidth]{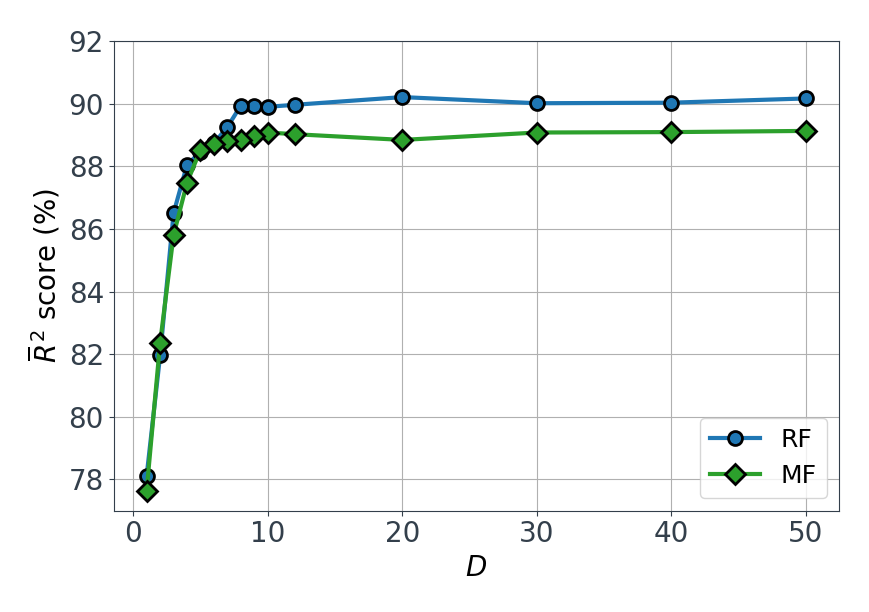}
        \caption{$\bar{R}^2$ score}
        \label{subfig:feature_selection_R2}
    \end{subfigure} \hspace{5pt}
    \begin{subfigure}[b]{0.4\linewidth}
        \includegraphics[width=\linewidth]{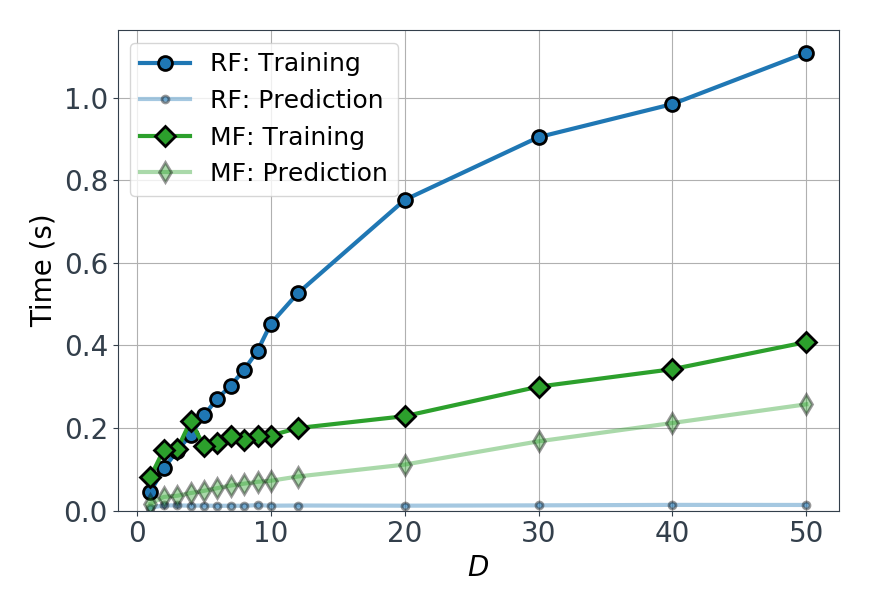}
        \caption{Training and prediction time}
        \label{subfig:feature_selection_time}
    \end{subfigure}
	\caption{Effect of number of features. Features are removed using backward elimination based on permutation importance. $\bar{R}^2$ scores are the mean scores from LOGO cross-validation. Results are averaged over three runs. Training and prediction times are measured on a personal laptop with a $8^{th}$ generation Intel\textsuperscript{\textregistered} Core\textsuperscript{\texttrademark} i5 processor.}
	\label{fig:feature_selection}
\end{figure}

Backward elimination is performed on Random forests and Mondrian forests, with features iteratively removed one at a time until $d=2$. For both forests the maximum depth is set at $\mathcal{D}_{max}=40$ and the number of trees at $M=160$. The mean $\bar{R}^2$ scores from LOGO cross-validation across all ten flow cases are plotted against $d$ in  Figure~\ref{subfig:feature_selection_R2}. The key points from this plot are as follows:

\begin{itemize}
\item The random forest (RF) generally has a $1-2\%$ higher $\bar{R}^2$ score than the Mondrian forest (MF), but the MF still returns competitive accuracy.

\item Many features can be removed without affecting the predictive accuracy of the MF or RF. This is perhaps unsurprising, since many of the 47 invariants are zero for two-dimensional flows. For $d<12$ the $\bar{R}^2$ scores begin to decrease as features containing important information are removed. 

\item The problem of irrelevant features degrading MF accuracy is not observed here. This may be because most of the non-zero features are relevant in the flow cases considered. 
\end{itemize}

Since the feature space can be reduced to $d=12$ with no significant loss of accuracy, this is done for the remainder of this paper. When backward elimination is halted at $d=12$ with the Mondrian forest, the following feature set is obtained:
\begin{equation} \label{eqn:feature_set_appen}
\boldsymbol{x} = \left\lbrace \widehat{\boldsymbol{S}}^2,\widehat{\boldsymbol{\Omega}}^2,\widehat{\mathbf{A}}_p^2, \widehat{\mathbf{A}}_k^2, \widehat{\boldsymbol{\Omega}}\widehat{\mathbf{A}}_k, \widehat{\boldsymbol{\Omega}}\widehat{\mathbf{A}}_k \widehat{\boldsymbol{S}}^2, \widehat{\boldsymbol{\Omega}}^2\widehat{\mathbf{A}}_k \widehat{\boldsymbol{S}}, \widehat{\mathbf{A}}_k^2\widehat{\mathbf{A}}_p\widehat{\boldsymbol{S}}, \widehat{\mathbf{A}}_k^2\widehat{\boldsymbol{S}}\widehat{\mathbf{A}}_p\widehat{\boldsymbol{S}}^2,Re_d, \hat{k},\hat{\omega} \right\rbrace^T,
\end{equation}
and it is used for all subsequent RF and MF computations in this paper. As shown in Figure~\ref{subfig:feature_selection_time}, the reduced feature space allows for reduced training/prediction times, with no appreciable loss in predictive accuracy.

\subsection{Hyper-parameter tuning}
\label{sub:hyperparams}

\begin{figure}[ht]
	\centering
	\includegraphics[width=0.28\linewidth]{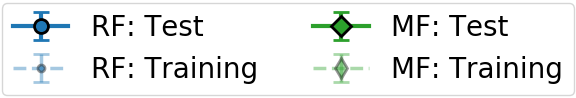} \\
	    \begin{subfigure}[b]{0.255\linewidth}
        \includegraphics[width=\linewidth]{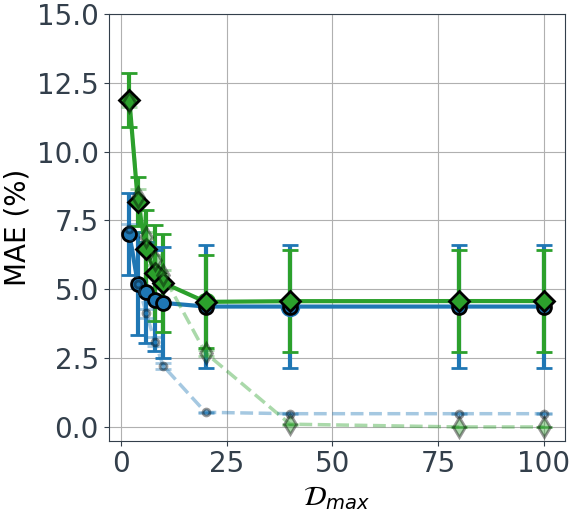}
        \caption{MAE, $M=100$}
        \label{subfig:gridsearch_Dmax}
    \end{subfigure} \hfil	
    \begin{subfigure}[b]{0.255\linewidth}
       	\includegraphics[width=\linewidth]{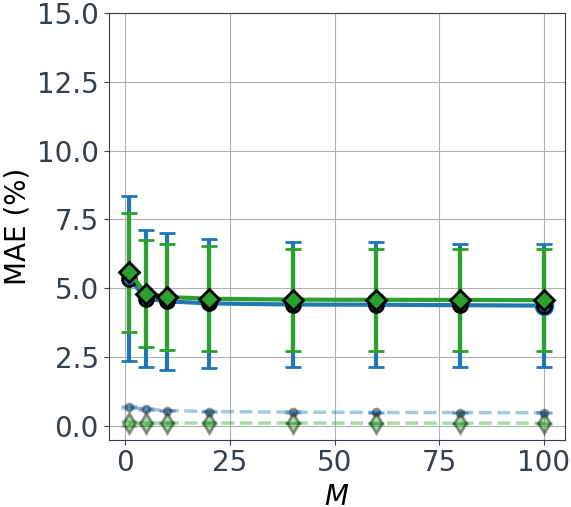}
        \caption{MAE, $\mathcal{D}_{max}=40$}
        \label{subfig:gridsearch_Ntrees}
    \end{subfigure} \hfil
        \begin{subfigure}[b]{0.225\linewidth}
        \includegraphics[width=\linewidth]{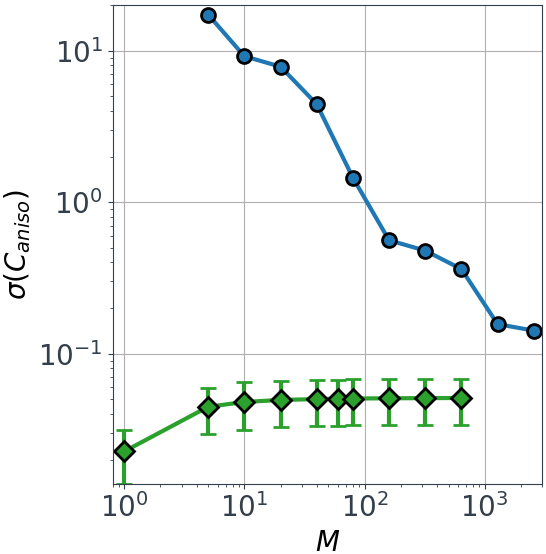}
        \caption{$\sigma(C_{aniso})$, $\mathcal{D}_{max}=40$}
        \label{subfig:gridsearch_Dmax40_std}
    \end{subfigure} \hfil
        \begin{subfigure}[b]{0.225\linewidth}
        \includegraphics[width=\linewidth]{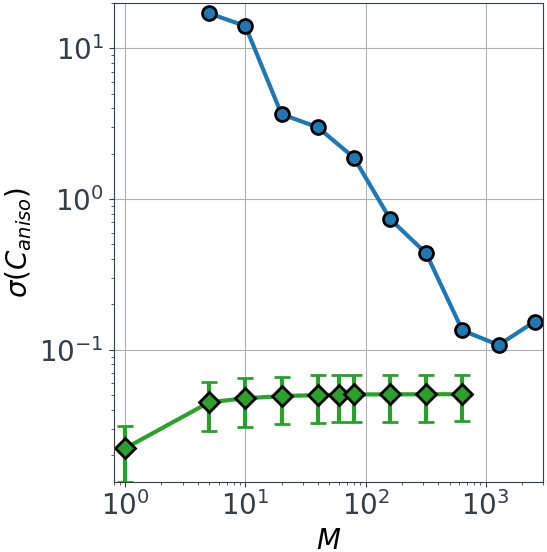}
        \caption{$\sigma(C_{aniso})$, $\mathcal{D}_{max}=\infty$}
        \label{subfig:gridsearch_Dmax999_std}
    \end{subfigure}
	\caption{Effect of hyper-parameters on MAE and predicted $\sigma(C_{aniso})$. Due to the computational cost of jackknifing, the RF results in c) and d) are for case 3b only. All other results are averaged across all ten folds of the LOGO cross-validation. Error bars represent standard deviation of MAE and predicted $\sigma(C_{aniso})$ across individual folds in cross-validation.}
	\label{fig:gridsearch}
\end{figure}

To determine suitable settings for the number of trees $M$ and maximum depth of trees $D_{max}$ an exhaustive grid search is performed. LOGO cross validation is performed for every combination of the hyperparameters in the sets $M=\left\lbrace 2,4,6,8,10,20,40,80,100 \right\rbrace$ and $\mathcal{D}_{max}=\left\lbrace 1,5,10,20,40,60,80,100 \right\rbrace$, and the resulting $\bar{R}^2$ scores are shown in Figures~\ref{subfig:gridsearch_Dmax} and~\ref{subfig:gridsearch_Ntrees}. Both RF's and MF's display similar trends here. The high training and test errors for low values of $\mathcal{D}_{max}$ in Figure~\ref{subfig:gridsearch_Dmax} indicate that the shallow trees suffer from under-fitting to the $C_{aniso}$ relationships in the data. Whereas the moderate difference between training and test errors at higher $\mathcal{D}_{max}$ values is possibly due to over-fitting and a lack of generalisation. Increasing $M$ (Fig.~\ref{subfig:gridsearch_Ntrees}) yields only a slight reduction in MAE, suggesting that if interpretability of the ML model is important single trees ($M=1$) could be used with only a slight increase in predictive errors.

A second grid search is performed over a larger range of $M$ and $\mathcal{D}_{max}$ to determine the sensitivity of the prediction uncertainty $\sigma(C_{aniso})$ to the hyperparameters. For the RF's $\sigma(C_{aniso})$ is estimated using the jackknife (Sec.~\ref{sub:jackknife}), while for the MF's $\sigma(C_{aniso})$ is naturally returned when predicting $C_{aniso}$ (see Sec.~\ref{sub:mondrian_forests}). Figures~\ref{subfig:gridsearch_Dmax40_std} and~\ref{subfig:gridsearch_Dmax999_std} show that the MF's achieve converged $\sigma$ estimates with a relatively low number of trees ($M > 160$). On the other hand, convergence isn't achieved with the RF's even with $M>1000$ trees. Increasing $M$ further (i.e. towards $N$) is impractical due to the significant computational cost of computing the infinitesimal jackknife for large values of $M$. 

Based on the aforementioned hyperparameter tuning, $M=160$ and $\mathcal{D}_{max}=40$ is chosen for all subsequent MF's to give low MAE's and converged $\sigma$ estimates. 

\section*{References}

\bibliography{library}

\begin{thebibliography}{49}
\expandafter\ifx\csname natexlab\endcsname\relax\def\natexlab#1{#1}\fi
\providecommand{\url}[1]{\texttt{#1}}
\providecommand{\href}[2]{#2}
\providecommand{\path}[1]{#1}
\providecommand{\DOIprefix}{doi:}
\providecommand{\ArXivprefix}{arXiv:}
\providecommand{\URLprefix}{URL: }
\providecommand{\Pubmedprefix}{pmid:}
\providecommand{\doi}[1]{\href{http://dx.doi.org/#1}{\path{#1}}}
\providecommand{\Pubmed}[1]{\href{pmid:#1}{\path{#1}}}
\providecommand{\bibinfo}[2]{#2}
\ifx\xfnm\relax \def\xfnm[#1]{\unskip,\space#1}\fi
\bibitem[{Wild(2015)}]{Wild2015}
\bibinfo{author}{J.~Wild},
\newblock \bibinfo{title}{{High-Performance High-Lift Design for Laminar
  Wings}},
\newblock in: \bibinfo{editor}{J.-P.~S. {D. Kn{\"{o}}rzer, C. Warsop, C.
  Diaconescu}} (Ed.), \bibinfo{booktitle}{Proc. Seventh Eur. Aeronaut. Days},
  \bibinfo{publisher}{Aviation in Europe - Innovation for Growth},
  \bibinfo{address}{London, UK}, \bibinfo{year}{2015}, pp.
  \bibinfo{pages}{305--310}. \DOIprefix\doi{10.2777/62810}.
\bibitem[{Tucker and DeBonis(2014)}]{Tucker2014a}
\bibinfo{author}{P.~G. Tucker}, \bibinfo{author}{J.~R. DeBonis},
\newblock \bibinfo{title}{Aerodynamics, computers and the environment},
\newblock \bibinfo{journal}{Philosophical Transactions of the Royal Society A:
  Mathematical, Physical and Engineering Sciences} \bibinfo{volume}{372}
  (\bibinfo{year}{2014}) \bibinfo{pages}{20130331}.
  \DOIprefix\doi{10.1098/rsta.2013.0331}.
\bibitem[{Brand et~al.(2011)Brand, Peinke, and Mann}]{Brand2011}
\bibinfo{author}{A.~J. Brand}, \bibinfo{author}{J.~Peinke},
  \bibinfo{author}{J.~Mann},
\newblock \bibinfo{title}{Turbulence and wind turbines},
\newblock \bibinfo{journal}{Journal of Physics: Conference Series}
  \bibinfo{volume}{318} (\bibinfo{year}{2011}) \bibinfo{pages}{072005}.
  \DOIprefix\doi{10.1088/1742-6596/318/7/072005}.
\bibitem[{Raynal et~al.(2016)Raynal, Augier, Bazer-Bachi, Haroun, and {Pereira
  Da Fonte}}]{Raynal2016}
\bibinfo{author}{L.~Raynal}, \bibinfo{author}{F.~Augier},
  \bibinfo{author}{F.~Bazer-Bachi}, \bibinfo{author}{Y.~Haroun},
  \bibinfo{author}{C.~{Pereira Da Fonte}}, \bibinfo{title}{{CFD Applied to
  Process Development in the Oil and Gas Industry - A Review}},
  \bibinfo{year}{2016}. \DOIprefix\doi{10.2516/ogst/2015019}.
\bibitem[{Scillitoe et~al.(2019)Scillitoe, Tucker, and Adami}]{Scillitoe2019}
\bibinfo{author}{A.~D. Scillitoe}, \bibinfo{author}{P.~G. Tucker},
  \bibinfo{author}{P.~Adami},
\newblock \bibinfo{title}{{Large Eddy Simulation of Boundary Layer Transition
  Mechanisms in a Gas-Turbine Compressor Cascade}},
\newblock \bibinfo{journal}{J. Turbomach.} \bibinfo{volume}{141}
  (\bibinfo{year}{2019}) \bibinfo{pages}{1--10}.
  \DOIprefix\doi{10.1115/1.4042023}.
\bibitem[{Mehta et~al.(2014)Mehta, van Zuijlen, Koren, Holierhoek, and
  Bijl}]{Mehta2014}
\bibinfo{author}{D.~Mehta}, \bibinfo{author}{A.~H. van Zuijlen},
  \bibinfo{author}{B.~Koren}, \bibinfo{author}{J.~G. Holierhoek},
  \bibinfo{author}{H.~Bijl},
\newblock \bibinfo{title}{{Large Eddy Simulation of wind farm aerodynamics: A
  review}},
\newblock \bibinfo{journal}{J. Wind Eng. Ind. Aerodyn.} \bibinfo{volume}{133}
  (\bibinfo{year}{2014}) \bibinfo{pages}{1--17}.
  \DOIprefix\doi{10.1016/j.jweia.2014.07.002}.
\bibitem[{Blocken(2018)}]{Blocken2018}
\bibinfo{author}{B.~Blocken},
\newblock \bibinfo{title}{{LES over RANS in building simulation for outdoor and
  indoor applications: A foregone conclusion?}},
\newblock \bibinfo{journal}{Build. Simul.} \bibinfo{volume}{11}
  (\bibinfo{year}{2018}) \bibinfo{pages}{821--870}.
  \DOIprefix\doi{10.1007/s12273-018-0459-3}.
\bibitem[{Hunt and Savill(2005)}]{Hunt2005}
\bibinfo{author}{J.~C. Hunt}, \bibinfo{author}{A.~M. Savill},
  \bibinfo{title}{{Guidelines and criteria for the use of turbulence models in
  complex flows}}, \bibinfo{year}{2005}.
  \DOIprefix\doi{10.1017/CBO9780511543227.008}.
\bibitem[{Tucker(2013)}]{Tucker2013}
\bibinfo{author}{P.~G. Tucker},
\newblock \bibinfo{title}{{Trends in turbomachinery turbulence treatments}},
\newblock \bibinfo{journal}{Prog. Aerosp. Sci.} \bibinfo{volume}{63}
  (\bibinfo{year}{2013}) \bibinfo{pages}{1--32}.
  \DOIprefix\doi{10.1016/j.paerosci.2013.06.001}.
\bibitem[{Duraisamy et~al.(2019)Duraisamy, Iaccarino, and Xiao}]{Duraisamy2019}
\bibinfo{author}{K.~Duraisamy}, \bibinfo{author}{G.~Iaccarino},
  \bibinfo{author}{H.~Xiao},
\newblock \bibinfo{title}{{Turbulence Modeling in the Age of Data}},
\newblock \bibinfo{journal}{Annu. Rev. Fluid Mech.} \bibinfo{volume}{51}
  (\bibinfo{year}{2019}) \bibinfo{pages}{357--377}.
  \DOIprefix\doi{10.1146/annurev-fluid-010518-040547}.
\bibitem[{Ling and Templeton(2015)}]{Ling2015}
\bibinfo{author}{J.~Ling}, \bibinfo{author}{J.~Templeton},
\newblock \bibinfo{title}{{Evaluation of machine learning algorithms for
  prediction of regions of high Reynolds averaged Navier Stokes uncertainty}},
\newblock \bibinfo{journal}{Phys. Fluids} \bibinfo{volume}{27}
  (\bibinfo{year}{2015}) \bibinfo{pages}{85103}.
  \DOIprefix\doi{10.1063/1.4927765}.
\bibitem[{Ling et~al.(2016)Ling, Kurzawski, and Templeton}]{Ling2016}
\bibinfo{author}{J.~Ling}, \bibinfo{author}{A.~Kurzawski},
  \bibinfo{author}{J.~Templeton},
\newblock \bibinfo{title}{{Reynolds averaged turbulence modelling using deep
  neural networks with embedded invariance}},
\newblock \bibinfo{journal}{J. Fluid Mech} \bibinfo{volume}{807}
  (\bibinfo{year}{2016}) \bibinfo{pages}{155--166}.
  \DOIprefix\doi{10.1017/jfm.2016.615}.
\bibitem[{Singh et~al.(2017)Singh, Matai, Mishra, Duraisamy, and
  Durbin}]{Singh2017}
\bibinfo{author}{A.~P. Singh}, \bibinfo{author}{R.~Matai},
  \bibinfo{author}{A.~Mishra}, \bibinfo{author}{K.~Duraisamy},
  \bibinfo{author}{P.~A. Durbin},
\newblock \bibinfo{title}{Data-driven augmentation of turbulence models for
  adverse pressure gradient flows},
\newblock in: \bibinfo{booktitle}{23rd {AIAA} Computational Fluid Dynamics
  Conference}, \bibinfo{publisher}{American Institute of Aeronautics and
  Astronautics}, \bibinfo{year}{2017}, pp. \bibinfo{pages}{1--20}.
  \DOIprefix\doi{10.2514/6.2017-3626}.
\bibitem[{Wu et~al.(2018)Wu, Xiao, and Paterson}]{Wu2018}
\bibinfo{author}{J.~L. Wu}, \bibinfo{author}{H.~Xiao},
  \bibinfo{author}{E.~Paterson},
\newblock \bibinfo{title}{{Physics-informed machine learning approach for
  augmenting turbulence models: A comprehensive framework}},
\newblock \bibinfo{journal}{Phys. Rev. Fluids} \bibinfo{volume}{7}
  (\bibinfo{year}{2018}) \bibinfo{pages}{1--42}.
  \DOIprefix\doi{10.1103/PhysRevFluids.3.074602}.
\bibitem[{Kaandorp and Dwight(2020)}]{Kaandorp2020}
\bibinfo{author}{M.~L. Kaandorp}, \bibinfo{author}{R.~P. Dwight},
\newblock \bibinfo{title}{Data-driven modelling of the reynolds stress tensor
  using random forests with invariance},
\newblock \bibinfo{journal}{Computers {\&} Fluids} \bibinfo{volume}{202}
  (\bibinfo{year}{2020}) \bibinfo{pages}{104497}.
  \DOIprefix\doi{10.1016/j.compfluid.2020.104497}.
\bibitem[{Wolpert(1996)}]{Wolpert1996}
\bibinfo{author}{D.~H. Wolpert},
\newblock \bibinfo{title}{{The Lack of a Priori Distinctions between Learning
  Algorithms}},
\newblock \bibinfo{journal}{Neural Comput.} \bibinfo{volume}{8}
  (\bibinfo{year}{1996}) \bibinfo{pages}{1341--1390}.
  \DOIprefix\doi{10.1162/neco.1996.8.7.1341}.
\bibitem[{Wager et~al.(2014)Wager, Hastie, and Efron}]{Wager2014}
\bibinfo{author}{S.~Wager}, \bibinfo{author}{T.~Hastie},
  \bibinfo{author}{B.~Efron},
\newblock \bibinfo{title}{{Confidence intervals for random forests: The
  jackknife and the infinitesimal jackknife}},
\newblock \bibinfo{journal}{J. Mach. Learn. Res.} \bibinfo{volume}{15}
  (\bibinfo{year}{2014}) \bibinfo{pages}{1625--1651}.
\bibitem[{Wu et~al.(2017)Wu, Wang, Xiao, and Ling}]{Wu2017}
\bibinfo{author}{J.~L. Wu}, \bibinfo{author}{J.~X. Wang},
  \bibinfo{author}{H.~Xiao}, \bibinfo{author}{J.~Ling},
\newblock \bibinfo{title}{{A Priori Assessment of Prediction Confidence for
  Data-Driven Turbulence Modeling}},
\newblock \bibinfo{journal}{Flow, Turbul. Combust.} \bibinfo{volume}{99}
  (\bibinfo{year}{2017}) \bibinfo{pages}{25--46}.
  \DOIprefix\doi{10.1007/s10494-017-9807-0}.
\bibitem[{Parish and Duraisamy(2016)}]{Parish2016}
\bibinfo{author}{E.~J. Parish}, \bibinfo{author}{K.~Duraisamy},
\newblock \bibinfo{title}{{A paradigm for data-driven predictive modeling using
  field inversion and machine learning}},
\newblock \bibinfo{journal}{J. Comput. Phys.} \bibinfo{volume}{305}
  (\bibinfo{year}{2016}) \bibinfo{pages}{758--774}.
  \DOIprefix\doi{10.1016/j.jcp.2015.11.012}.
\bibitem[{Geneva and Zabaras(2019)}]{Geneva2019}
\bibinfo{author}{N.~Geneva}, \bibinfo{author}{N.~Zabaras},
\newblock \bibinfo{title}{{Quantifying model form uncertainty in
  Reynolds-averaged turbulence models with Bayesian deep neural networks}},
\newblock \bibinfo{journal}{J. Comput. Phys.} \bibinfo{volume}{383}
  (\bibinfo{year}{2019}) \bibinfo{pages}{125--147}.
  \DOIprefix\doi{10.1016/j.jcp.2019.01.021}.
\bibitem[{Blauw and Dwight(2019)}]{blauw2019}
\bibinfo{author}{C.~A. Blauw}, \bibinfo{author}{R.~P. Dwight},
  \bibinfo{title}{{Bayesian Additive Regression Trees for data-driven RANS
  turbulence modelling}}, Ph.D. thesis, Delft University of Technology,
  \bibinfo{year}{2019}.
\bibitem[{Hensman et~al.(2013)Hensman, Fusi, and Lawrence}]{Hensman2013}
\bibinfo{author}{J.~Hensman}, \bibinfo{author}{N.~Fusi}, \bibinfo{author}{N.~D.
  Lawrence},
\newblock \bibinfo{title}{Gaussian processes for big data},
\newblock in: \bibinfo{booktitle}{Proceedings of the Twenty-Ninth Conference on
  Uncertainty in Artificial Intelligence}, UAI'13, \bibinfo{publisher}{AUAI
  Press}, \bibinfo{address}{Arlington, Virginia, USA}, \bibinfo{year}{2013}, p.
  \bibinfo{pages}{282–290}.
\bibitem[{Lakshminarayanan et~al.(2017)Lakshminarayanan, Pritzel, and
  Blundell}]{Lakshminarayanan2017}
\bibinfo{author}{B.~Lakshminarayanan}, \bibinfo{author}{A.~Pritzel},
  \bibinfo{author}{C.~Blundell},
\newblock \bibinfo{title}{Simple and scalable predictive uncertainty estimation
  using deep ensembles},
\newblock in: \bibinfo{booktitle}{Proceedings of the 31st International
  Conference on Neural Information Processing Systems}, NIPS'17,
  \bibinfo{publisher}{Curran Associates Inc.}, \bibinfo{address}{Red Hook, NY,
  USA}, \bibinfo{year}{2017}, p. \bibinfo{pages}{6405–6416}.
\bibitem[{Lakshminarayanan et~al.(2015)Lakshminarayanan, Roy, and
  Teh}]{Lakshminarayanan2015}
\bibinfo{author}{B.~Lakshminarayanan}, \bibinfo{author}{D.~Roy},
  \bibinfo{author}{Y.~W. Teh},
\newblock \bibinfo{title}{Particle gibbs for bayesian additive regression
  trees},
\newblock in: \bibinfo{booktitle}{Artificial Intelligence and Statistics},
  \bibinfo{year}{2015}, pp. \bibinfo{pages}{553--561}.
\bibitem[{Lakshminarayanan et~al.(2016{\natexlab{a}})Lakshminarayanan, Roy, and
  Teh}]{Lakshminarayanan2016a}
\bibinfo{author}{B.~Lakshminarayanan}, \bibinfo{author}{D.~M. Roy},
  \bibinfo{author}{Y.~W. Teh},
\newblock \bibinfo{title}{{Mondrian Forests: Efficient Online Random Forests}},
\newblock in: \bibinfo{booktitle}{Proc. 19th Int. Conf. Artif. Intell. Stat.},
  \bibinfo{address}{Cadiz, Spain}, \bibinfo{year}{2016}{\natexlab{a}}, pp.
  \bibinfo{pages}{1--15}.
\bibitem[{Lakshminarayanan et~al.(2016{\natexlab{b}})Lakshminarayanan, Roy, and
  Teh}]{Lakshminarayanan2016b}
\bibinfo{author}{B.~Lakshminarayanan}, \bibinfo{author}{D.~M. Roy},
  \bibinfo{author}{Y.~W. Teh},
\newblock \bibinfo{title}{{Mondrian forests for large-scale regression when
  uncertainty matters}},
\newblock \bibinfo{journal}{Proc. 19th Int. Conf. Artif. Intell. Stat. AISTATS
  2016} \bibinfo{volume}{51} (\bibinfo{year}{2016}{\natexlab{b}})
  \bibinfo{pages}{1478--1487}.
\bibitem[{Xiao and Cinnella(2019)}]{Xiao2019}
\bibinfo{author}{H.~Xiao}, \bibinfo{author}{P.~Cinnella},
\newblock \bibinfo{title}{{Quantification of Model Uncertainty in RANS
  Simulations: A Review}},
\newblock \bibinfo{journal}{Prog. Aerosp. Sci.} \bibinfo{volume}{108}
  (\bibinfo{year}{2019}) \bibinfo{pages}{1--31}.
  \DOIprefix\doi{10.1016/j.paerosci.2018.10.001}.
\bibitem[{Tucker(2014)}]{Tucker2014}
\bibinfo{author}{P.~Tucker}, \bibinfo{title}{Unsteady Computational Fluid
  Dynamics in Aeronautics}, \bibinfo{publisher}{Springer Netherlands},
  \bibinfo{year}{2014}. \DOIprefix\doi{10.1007/978-94-007-7049-2}.
\bibitem[{Wu et~al.(2019)Wu, Xiao, Sun, and Wang}]{Wu2019}
\bibinfo{author}{J.~Wu}, \bibinfo{author}{H.~Xiao}, \bibinfo{author}{R.~Sun},
  \bibinfo{author}{Q.~Wang},
\newblock \bibinfo{title}{{Reynolds-averaged Navier-Stokes equations with
  explicit data-driven Reynolds stress closure can be ill-conditioned}},
\newblock \bibinfo{journal}{J. Fluid Mech.} \bibinfo{volume}{869}
  (\bibinfo{year}{2019}) \bibinfo{pages}{553--586}.
  \DOIprefix\doi{10.1017/jfm.2019.205}.
\bibitem[{Breiman(2001)}]{Breiman2001}
\bibinfo{author}{L.~Breiman},
\newblock \bibinfo{title}{{Random Forests}},
\newblock \bibinfo{journal}{Mach. Learn.} \bibinfo{volume}{45}
  (\bibinfo{year}{2001}) \bibinfo{pages}{5--32}.
  \DOIprefix\doi{10.3390/rs10060911}.
\bibitem[{Breiman et~al.(1984)Breiman, Friedman, Stone, and
  Olshen}]{Breiman1984}
\bibinfo{author}{L.~Breiman}, \bibinfo{author}{J.~Friedman},
  \bibinfo{author}{C.~J. Stone}, \bibinfo{author}{R.~A. Olshen},
  \bibinfo{title}{{Classification and Regression Trees}},
  \bibinfo{edition}{1st} ed., \bibinfo{publisher}{Chapman {\&} Hall},
  \bibinfo{year}{1984}.
\bibitem[{Economon(2018)}]{Economon2018}
\bibinfo{author}{T.~D. Economon},
\newblock \bibinfo{title}{Simulation and adjoint-based design for variable
  density incompressible flows with heat transfer},
\newblock in: \bibinfo{booktitle}{2018 Multidisciplinary Analysis and
  Optimization Conference}, \bibinfo{publisher}{American Institute of
  Aeronautics and Astronautics}, \bibinfo{year}{2018}, pp.
  \bibinfo{pages}{1--24}. \DOIprefix\doi{10.2514/6.2018-3111}.
\bibitem[{Menter et~al.(2003)Menter, Kuntz, and Langtry}]{Menter2003a}
\bibinfo{author}{F.~R. Menter}, \bibinfo{author}{M.~Kuntz},
  \bibinfo{author}{R.~Langtry},
\newblock \bibinfo{title}{{Ten Years of Industrial Experience with the SST
  Turbulence Model}},
\newblock \bibinfo{journal}{Turbul. Heat Mass Transf. 4} \bibinfo{volume}{4}
  (\bibinfo{year}{2003}) \bibinfo{pages}{625--632}.
\bibitem[{Bentaleb et~al.(2012)Bentaleb, Lardeau, and
  Leschziner}]{Bentaleb2012}
\bibinfo{author}{Y.~Bentaleb}, \bibinfo{author}{S.~Lardeau},
  \bibinfo{author}{M.~A. Leschziner},
\newblock \bibinfo{title}{{Large-eddy simulation of turbulent boundary layer
  separation from a rounded step}},
\newblock \bibinfo{journal}{J. Turbul.}  (\bibinfo{year}{2012}).
  \DOIprefix\doi{10.1080/14685248.2011.637923}.
\bibitem[{Fr{\"{o}}hlich et~al.(2005)Fr{\"{o}}hlich, Mellen, Rodi, Temmerman,
  and Lescheziner}]{Frohlich2005}
\bibinfo{author}{J.~Fr{\"{o}}hlich}, \bibinfo{author}{C.~P. Mellen},
  \bibinfo{author}{W.~Rodi}, \bibinfo{author}{L.~Temmerman},
  \bibinfo{author}{M.~A. Lescheziner},
\newblock \bibinfo{title}{{Highly resolved large-eddy simulation of separated
  flow in a channel with streamwise periodic constrictions}},
\newblock \bibinfo{journal}{J. Fluid Mech.} \bibinfo{volume}{526}
  (\bibinfo{year}{2005}) \bibinfo{pages}{19--66}.
  \DOIprefix\doi{10.1017/S0022112004002812}.
\bibitem[{Laval and Marquillie(2011)}]{Laval2011}
\bibinfo{author}{J.~P. Laval}, \bibinfo{author}{M.~Marquillie},
\newblock \bibinfo{title}{{Direct Numerical Simulations of
  Converging–Diverging Channel Flow}},
\newblock \bibinfo{journal}{Prog. Wall Turbul. Underst. Model. ERCOFTAC Ser.}
  \bibinfo{volume}{14} (\bibinfo{year}{2011}) \bibinfo{pages}{203--209}.
\bibitem[{Schiavo et~al.(2015)Schiavo, Jesus, Azevedo, and Wolf}]{Schiavo2015}
\bibinfo{author}{L.~A. Schiavo}, \bibinfo{author}{A.~B. Jesus},
  \bibinfo{author}{J.~L. Azevedo}, \bibinfo{author}{W.~R. Wolf},
\newblock \bibinfo{title}{{Large Eddy Simulations of convergent–divergent
  channel flows at moderate Reynolds numbers}},
\newblock \bibinfo{journal}{Int. J. Heat Fluid Flow} \bibinfo{volume}{56}
  (\bibinfo{year}{2015}) \bibinfo{pages}{137--151}.
  \DOIprefix\doi{10.1016/J.IJHEATFLUIDFLOW.2015.07.006}.
\bibitem[{Vinuesa et~al.(2018)Vinuesa, Negi, Atzori, Hanifi, Henningson, and
  Schlatter}]{Vinuesa2018}
\bibinfo{author}{R.~Vinuesa}, \bibinfo{author}{P.~S. Negi},
  \bibinfo{author}{M.~Atzori}, \bibinfo{author}{A.~Hanifi},
  \bibinfo{author}{D.~S. Henningson}, \bibinfo{author}{P.~Schlatter},
  \bibinfo{title}{{Turbulent boundary layers around wing sections up to
  Rec=1,000,000}}, \bibinfo{year}{2018}.
  \DOIprefix\doi{10.1016/j.ijheatfluidflow.2018.04.017}.
\bibitem[{Tyacke and Tucker(2012)}]{Tyacke2012}
\bibinfo{author}{J.~Tyacke}, \bibinfo{author}{P.~Tucker},
\newblock \bibinfo{title}{{LES} of heat transfer in electronics},
\newblock \bibinfo{journal}{Applied Mathematical Modelling}
  \bibinfo{volume}{36} (\bibinfo{year}{2012}) \bibinfo{pages}{3112--3133}.
  \DOIprefix\doi{10.1016/j.apm.2011.09.072}.
\bibitem[{Pope(2000)}]{Pope2000}
\bibinfo{author}{S.~B. Pope}, \bibinfo{title}{{Turbulent Flows}},
  volume~\bibinfo{volume}{1}, \bibinfo{year}{2000}.
  \DOIprefix\doi{10.1088/1468-5248/1/1/702}.
\bibitem[{Banerjee et~al.(2007)Banerjee, Krahl, Durst, and
  Zenger}]{Banerjee2007}
\bibinfo{author}{S.~Banerjee}, \bibinfo{author}{R.~Krahl},
  \bibinfo{author}{F.~Durst}, \bibinfo{author}{C.~Zenger},
\newblock \bibinfo{title}{Presentation of anisotropy properties of turbulence,
  invariants versus eigenvalue approaches},
\newblock \bibinfo{journal}{Journal of Turbulence} \bibinfo{volume}{8}
  (\bibinfo{year}{2007}) \bibinfo{pages}{1--32}.
  \DOIprefix\doi{10.1080/14685240701506896}.
\bibitem[{Shih et~al.(1995)Shih, Zhu, and Lumley}]{Shih1995}
\bibinfo{author}{T.-H. Shih}, \bibinfo{author}{J.~Zhu}, \bibinfo{author}{J.~L.
  Lumley},
\newblock \bibinfo{title}{A new reynolds stress algebraic equation model},
\newblock \bibinfo{journal}{Computer Methods in Applied Mechanics and
  Engineering} \bibinfo{volume}{125} (\bibinfo{year}{1995})
  \bibinfo{pages}{287--302}. \DOIprefix\doi{10.1016/0045-7825(95)00796-4}.
\bibitem[{Hellsten and Wallin(2009)}]{Hellsten2009}
\bibinfo{author}{A.~Hellsten}, \bibinfo{author}{S.~Wallin},
\newblock \bibinfo{title}{Explicit algebraic reynolds stress and non-linear
  eddy-viscosity models},
\newblock \bibinfo{journal}{International Journal of Computational Fluid
  Dynamics} \bibinfo{volume}{23} (\bibinfo{year}{2009})
  \bibinfo{pages}{349--361}. \DOIprefix\doi{10.1080/10618560902776828}.
\bibitem[{Lundberg and Lee(2017)}]{Lundberg2017}
\bibinfo{author}{S.~M. Lundberg}, \bibinfo{author}{S.-I. Lee},
\newblock \bibinfo{title}{A unified approach to interpreting model
  predictions},
\newblock in: \bibinfo{editor}{I.~Guyon}, \bibinfo{editor}{U.~V. Luxburg},
  \bibinfo{editor}{S.~Bengio}, \bibinfo{editor}{H.~Wallach},
  \bibinfo{editor}{R.~Fergus}, \bibinfo{editor}{S.~Vishwanathan},
  \bibinfo{editor}{R.~Garnett} (Eds.), \bibinfo{booktitle}{Advances in Neural
  Information Processing Systems}, volume~\bibinfo{volume}{30},
  \bibinfo{publisher}{Curran Associates, Inc.}, \bibinfo{year}{2017}, pp.
  \bibinfo{pages}{4765--4774}.
\bibitem[{Lundberg et~al.(2020)Lundberg, Erion, Chen, DeGrave, Prutkin, Nair,
  Katz, Himmelfarb, Bansal, and Lee}]{Lundberg2020}
\bibinfo{author}{S.~M. Lundberg}, \bibinfo{author}{G.~Erion},
  \bibinfo{author}{H.~Chen}, \bibinfo{author}{A.~DeGrave},
  \bibinfo{author}{J.~M. Prutkin}, \bibinfo{author}{B.~Nair},
  \bibinfo{author}{R.~Katz}, \bibinfo{author}{J.~Himmelfarb},
  \bibinfo{author}{N.~Bansal}, \bibinfo{author}{S.-I. Lee},
\newblock \bibinfo{title}{From local explanations to global understanding with
  explainable {AI} for trees},
\newblock \bibinfo{journal}{Nature Machine Intelligence} \bibinfo{volume}{2}
  (\bibinfo{year}{2020}) \bibinfo{pages}{56--67}.
  \DOIprefix\doi{10.1038/s42256-019-0138-9}.
\bibitem[{Gorl{\'{e}} et~al.(2019)Gorl{\'{e}}, Zeoli, Emory, Larsson, and
  Iaccarino}]{Gorle2019}
\bibinfo{author}{C.~Gorl{\'{e}}}, \bibinfo{author}{S.~Zeoli},
  \bibinfo{author}{M.~Emory}, \bibinfo{author}{J.~Larsson},
  \bibinfo{author}{G.~Iaccarino},
\newblock \bibinfo{title}{{Epistemic uncertainty quantification for
  Reynolds-averaged Navier-Stokes modeling of separated flows over streamlined
  surfaces}},
\newblock \bibinfo{journal}{Phys. Fluids} \bibinfo{volume}{31}
  (\bibinfo{year}{2019}). \DOIprefix\doi{10.1063/1.5086341}.
\bibitem[{Xiao et~al.(2017)Xiao, Wang, and Ghanem}]{Xiao2017}
\bibinfo{author}{H.~Xiao}, \bibinfo{author}{J.~X. Wang}, \bibinfo{author}{R.~G.
  Ghanem},
\newblock \bibinfo{title}{{A random matrix approach for quantifying model-form
  uncertainties in turbulence modeling}},
\newblock \bibinfo{journal}{Comput. Methods Appl. Mech. Eng.}
  (\bibinfo{year}{2017}). \DOIprefix\doi{10.1016/j.cma.2016.10.025}.
\bibitem[{Mourtada et~al.(2019)Mourtada, Ga{\"{i}}ffas, and
  Scornet}]{Mourtada2019}
\bibinfo{author}{J.~Mourtada}, \bibinfo{author}{S.~Ga{\"{i}}ffas},
  \bibinfo{author}{E.~Scornet},
\newblock \bibinfo{title}{{AMF: Aggregated Mondrian Forests for Online
  Learning}},
\newblock \bibinfo{journal}{arXiv:1906.10529v1 [stat.ML]}
  (\bibinfo{year}{2019}) \bibinfo{pages}{1--32}.
\bibitem[{Roberts et~al.(2017)Roberts, Bahn, Ciuti, Boyce, Elith,
  Guillera-Arroita, Hauenstein, Lahoz-Monfort, Schr{\"{o}}der, Thuiller,
  Warton, Wintle, Hartig, and Dormann}]{Roberts2017}
\bibinfo{author}{D.~R. Roberts}, \bibinfo{author}{V.~Bahn},
  \bibinfo{author}{S.~Ciuti}, \bibinfo{author}{M.~S. Boyce},
  \bibinfo{author}{J.~Elith}, \bibinfo{author}{G.~Guillera-Arroita},
  \bibinfo{author}{S.~Hauenstein}, \bibinfo{author}{J.~J. Lahoz-Monfort},
  \bibinfo{author}{B.~Schr{\"{o}}der}, \bibinfo{author}{W.~Thuiller},
  \bibinfo{author}{D.~I. Warton}, \bibinfo{author}{B.~A. Wintle},
  \bibinfo{author}{F.~Hartig}, \bibinfo{author}{C.~F. Dormann},
  \bibinfo{title}{{Cross-validation strategies for data with temporal, spatial,
  hierarchical, or phylogenetic structure}}, \bibinfo{year}{2017}.
  \DOIprefix\doi{10.1111/ecog.02881}.

\end{thebibliography}

\end{document}